**Perspiration vapor lightens near-skin air—but hinders human evaporative cooling in arid heat**

Shri H. Viswanathan[1], Ankit Joshi,[1,2,] Isabella DeClair,[1] Bryce Twidwell,[1] Muhammad Abdullah,[3] Lyle Bartels,[1,4] Faisal Abedin,[1] Joseph Rotella,[1] Cibin T. Jose,[1] and Konrad Rykaczewski[1,2]

1. School for Engineering of Matter, Transport and Energy, Arizona State University, Tempe, AZ 85287, USA
2. Julie Ann Wrigley Global Futures Laboratory, Arizona State University, Tempe, AZ 85287, USA
3. School of Sustainability, Arizona State University, Tempe, AZ, USA
4. Thermetrics, LLC, Seattle, WA, USA

**Corresponding author:** Konrad Rykaczewski
**Email:**konradr@asu.edu

**Author Contributions:**
K.R. and S.H.V. designed research; S.H.V., A.J., I.D.C., B.T., M.A., L.B., C.T.J., F.A., J.R. performed research; S.H.V. and K.R., analyzed data and wrote the paper; all authors performed review and editing of the paper

**Abstract**
Sweat evaporation is the body's primary cooling mechanism, yet the physical factors governing it are not fully understood. We identify a "dueling buoyancy" effect in the context of the human body, in which perspiration vapor reduces the near-skin air density, counteracting the downward flow driven by cooling of warm air upon contact with the skin. In hot, arid, stagnant environments, this opposing buoyancy suppresses free convection and can reduce sweat evaporation by more than half. As a result, commonly used thermoregulation models can substantially underpredict body temperature (e.g., by 1 °C after 2-hours of exposure to typical Arizona summer conditions). We develop compact, physics-informed models for free-convective heat transfer coefficients across wide temperature and humidity ranges, enabling improved thermoregulation modeling and thermal audits. These results enhance understanding of human heat balance and support more accurate heat-stress assessment to inform behavioral, infrastructural, and policy decisions for extreme-heat adaptations.

**Acknowledgements**
We sincerely thank Assistant Professor Yihang Lu of Zhejiang University of Water Resources and Electric Power for identifying temperature conversion and indexing errors in the thermoregulation code in the original publication. Correcting these errors resulted in a minor 2 °C shift in the ambient temperature at which the physiological responses shown in Figure 4 occur; these corrections have been incorporated into the version below. Figures S22 and S23 and related numerical values in the discussion were also updated. We thank and gratefully acknowledge Research Computing at Arizona State University for providing high-performance computing resources that have contributed to the research results reported within this paper.

## Introduction

Overheating poses an escalating threat to human health and productivity. Sweat evaporation is the body's principal cooling pathway, yet we will show that quantitative understanding of this process in free convective conditions remains incomplete. This knowledge gap follows a historical trend: comprehension of sweat evaporation and its key role in thermoregulation often lagged other scientific advancements. For instance, in 1912, while Einstein was formulating the general theory of relativity, the Portuguese marathoner Francisco Lázaro (*1*) believed that sweating hindered performance and coated his body with wax and fat, fatally succumbing to heatstroke during the Olympic race. A decade later, Lewis (*2*) introduced the classical theory linking convective heat and mass transfer (see **Supplementary Materials** for details), but it took two decades for it to be applied to estimate human sweat evaporation rates (*3*, *4*). Assuming sweat to be an isothermal film at the temperature of the underlying skin, which is valid for low to moderate evaporation rates after about 15 minutes from the onset of sweating (*5*, *6*)—the Lewis analogy states that the evaporative heat flux ($Q_e$, W·m$^{-2}$) can be expressed as (*7*, *8*):

$$Q_e = h_e(p_{sk} - p_{air})\omega = 16.5h_c(p_{sk} - p_{air})\omega \qquad \textbf{Eq.1}$$

Where $\omega$ is the fraction of wet skin, $h_c$ (W·m$^{-2}$·K$^{-1}$) is the convective heat transfer coefficient, and $h_e$ (W·m$^{-2}$·kPa$^{-1}$) is the corresponding evaporative mass transfer coefficient with units adjusted for the driving humidity difference expressed in terms of partial pressure of water vapor (kPa) of skin ($p_{sk} = C_{sk}R(T_{sk} + 273.15)/M_w$ where $C_{sk}$ is the vapor concentration (kg·m$^{-3}$), $R$ = 8.314 J·mol$^{-1}$·K$^{-1}$ is the universal gas constant, and $M_w$ = 18.02 g·mol$^{-1}$ is the molecular weight of water) and ambient air ($p_{air}$). For forced convection, this direct relationship between $h_c$ and $h_e$ can be readily confirmed using a thermal manikin in a climate-controlled chamber with a wind tunnel or its digital twin (**Fig. 1A-B**, see **Supplementary Materials** for details**, Fig. S1A–F**, and **Fig. S2**). In contrast, we will show that, under free convection in arid heat conditions, the standard "textbook" models for the human-specific heat transfer coefficients employed for over seven decades diverge substantially from experimental observations.

At negligible wind speeds, the difference in air density at the skin surface and in the environment creates a buoyant force that drives a near-skin airflow and free convective heat and mass transfer. Since the 1950s (*7*, *9*, *10*), free-convective heat transfer, and therefore $Q_e$, from the human body has been modeled solely as a consequence of thermal buoyancy, driven by skin-to-air temperature differences ($\Delta T_{sk-air} = T_{sk} - T_{air}$). When ambient air temperature exceeds dry skin temperature, the cooled air adjacent to the skin becomes denser and descends, generating downward convection. However, in hot and dry conditions, humidity from evaporating sweat exerts an equally important but opposing influence on air density near wet skin (**Fig. 1C**). Specifically, humid air is lighter than dry air—saturated air at 35°C near wet skin is 1.9% lighter than dry air at the same temperature but 10% relative humidity ($RH_{air}$)—producing upward buoyant flow. To thermally produce an equivalent decrease in density while maintaining the absolute water vapor content, the 35°C, 10% $RH_{air}$ air would have to be heated by 6°C. For wet skin surrounded by hotter, drier air, these thermal (downward) and humidity (upward)- driven buoyancies oppose one another, at times canceling to yield neutral buoyancy and flow stagnation (e.g., 41 °C air at 10% $RH_{air}$). In contrast, at lower air temperatures, the two buoyancies can be aligned. Such "dueling" or "reinforcing" buoyancy phenomena are well recognized in engineering (*11–14*). However, despite numerous prior efforts to model free convective flow, sweat evaporation, and moisture transport around the human body (e.g., in the contexts of indoor thermal comfort (*15*, *16*), personal ventilation (*17*), ventilated garments (*18*, *19*), helmets (*20*), and breathing (*21*)), human physiological models and heat stress assessments have largely overlooked humidity-driven buoyancy in heat and mass transfer processes (*8*).

Although forced convection dominates in outdoor heat-stress scenarios (*22*, *23*), free or mixed convection is more relevant in most indoor environments, where air is often considered still (i.e., air speed below 0.15 to 0.2 m·s$^{-1}$, depending on the standard considered (*24*)) and people spend ~90% of their time (*25*). In practice, human motion and variable airflows likely cause indoor conditions to fluctuate between free and mixed convection regimes; however, the latter cannot be properly quantified without a clear understanding of the former. Furthermore, indoor temperatures often exceed ASHRAE comfort limits (20–27.5 °C) (*24*, *26*), making indoor overheating a major concern due to its negative impacts on cognition (*27*, *28*), productivity, and heat-related illness (*28*, *29*). For example, in 2024, Maricopa County, Arizona, reported 138 indoor heat-related deaths (out of 608 total), 88% of which occurred in homes without functional cooling (*30*). Many outdoor heat-related deaths occur among the unhoused population, who often shelter in tents where air speeds are reduced to levels comparable to indoor conditions (*31*). While in Arizona, where air temperatures exceeding 40 °C occur from March to November, most of the population has access to air conditioning, environmental conditions in which humidity-driven buoyancy is significant are common in lower-income regions that lack such a cooling infrastructure. In particular, drylands currently cover about 40–45% of the global land surface and are expanding, with their population of roughly 2.3 billion projected to double potentially, mostly in developing countries, by the end of the century (*32*, *33*). In these settings, people primarily shelter from the heat indoors, where, in the absence of air conditioning, heat loss is dominated by free or mixed convection and sweat evaporation. These processes are critical for preventing overheating but remain poorly characterized.

Here, we quantify the dueling, and at times reinforcing, buoyancy effects on free convection and sweat evaporation from the human body using comprehensive sweating thermal manikin experiments and digital-twin multiphysics simulations. In experiments, multiple heat transfer mechanisms—namely, free convection, evaporation, and radiation—contribute simultaneously to the net heat flux measured by the sweating thermal manikin. Consequently, matching digital twin simulations are required to isolate and quantify each mechanism's contribution. The simulations are comprehensively validated using multiple experimental approaches that isolate one or two transport mechanisms, cover a broad range of conditions, and incorporate airflow measurements. We highlight that while this manikin represents an anthropometrically average male, we previously found that body shape has a minor impact on the convective heat transfer coefficient (*34*, *35*). We further show that standard "temperature difference-only" models can overpredict human free convective heat transfer and sweat evaporation rates, leading to a substantial underestimation of physiological heat stress in arid heat. To enable more accurate heat-stress assessments and improved design of mitigation strategies for extreme-heat environments, we develop compact, physics-informed models for human free-convective heat transfer valid across wide temperature and humidity ranges.

## Results

### *Multiphysics digital twin matches thermal manikin measurements across diverse conditions*

To validate our digital manikin twin, we performed a comprehensive series of thermal manikin experiments under conditions designed to generate buoyant near-body flow through temperature-only, humidity-only, and combined temperature–humidity effects. A multiphysics digital twin replicating the experimental setup was developed using COMSOL Multiphysics software by coupling "Turbulent Fluid Flow", "Heat Transfer in Fluids", "Surface-to-Surface Radiation" and "Moisture Transport" physics modules. This series of experiments includes conditions that isolate individual transport mechanisms, enabling direct comparison of measured and simulated heat fluxes (e.g., isothermal cases in which free convection is driven solely by humidity variations). In experiments isolating temperature-driven free convection, radiative exchange is accounted for using simulations previously validated under complex irradiation conditions (*22*, *36*–*38*). In addition to matching whole-body and segmental heat fluxes, we also compare measured and simulated air velocities above the manikin. The experiments were conducted inside a nearly sealed enclosure that limited ambient air motion to below 0.02 m·s$^{-1}$ (**Fig. 1A**) located within a climate-controlled chamber where $T_{air}$ and $RH_{air}$ were systematically varied (see **Methods** and **Text S3 in Supplementary Materials** for details). The manikin's surface temperature ($T_{sk}$) was set to 35 °C, representative of typical human skin (*7*). To isolate thermally driven free convection, the manikin surface was kept dry (no sweating) while $T_{air}$ was varied (40.3–44.8 °C) at constant absolute humidity. Unlike earlier studies focused only on cold-to-moderate environments ($T_{air}$: 20.1–29.1 °C) (*8*, *34*, *39*)—where the body generates an upward buoyant plume—our conditions also featured $T_{air} > T_{sk}$ (moderate-to-hot), resulting in downward free convection (**Fig. 1B**). To isolate only humidity-driven evaporation, the temperatures of the air and water-saturated manikin textile "skin" were set to 35 °C, while only $RH_{air}$ was varied (**Fig. 1A, Inset**). Finally, combined temperature and humidity effects were explored by increasing $T_{air}$ by 4.6 °C or 7.5 °C above the water-saturated manikin surface temperature while varying humidity. The results in **Fig. 1D** show that the whole-body net heat flux predicted by the digital twin ($Q_{Sim-WB}$, W·m$^{-2}$) closely matches the measured values ($Q_{Exp-WB}$, W·m$^{-2}$) across all tested conditions. Comparable agreement was observed for air velocity measurements and heat fluxes on individual body segments, with any moderate deviations and their causes detailed in **Supplementary Materials** (see **Text S3**, **Fig. S3A–F**, **Fig. S4A–B**, **Fig. S5A–D**, **Fig. S6A–D**, and **Fig. S7A–D**). Having validated the model, we next use the digital twin to quantify the coupled influence of temperature and humidity on free-convection coefficients and sweat evaporation rates across a broad range of environmental conditions.

### *Temperature and humidity govern evaporation rates and free convection coefficients*

With the experimentally validated digital twin, we systematically explored the influence of dueling or reinforcing buoyancies on free convection and sweat evaporation for an average western male by parametrically varying $T_{air}$ (20–50 °C) and $RH$ (10–90%) (100 cases total, simulated using High Performance Research Computing facilities at Arizona State University (*40*)), while maintaining a water-saturated skin surface at 35 °C (**Fig. S8A–C**). **Fig. 2A** compares the simulated whole-body free convective coefficients ($h_{sim}$, W·m$^{-2}$·K$^{-1}$) against the "temperature-difference only" free convective empirical formulations, $h_{\Delta T-only} = 1.36\ |\Delta T_{sk-air}|^{0.3}$ (W·m$^{-2}$·K$^{-1}$), derived from our prior cold-to-moderate studies (*34*) and validation cases (moderate-to-hot). Additionally, the individual buoyant forces arising from the temperature ($\Delta T_{sk-air} = T_{sk} - T_{air}$) and humidity concentration ($\Delta C_{sk-air} = C_{sk} - C_{air}$) differences, namely, $B_{\Delta T} = g\rho\beta_T\Delta T_{sk-air}$ and $B_{\Delta C} = g\rho\beta_C\Delta C_{sk-air}$, were also computed (see **Supplementary Materials** for details**, Fig. S9A–C**, **Fig. S10A–C**, and **Fig. S11A–C**). Here $g$ is the gravitational acceleration (9.81 m·s$^{-2}$), $\beta_T$ is the thermal coefficient of expansion (reciprocal of the absolute average of $T_{air}$ and $T_{sk}$), and $\beta_C$ is the coefficient of expansion due to compositional changes. $\beta_C$ can be obtained from the density of air ($\rho$), the average molecular weight of air ($MW_a$ =29 g·mol$^{-1}$), and that of water vapor ($MW_C$ =18.02 g·mol$^{-1}$) as $(MW_a/MW_C - 1)/\rho$ or 0.607/$\rho$ (*11*). We note that this study was restricted to cases with a positive skin-to-air moisture difference ($C_{sk} > C_{air}$); scenarios with $C_{sk} < C_{air}$, which would cause condensation on the skin, were omitted.

Depending on ambient conditions, the coupled thermal- and humidity-driven buoyancies can either enhance or suppress free convection and sweat evaporation. When air is cooler than skin, both buoyancies are positive (**Table S4**) and act mutually to drive a stronger ascending plume ($\rho_{sk} < \rho_{air}$), leading to enhanced heat and mass transfer, i.e., the coupled temperature- and humidity-differences driven $h_{sim}$ exceeds the temperature difference-only driven $h_{\Delta T-only}$ by up to 70.7% (34 °C air at 10% RH). In contrast, when $T_{air}$ exceeds $T_{sk}$, the buoyancies oppose each other. For small temperature differences, the humidity-driven $B_{\Delta C}$ dominates, driving an upward plume and leading to $h_{sim} > h_{\Delta T-only}$. As $T_{air}$ becomes substantially hotter than the skin, the opposition between $B_{\Delta T}$ and $B_{\Delta C}$ intensifies, thereby progressively weakening the plume and hindering the heat transfer. Eventually, the two buoyancies nearly balance out each other ($B_{\Delta T} \approx B_{\Delta C}$), thus creating a near-stagnant flow regime. Here, the simultaneous occurrence of moisture transport leads to maximum suppression of the heat transfer, making $h_{sim}$ reach its lowest value. For instance, when air is at 41 °C and 10% $RH_{air}$, $h_{sim}$ (1.01 W·m$^{-2}$·K$^{-1}$) is 56.4% lower than $h_{\Delta T-only}$ (2.29 W·m$^{-2}$·K$^{-1}$). Across all cases, near-stagnation conditions produced $h_{sim}$ values between 0.71 W·m$^{-2}$·K$^{-1}$ (35.5 °C air at 90% $RH_{air}$) and 1.26 W·m$^{-2}$·K$^{-1}$ (39 °C air at 30% $RH_{air}$). As the $RH_{air}$ increases, the temperature at which the buoyancies balance each other shifts closer to the skin temperature, consistent with the fact that air at 35 °C and 100% $RH_{air}$ (assuming mean $T_{sk}$ across the body surface is also 35 °C), nullifies both temperature and humidity differences, yielding an "ultimate" near-stagnation condition, with no free convection or sweat evaporation. With $T_{air}$ rising beyond the stagnation condition, $B_{\Delta T}$ overtakes $B_{\Delta C}$, thus reversing the flow to a descending plume. Yet, as $B_{\Delta C}$ continues to oppose, $h_{sim}$ remains suppressed relative to $h_{\Delta T-only}$. Overall, these findings highlight the critical transition from supportive to opposing interactions between the temperature- and humidity-driven buoyancies and their suppressive effect on the free convection in hot, arid conditions–effects overlooked by standard empirical formulations.

The coupled buoyancies influenced sweat evaporation in parallel. **Fig. 2B** compares the simulated evaporative heat fluxes ($Q_{sim}$, W·m$^{-2}$) with the standard "textbook" estimate of evaporative heat flux ($Q_{\Delta T-only}$, W·m$^{-2}$), which is conventionally computed as the product of $h_{\Delta T-only}$, Lewis factor (16.5 K·kPa$^{-1}$, (*7*)) and the skin-to-air vapor pressure differences, as prescribed by the Chilton-Colburn analogy (*41*) and Lewis relation (2). Across all $RH_{air}$, when the air temperature is cooler than 25 °C, $Q_{sim}$ is slightly lower than $Q_{\Delta T-only}$ (2.4–14.3%), by an average of 18 W·m$^{-2}$. Between 30–35 °C, since the buoyancies reinforce each other, sweat evaporation is enhanced. For instance, air at 34 °C and 10% $RH_{air}$, $Q_{sim}$ (188.7 W·m$^{-2}$), is 65.2% higher than $Q_{\Delta T-only}$ (114.2 W·m$^{-2}$). However, as $T_{air}$ becomes hotter, the competing temperature and humidity differences increasingly hinder the sweat evaporation, eventually bringing the plume to a near-stagnant condition and significantly suppressing the evaporation rates by almost 56% at 41 °C and 10% $RH_{air}$ relative to the standard values ($Q_{sim}$ = 82.4 W·m$^{-2}$, $Q_{\Delta T-only}$ = 182.7 W·m$^{-2}$). Beyond the stagnation conditions, at higher humidities (lower vapor pressure differences), both $Q_{sim}$ and $Q_{\Delta T-only}$ are smaller, and the differences between them are smaller as well (see **Supplementary Materials** for details, **Fig. S12A–F, Fig. S13A–F**, and **Fig. S14A–F** for individual $RH_{air}$ plots).

## Discussion

### ***Physics-informed free convection model accounting for humidity-induced buoyancy***

For water vapor diffusion in air at 35 °C, the Prandtl number (ratio of momentum to heat diffusivities, $\nu/\alpha = 0.70$) and the Schmidt number (ratio of momentum to mass diffusivities, $\nu/D = 0.62$) are approximately equal. Therefore, the Nusselt number ($Nu = h_c L/k$, where $L$ (m) is the characteristic dimension and $k$ (W·m$^{-1}$·K$^{-1}$) is the thermal conductivity of air), representing the non-dimensional free convective coefficient during simultaneous heat and water vapor transfer, can be calculated from the classical correlations using the total Rayleigh ($Ra_T$) number, which is the sum of its thermal ($Ra_{\Delta T}$) and concentration ($Ra_{\Delta C}$) components (*42*):

$$Ra_T = Ra_{\Delta T} + Ra_{\Delta C} = \frac{g\beta_T \Delta T_{sk-air} L^3}{\nu\alpha} + \frac{g\beta_C \Delta C_{sk-air} L^3}{\nu D} \quad \textbf{Eq.2}$$

The properties in **Eq.2** are calculated at the film temperature and water vapor concentration (i.e., the average of skin and air values). For free convection from a vertical plate, the Nusselt number follows (*43*):

$$Nu = 0.59 Ra_T^{\frac{1}{4}} (10^4 < Ra_T < 10^9, \text{laminar}) \quad \textbf{Eq.3}$$

$$Nu = 0.1 Ra_T^{\frac{1}{3}} (10^9 < Ra_T < 10^{13}, \text{turbulent}) \quad \textbf{Eq.4}$$

Our previous work showed that the whole-body free convective coefficients under thermally driven conditions are about 10% higher than predicted for turbulent free convection over a vertical plate of height 1.8 m (i.e., the manikin height) (*34*). Based on this, we recast all simulation outcomes in terms of the Nusselt number and corresponding $Ra_T$. **Fig. 3A** demonstrates that the non-dimensionalized simulation results align well with the classical vertical-plate correlations, especially for laminar flow. Our earlier studies based on thermally driven free convection covered $Ra_{\Delta T}$ of 2×10$^9$ to 7×10$^9$, where the data overlaps both the laminar and turbulent correlations (*34*). Expanding this range to $|Ra_T|$ of 4×10$^6$ to 1.1×10$^{10}$ reveals that the laminar correlation with 1/4 exponent provides a substantially more universal fit.

**Fig. 3A** also demonstrates an interesting feature of the simultaneous heat and mass transfer from the human body in a free convective regime: even as $Ra_T \to 0$, the $Nu$ retains an albeit small, but finite value. This phenomenon occurs when $\Delta T_{sk-air}$ and $\Delta C_{sk-air}$ are non-zero but generate opposing buoyancy forces that cancel each other, creating a stagnant medium ($Ra_T \to 0$), so heat and mass transfer occur through diffusion only (corresponding to a baseline $Nu_0$ that depends on the domain size). Therefore, we propose the following functional form for the Nusselt number for free convection from a sweaty human body:

$$Nu = Nu_0 + C|Ra_T|^{1/4} \quad \textbf{Eq.5}$$

Fitting our simulation data to **Eq. 5**, matching the laminar coefficient, yields $C$ = 0.574 ± 0.018 (68% CI) and $Nu_0$ =17.27 ± 3.91 (68% CI; corresponding to heat transfer coefficient of 0.27 W·m$^{-2}$·K$^{-1}$ at 35 °C) with $R^2$ value of 0.994. Supporting our physical reasoning for the stagnant flow case, pure heat diffusion simulations for a range of enclosure sizes and geometries yield a domain-averaged $\mathrm{Nu}_0$ = 23.47 ± 2.88 (68% CI; corresponding to heat transfer coefficient of 0.34 W·m$^{-2}$·K$^{-1}$ at 35 °C), close to the above fitted value. The diffusive baseline $\mathrm{Nu}_0$ should be interpreted as the effective heat transfer in a bounded, room-scale quiescent enclosure in the absence of buoyancy-driven flow, and not as a free-space conduction limit. Full details of the domain-size sensitivity analysis for the pure heat diffusion simulations and the resulting $\mathrm{Nu}_0$ are provided in the **Supplementary Materials (see Text S7** and **Fig. S20A–D**).

The advantage of the dimensionless form of **Eq.5** is its reduced dependence on air properties, which vary with environmental conditions. However, this formulation might not be familiar to practitioners within thermophysiology and related fields. Therefore, to facilitate the application and interpretation of **Eq. 5**, we will express it in terms of the heat transfer coefficient. Specifically,

$$h_c = \frac{k}{L} Nu = \frac{k}{L} Nu_0 + \frac{kCL^{3/4} g^{1/4} \beta_T^{1/4}}{L\nu^{1/4}\alpha^{1/4}} \left| \Delta T_{sk-air} + \frac{\beta_C \alpha}{\beta_T D} \Delta C_{sk-air} \right|^{1/4} \quad \textbf{Eq.6}$$

A generalized form of the equation can be written as:

$$h_c = h_0 + B \left| \Delta T_{sk-air} + \frac{\beta_C \alpha}{\beta_T D} \Delta C_{sk-air} \right|^{1/4} \quad \textbf{Eq.7}$$

Here, $h_0$ is the diffusion-only value (units of W·m$^{-2}$·K$^{-1}$), $B$ is the coefficient specifying geometrical, flow, and fluid property impacts, and $\Delta T_{sk-air} + (\beta_C \alpha/\beta_T D) C_{sk-air}$ is the combined temperature difference (input unit of °C) and water vapor concentration difference (input units of kg·m$^{-3}$) buoyancy driving term. In the fitting process, we implicitly neglect the temperature variation in the $h_0$ and the $B$ coefficient by assuming them to be constant. The average $\beta_C \alpha/\beta_T D$ coefficient for the wide range of environmental conditions in the simulated cases is 145.7 ± 3.69 (68% CI), which also nearly exactly corresponds to one calculated with all properties at 35 °C (a switch to higher skin temperature of 37 °C common in hot conditions only increases the coefficient value to 146.9). Fitting of the simulated heat transfer coefficients to **Eq.7** with air conditions specific $\beta_C \alpha/\beta_T D$ or its average value (145.7) yields nearly identical coefficients of 1.26 ± 0.06 and 1.26 ± 0.04 (68% CI) with both $R^2$ values of 0.994 (see **Supplementary Materials**, **Text S8**). Illustrating a minor impact of considering specific environmental conditions, these coefficients can also be obtained from the dimensionless equation form using air properties at 35 °C (i.e., $Nu_0 k/L$ =0.269 and $(kCL^{3/4} g^{1/4} \beta_T^{1/4})/(L\upsilon^{1/4}\alpha^{1/4})$ =1.255). Based on the "diffusion-only" simulations and with an emphasis on simplicity, we propose the following simplified whole-body expression using rounded coefficients:

$$h_c = 0.3 + 1.25|\Delta T_{sk-air} + 146\Delta C_{sk-air}|^{1/4} \quad \textbf{Eq.8}$$

As shown in **Fig. 3B**, the model from **Eq. 8** accurately predicts the 100 simulated cases. We note that the impact of the simplifying rounding of the coefficient was negligible, below 0.5%. For the entire range of heat transfer coefficient values, the mean error, mean absolute error, and root mean square errors between the simulated values and **Eq.8** predictions expressed in terms of percent are -2.6 ± 13.3% (68% CI), 7.7 ± 11.2 % (68% CI), and 13.3%, respectively. If the range is reduced to values above 1.75 W·m$^{-2}$·K$^{-1}$, the mean error, mean absolute error, and root mean square errors reduce to -0.3 ± 3.9% (68% CI), 3.1 ± 2.4 % (68% CI), and 3.9%, respectively. To provide a conservative estimate, we specify a 14% (68% CI) uncertainty for **Eq.8** which is typical of convection coefficient correlations (*43*).

In contrast to the whole-body coefficients, we observed substantial differences for regional values depending on flow direction. This observation reflects the fact that the boundary layer grows in the flow direction, leading to the highest heat transfer near the start of the flow or at the point where the flow transitions to turbulence. To maintain continuity in $h_c$ when $Ra_T$ changes sign, we first fitted **Eq. 7** to the positive $Ra_T$ values. We used the resulting $h_0$ values as inputs into corresponding fits for the negative $Ra_T$ values (leaving *B* as the only fitting parameter). In either case, **Table 1** and plots described in the **Supplementary Materials** (**Fig. S15A–F**, **Fig. S16A–F**, and **Fig. S17A–F**) show that the fitted models provide excellent match with the simulated segmental $h_c$. While the results discussed so far pertain to standing posture, equivalent analyses were conducted for seated and supine postures, yielding $h_c = 0.48 + 1.19\ |\Delta T_{sk-air}|^{1/4}$ and $h_c = 0.34 + 1.46\ |\Delta T_{sk-air}|^{1/4}$ (see **Supplementary Materials**, **Fig. S18C**).

***Threshold air speed for forced convection surpassing mixed convection***

To assess how humidity-induced buoyancy influences the mixed convection regime, we estimate the air speed at which forced convection effectively dominates over mixed convection under two cases: (i) temperature-driven buoyancy only and (ii) combined temperature and humidity-driven buoyancy. Mixed convection, where free ($h_c$, W·m$^{-2}$·K$^{-1}$) and forced ($h_F$, W·m$^{-2}$·K$^{-1}$) convective heat transfer coefficients are of comparable magnitude, occurs under many realistic low air speed conditions. The mixed regime is often neglected in human convection measurements (*39*, *44*), human heat balance models, and thermoregulation models, with the simplifying assumption of a sharp transition between free and forced convection at a single threshold air speed (e.g., 0.2 m·s$^{-1}$ for thermal audits (*7*)). In fluid mechanics, regime classification is typically based on the ratio of the Grashof (Rayleigh number divided by Prandtl number) and the square of the Reynolds number, which compares buoyancy and inertial forces. When this ratio is near unity, mixed convection occurs. In this regime, the combined convection heat transfer coefficient ($h_{Mix}$, W·m$^{-2}$·K$^{-1}$) is commonly estimated using a power-law superposition (*43*):

$$h_{Mix} = (h_c^n \pm h_F^n)^{1/n} \quad \textbf{Eq.9}$$

where the empirical exponent $n$ is typically taken as 3. This formulation implies a smooth transition between regimes rather than a sharp switch. **Eq. 9** also allows for reinforcement of the two mechanisms such that $h_{\mathrm{Mix}}$ can exceed either $h_c$ or $h_F$ alone when the flows do not oppose each other. **Eq.9** can be recast as the ratio of mixed and forced convective coefficients, $h_{Mix}/h_F = (1 + (h_c/h_F)^n)^{1/n}$, which provides a convenient $h_c/h_F$ criterion for defining practical transition thresholds. For example, requiring $h_{\mathrm{Mix}}/h_F \leq 1.05$ or $\leq 1.1$ yields $h_c/h_F \leq 0.54$ and $\leq 0.69$, respectively. Using the widely adopted de Dear (*39*) correlation $h_F = 10.4\ V_{\mathrm{air}}^{0.56}$ (we highlight that this correlation has not been validated for air speeds below 0.2 m·$s^{-1}$), these ratios can be translated into a threshold air speed $V_{\mathrm{air},t}$ at which forced convection effectively dominates.

**Fig. 3C** presents contours of $V_{\mathrm{air},t}$ for the $h_{\mathrm{Mix}}/h_F \leq 1.05$ criterion computed using (i) a temperature-only natural convection correlation, $h_c = 1.36\ |\Delta T_{sk-air}|^{0.3}$ and (ii) a formulation that also accounts for humidity-driven buoyancy (**Eq. 8**). For both approaches, most of the threshold air speeds are within the commonly cited range below 0.3 m·s$^{-1}$ (*24*), with a change of $h_{\mathrm{Mix}}/h_F \leq 1.1$ criterion decreasing the top of the range to 0.225 m·s$^{-1}$ (see **Supplementary Materials Fig. S21**). In the temperature-only case, $V_{\mathrm{air},t}$ is symmetric about the assumed skin temperature (35 °C), where the temperature difference, and thus buoyancy, vanishes. Under these conditions, any non-zero airflow nominally dominates, although the net heat transfer is zero. When humidity-induced buoyancy is included, the minimum threshold shifts away from 35 °C as relative humidity decreases. A finite minimum air speed of approximately 0.03 m·s$^{-1}$ emerges along the neutral-buoyancy (stagnation) line. However, $V_{\text{air},t}$ increases rapidly with small deviations from this line, with even a 2–3 °C change in air temperature raising the threshold to the typical range of 0.15–0.2 m·s$^{-1}$. Overall, humidity shifts the neutral buoyancy condition, increasing the minimum threshold air speed at a given location, but does not substantially reduce the air speed required for a clear transition to forced convection. Rather than lowering the threshold, humidity primarily redistributes it across temperature–humidity space. It is also worthwhile pointing out that besides imposed air flow from fans, the thresholds for forced convection can be exceeded on some body parts due to motion (e.g., even walking at 0.9 m·s$^{-1}$ on a treadmill or stationary cycling at 50 revolutions per minute leads to the lowest local air speeds of 0.25 to 0.29 m·s$^{-1}$ (*8*)).

***Neglecting humidity-induced buoyancy and mixed convection can cause substantial errors in thermoregulation model predictions***

To assess the effects of humidity-driven buoyancy on thermophysiological responses, we employed our recently updated version of the classical six-segment Stolwijk thermoregulation model (*10*, *45*). Simulations were performed under conditions representative of a semi-outdoor environment typical of Phoenix summers, using three approaches to calculate sweat evaporation under free convection conditions. We first evaluate the thermophysiological responses using the common simplifying assumption of a sharp transition between free and forced convection at under 0.2 m·s$^{-1}$ (*7*, *8*, *39*, *44*). Specifically, we simulated the thermophysiological responses of a lightly dressed (0.25 clo) average young male sheltering from the heat (1.5 Met typical of standing or eating and drinking (*46*)) for two midday hours inside a canopy-covered tent, with air at 43 °C and 20% $RH_{air}$, a moderate mean radiant temperature ($T_{mrt}$) of 55 °C (typical outdoor values in Phoenix are above 70 °C (*47*)), and air velocity ($U_{air}$, m·s$^{-1}$) under 0.2 m·s$^{-1}$ (*31*). The three ways to compute sweat evaporation are, (A) constant $h_c$ approach used in both versions of the Stolwijk model (*10*, *45*), (B) temperature-difference driven approach ($h_{\Delta T} = A(\Delta T_{sk-air})^{1/4}$), and (C) combined temperature- and humidity-difference driven approach ($h_{\Delta T,\Delta C} = h_0 + B|\Delta T_{sk-air} + 146\Delta C_{sk-air}|^{1/4}$) (see **Methods** for details). Predictions of the core temperature ($T_{core}$) and mean skin temperature ($\bar{T}_{sk}$) using the first two approaches differed by less than 0.15 °C; thus, we focus on comparing the temperature-difference ($h_{\Delta T}$) and combined temperature- and humidity-difference driven ($h_{\Delta T,\Delta C}$) approaches.

**Fig. 4A–B** show that neglecting humidity-driven buoyancy contributions results in a significant underprediction of thermophysiological responses throughout the 2 hours of free convective conditions. The $h_{\Delta T}$ approach predicts that after 40 minutes, the $T_{core}$ and $\bar{T}_{sk}$ stabilize at about 37.9 °C and 37.0 °C, respectively. In contrast, the $h_{\Delta T,\Delta C}$ approach predicts that the $T_{core}$ and $\bar{T}_{sk}$ continue to rise reaching 38.4 °C and 38.0 °C after 60 minutes, and 38.9 °C and 38.5 °C after 120 minutes. In other words, neglecting humidity-driven buoyancy in these harsh and windless conditions leads to a dangerous underprediction of $T_{core}$ by 1 °C and $\bar{T}_{sk}$ by 1.5 °C.

The underlying physical mechanisms are clarified in **Fig. 4C–D**, which depict convective, radiative, evaporative, and metabolic heat rates, as well as net body heat storage. The mean buoyancy-driving terms ($\overline{\Delta T}_{sk-air}$ vs. $\overline{\Delta T}_{sk-air} + 146\overline{\Delta C}_{sk-air}$) for both approaches are also shown after scaling them by a factor of 50 for better visualization. In both cases, convective heat rates remain between 5 and 33 W, while radiative and metabolic rates remain near 160 W. During the first 10–20 minutes, sweat evaporation rapidly increases from approximately -15 W to -190 W for both models. In the $h_{\Delta T}$ approach, the sweat evaporation rate continues to gradually increase to -290 W, with buoyancy driven by nearly constant $\overline{\Delta T}_{sk-air}$ of -6 °C. As a result, the body heat storage rate drops below 55 W after 40 minutes and is almost eliminated after 60 minutes of heat exposure, leading to stabilization of the body temperatures. In contrast, in the $h_{\Delta T,\Delta C}$ approach, the evaporation rate stalls due to nearly vanishing buoyancy with $\overline{\Delta T}_{sk-air} + 146\overline{\Delta C}_{sk-air}$ between -1 °C to 0.5 °C. As a result, the body heat storage rate remains around 100 W for the first 60 minutes and does not drop below 50 W until 100 minutes of heat exposure, leading to increasing body temperatures. We note that the minor peaks in the evaporation rate in this case are due to the switching of the $\overline{\Delta T}_{sk-air} + 146\overline{\Delta C}_{sk-air}$ sign, and therefore a shift in the overall flow direction and the employed models from **Table 1**.

The simulated stagnant conditions and associated physiological responses represent a severe scenario with nearly diminished buoyancy. In practice, these conditions can be altered by several factors, including variations in mean radiant temperature, air temperature, relative humidity, and airflow driven by environmental or body motion. Although the mean radiant temperature corresponds to in-tent measurements and is moderate by Phoenix standards (*31*), it remains a dominant contributor to heat load. With $T_{air} = 43$ °C and $RH_{air} = 20\%$, reducing the $T_{mrt}$ to match that of air results in conditions that can be mitigated by evaporative cooling, regardless of how buoyancy is modeled. In particular, after two hours under these conditions, the $T_{core}$ predicted using both approaches are about 37.7 °C, while the $\bar{T}_{sk}$ predictions only differ by 0.3°C.

To provide a broader perspective, we simulated a 2-hour human exposure under stagnant air, with fixed activity and clothing levels, across $RH_{air}$ values from 5–100% and $T_{air}$ from 30–50 °C, within survivability limits defined by Vanos et al.(*48*), and for both $T_{mrt} = T_{air}$ or $T_{mrt} = 55$ °C (see **Supplementary Materials Fig. S22A–L**). Because temperature- and humidity-driven buoyancy forces can either oppose or reinforce each other, thermoregulation simulations using the $h_{\Delta T,\Delta C}$ formulation may predict either higher or lower body temperatures compared to the $h_{\Delta T}$ approach. Simulations across this broad condition set show that even when $T_{mrt} = T_{air}$, neglecting humidity induced buoyancy can lead to underprediction of $T_{core}$ by more than 0.6 °C and $\bar{T}_{sk}$ by over 1.25 °C when the two buoyancy effects oppose each other. With $T_{mrt} = 55$ °C, these underpredictions increase to over 1.5 °C for $T_{core}$ and 2.0°C for $\bar{T}_{sk}$. Conversely, when the two buoyancy forces reinforce each other, neglecting the humidity-induced buoyancy leads to thermophysiological overpredictions, with $T_{core}$ differences of -0.2 °C to -0.75 °C and $\bar{T}_{sk}$ differences of -0.5 to -1°C (in full shade and at $T_{mrt} = 55$ °C).

Lastly, we implemented **Eq.9** for mixed convection in the thermoregulation model and evaluated the impact of both constant and fluctuating air speed on the results presented in **Fig. 4** (i.e., at $T_{air} = 43$ °C, $RH_{air} = 20\%$, $T_{mrt} = 55$ °C). Under these conditions, the two buoyancy effects nearly cancel each other. As constant air speed approaches the low threshold for forced convection significance (about 0.03 m·s$^{-1}$; see **Fig.3C**), $T_{core}$ and $\bar{T}_{sk}$ predicted using the $h_{\Delta T,\Delta C}$ approach decrease rapidly with increasing airflow, converging to within 0.1 °C of the $h_{\Delta T}$ predictions at air speeds above about 0.1 m·s$^{-1}$ (see **Supplementary Materials, Fig. S23A**). However, perfectly stagnant or steady airflow conditions are unlikely in real settings; instead, airflow typically fluctuates. To capture this, we simulated alternating airflow between stagnant and a constant value at one-minute intervals. In this fluctuating air flow scenario, the decrease of $T_{core}$ and $\bar{T}_{sk}$ predicted by the $h_{\Delta T,\Delta C}$ approach is more gradual with increasing air flow, remaining approximately 0.8 °C (core) and 1.3 °C (mean skin) higher than $h_{\Delta T}$ predictions at the ~0.03 m·s$^{-1}$ threshold (see **Supplementary Materials Fig. S23B**). When airflow fluctuates between stagnant and 0.1 m·s$^{-1}$, the differences between the two approaches remain substantial, at approximately 0.4 °C for $T_{core}$ and 0.9 °C for $\bar{T}_{sk}$. Notably, even these reduced body temperature differences are comparable to the maximum thermophysiological effects of high turbulence under strong forced convection (*49*). Therefore, depending on environmental conditions, humidity-induced buoyancy can have moderate to strong impacts on thermophysiological predictions of indoor overheating and should be accounted for in such assessments.

### *Limitations*

While this work provides a quantitative understanding of the coupled buoyancy effects, the findings rely on several assumptions that should be reiterated. First, the skin was modeled as uniformly water-saturated at 35 °C. In contrast, spatial variations in sweat rates and skin temperature (*50*) could generate heterogeneous vapor concentrations that, if accounted for, could further refine predicted convective coefficients. Partially wet skin ($\omega$) linearly reduces the contribution of humidity-induced buoyancy and can be incorporated into the buoyancy term as $|\Delta T_{sk-air} + 146\omega\Delta C_{sk-air}|$. However, this effect is likely to be minor, given its transient nature. Under hot conditions with low evaporation rates, the skin rapidly becomes covered by a sweat film. For instance, in the scenario simulated in **Fig.4**, $\omega$ reaches 0.9 within approximately 15 minutes of sweating onset. Second, our physics-informed models were derived for static postures (standing, seating, and supine). Local or whole-body movements can locally induce mixed convection, highlighting the need to quantify motion effects on convective coefficients in the future. Third, our analysis of the thermophysiological impacts of mixed convection relies on a generic empirical heat transfer correlation that has not been validated for the human body. Accordingly, future experiments should explore humidity-induced buoyancy effects in mixed convection conditions. Fourth, our model considers only direct skin-to-ambient heat exchange (a shamelessly naked manikin). Clothing can substantially affect the heat and mass transfer, particularly through complex processes occurring within the skin-to-fabric microclimate (*7*, *8*, *51–54*). Therefore, in the future, the complex buoyancy effects arising from clothing should be investigated. Finally, the findings are based on an average western male body shape; therefore, exploring the impact of body shapes (including children, elderly, pregnant women, etc.) on the free convective coefficients, accounting for both thermal- and humidity-driven buoyancy effects, would further improve our understanding of human heat transfer.

***Implications***

Our findings demonstrate that, at air temperatures above ~30 °C, humidity-driven buoyancy can affect human free convective heat transfer and sweat evaporation in ways comparable to those of temperature-induced buoyancy. These two buoyancy forces may either reinforce or oppose each other; however, decreases in flow under opposing conditions have more pronounced thermophysiological consequences. For example, under hot (>37–38 °C) arid conditions with negligible airflow, the competing temperature- and humidity-driven buoyancies can suppress sweat evaporation by up to 56% (for air 41 °C with 10% $RH_{air}$). In the absence of background airflow, this phenomenon leads to substantially increased body heat storage and significantly elevated mean skin and core temperatures. Using classical temperature-difference-based free-convection heat transfer coefficient correlations results in thermoregulation models substantially underpredicting this effect. As expected, reduced net buoyancy lowers the airflow threshold at which forced convection overtakes mixed convection. However, while mixed convection moderates the thermophysiological impacts of humidity-driven buoyancy, these impacts remain substantial and can be comparable to those induced by air turbulence under forced convection. Accordingly, humidity-induced buoyancy should be considered when assessing risks of indoor overheating and heat-related illness, particularly for vulnerable populations, including individuals without access to air conditioning and residents of informal or makeshift housing.

The proposed physics-informed models of free and mixed convection can be readily integrated into thermoregulation models(*45*, *46*, *55*), thermal audit systems, and other heat stress assessment tools (*8*, *48*). Improved characterization of human thermophysiological responses in hot environments using these models can help guide the design of enhanced heat-resilient spaces, heat-mitigation strategies, and targeted whole-body and local cooling solutions, thereby potentially helping to avert the rise in indoor heat-related health issues. The suppression of sweat evaporation during free convection in hot, arid conditions also provides additional motivation to use fans, potentially in a cyclical on-off mode in very hot conditions when continual forced convective heat gains exceed the benefits of evaporation (*56*, *57*). Overall, these insights into the physical processes controlling sweat evaporation rate have substantial implications for thermophysiological simulations that can help guide decisions affecting human health, safety, and productivity in hot, arid climates.

## Materials and Methods

***Thermal manikin experimentation***

To build a validated multiphysics model, a series of experiments (15 cases) were conducted inside a nearly sealed enclosure located within a climatic chamber (*34*) (**Fig. 1A** and **Fig. S3A–F**), using a 35-segmented sweating thermal manikin (ANDI from Thermetrics, LLC) representative of an average young western male (50th percentile height (m) and BMI (kg·m$^{-2}$)). Three complementary sets of experiments—examining the individual effects of temperature differences, humidity differences, and combined temperature and humidity differences on free convection and sweat evaporation—were conducted by systematically varying the enclosure's air temperature ($T_{air}$: 35–44.8 °C) and relative humidity ($RH_{air}$: 23.5–81.4%) (see **Table S3** for details).The manikin skin temperature was held uniformly at 35 °C with exception of 31 or 34 °C for temperature-difference only cases. For experiments involving sweat evaporation, a water-saturated (at 35 °C) textile "skin" (**Fig.1A** Inset) was used to maintain a wet skin fraction of 1. All 15 cases were repeated thrice, each lasting 75 minutes, and steady-state data collected from the final 30 minutes (segmental skin temperature fluctuations remained < 0.05 °C) were used to determine the whole-body ($Q_{Exp-WB}$, W·m$^{-2}$) and zonal ($Q_{Exp-Z}$, W·m$^{-2}$) heat fluxes (see **Supplementary Materials** for details).

***Multiphysics simulations***

A digital twin of the experimental setup was developed in COMSOL Multiphysics 6.3. Since the entire setup was symmetrical, only one-half of the geometry was simulated to reduce computational cost. The model comprised the "Turbulent Fluid Flow", "Heat Transfer in Fluids", "Surface-to-Surface Radiation" and “Moisture Transport” physics modules coupled through “Non-Isothermal flow”, “Heat and Moisture Transfer”, “Heat transfer with Surface–to–Surface radiation”, and “Moisture flow” multiphysics interfaces. A Low-Re k-ε Turbulence model (with Low-Re wall functions) was used to capture near-skin flow features. Moist air was treated as a weakly compressible flow, and the gravity module was enabled to resolve the buoyancy-driven flow (*34*). Experimental boundary conditions were virtually replicated within the digital twin Following the mesh refinement study (**Fig. S4B**), the model was solved transiently for 300 s. A quasi-steady state was achieved between 240 and 300 s, with net heat flux variations under 5% sensitivity tests comparing the Low-Re k-ε model against laminar formulation simulations yielded differences in the computed free convection coefficient of less than 5%, confirming negligible sensitivity to the turbulence closure choice (**S8D–E**). The optimized mesh was initialized using a wall-distance method and solved transiently for 50 s, to compute the whole-body and regional free convective coefficients and evaporative fluxes. All cases reached stable steady-state values within 30–40 s, well within the original 50 s simulation window (**Fig. S8E**).. The simulated net whole-body ($Q_{Sim-WB}$, W·m$^{-2}$) and zonal ($Q_{Sim-Z}$, W·m$^{-2}$) heat fluxes were compared to experimental measurements for validation (see **Fig. 1D**, **Supplementary Materials, Fig. S5A–D, Fig. S6A–D**, and **Fig. S7A–D**).

***Parametric study using the validated multiphysics model***

The validated numerical model was used to investigate the coupled buoyancy effects by parametrically varying inlet $T_{air}$ (20–50 °C) and $RH_{air}$ (10–90%). The parametric study was performed within a simplified virtual enclosure (2 m length, 1 m width, 3 m height) with a watertight 3D geometry of an average young western male (50$^{th}$ percentile height and BMI) positioned at the geometric center (see **Supplementary Materials, Fig. S8A–B**). Radiative heat exchange was not analyzed for this study. To avoid numerical convergence issues (*34*), the inlet and outlet boundaries of the enclosure were dynamically assigned along the direction of the buoyant flow. For an ascending buoyant plume, the enclosure floor was set as the inlet and the ceiling as the outlet (vice versa for a descending plume). A mesh refinement analysis was performed as shown in **Fig. S8C**. Sensitivity analyses confirmed that the model results are not affected by the dynamic inlet/outlet boundary assignment and the virtual enclosure size (see **Supplementary Materials**, **Fig. S8D**). Furthermore,

***Thermoregulation modeling***

To analyze the impact of coupled thermal- and humidity-induced buoyancies on thermophysiological responses, we used the updated Stolwijk-2024 thermoregulation model, previously described and validated for multiple human heat exposure trials by Joshi et al. (*45*). The Matlab code for replicating **Fig.4** results is included at the end of the **Supplementary Materials**. In the present work, we quantified the effect of three approaches for modeling regional free convective coefficients ($h_c$, W·m$^{-2}$·K$^{-1}$). First approach used a constant $h_c$ values employed in the original Stolwijk model (*10*) and the newer ones introduced in JOS-2 and JOS-3 models (*46*) (**Table 2**). Second, a temperature-difference-based approach was employed, where $h_c$ was defined as $A(\Delta T_{sk-air})^{1/4}$. We used regional $A$ values proposed by Wissler (*8*), representing averages of measurements reported by Quintela et al. (*58*) and de Dear et al. (*39*) (**Table 2**). $h_c$ for hands and feet were unavailable in $A(\Delta T_{sk-air})^{1/4}$ form. Hence, we fitted this form to the average of values predicted by the two correlations (right and left halves) reported by Quintela et al. (*58*). This yielded $A = 2.7$ for hands and $A = 1.57$ for legs (see **Supplementary Materials** for details**,** and **Fig. S18A–B**).

Thirdly, a combined temperature- and humidity-difference-based approach developed from the current study was used, where $h_c$ was defined as $h_0 + B|\Delta T_{sk-air} + 146\Delta C_{sk-air}|^{1/4}$ (**Eq. 8**) with regional fitted parameters outlined in **Table 2**. We note that, hands were excluded from the validation simulations for the following reasons. Our physical manikin does not have heated fingers; therefore, free convective coefficients for the hands were not measured. Moreover, simulating convection around hands requires a dense mesh, specifically around the fingers (*35*) and the convective coefficients for the hands are highly dependent on pose and orientation. Hence, hands were not simulated in the convection models. Instead of our own measurements or simulations, for the hands we fitted **Eq. 8** to the average of the two correlations given by Quintela et al. (*58*), obtaining $h_0 = 0.5$ and $B = 2.4$ (**Fig. S18A** shows reasonable agreement between our fitted correlation and the average results of Quintela et al. (*58*)).

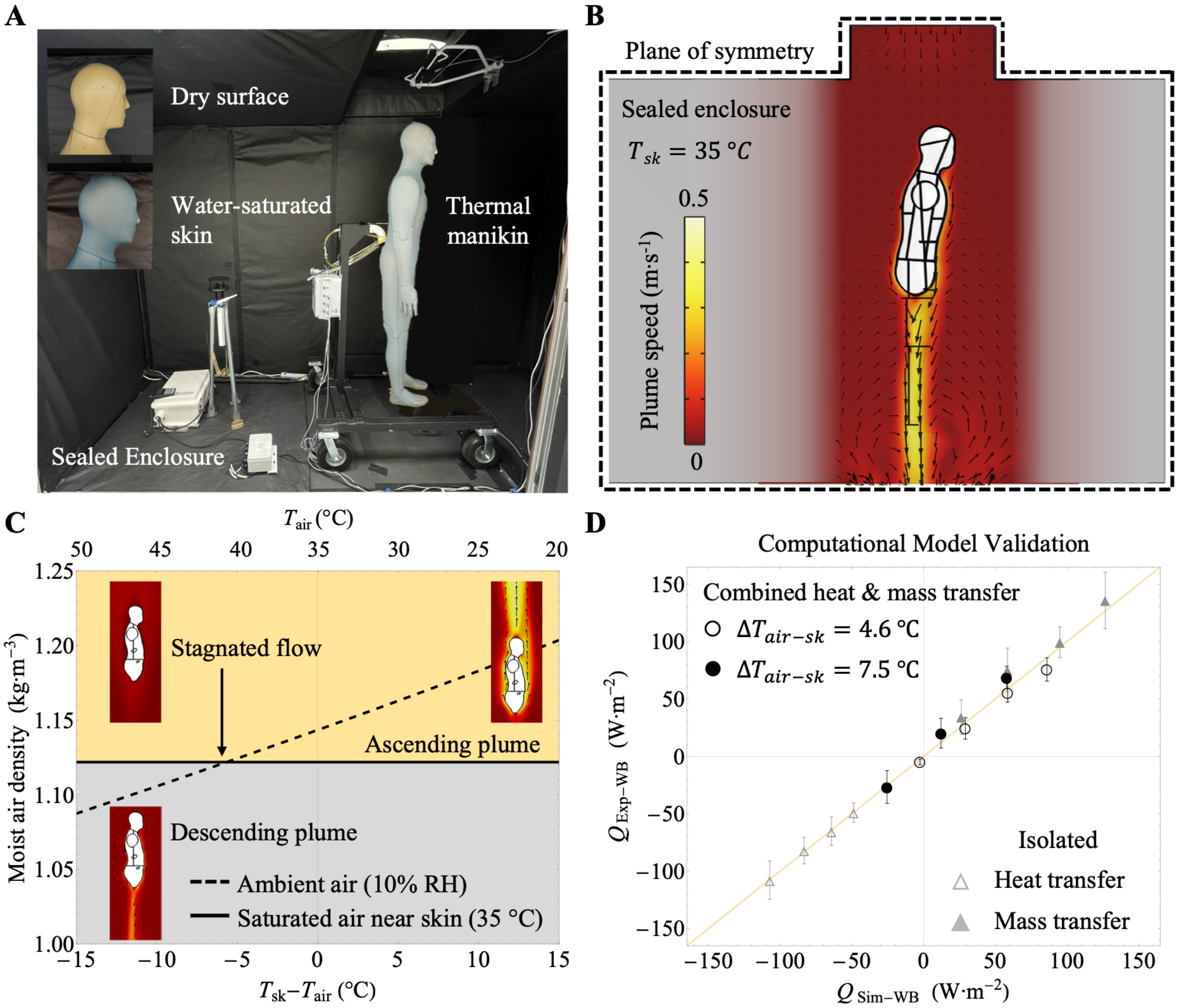

**Fig. 1. Experimental validation of the multiphysics numerical model. (A)** Experimental setup for investigating coupled-buoyancy effects on heat exchange with a sweating thermal manikin inside a nearly sealed enclosure located within a climatic chamber. Inset: Manikin surface fitted with a water-saturated textile "skin" during sweat evaporation experiments and left dry otherwise. **(B)** Cross-sectional side view of the manikin's digital twin developed using a multiphysics numerical model. Inset: Partially illustrated descending buoyant plume simulated around the manikin when ambient air ($T_{air}$, °C) is hotter than the skin ($T_{sk}$, °C) with plume speeds ranging from 0 to 0.5 m·s$^{-1}$. **(C)** Comparison of moist air density near and away from the water-saturated skin ($T_{sk}$ at 35 °C), under varying $T_{air}$ at 10% RH. Inset: Coupled thermal- and humidity-difference induced density variations leading to ascending, stagnant, and descending buoyant plume. **(D)** Simulated whole-body net heat fluxes ($Q_{Sim-WB}$, W·m$^{-2}$) closely match experimental measurements ($Q_{Exp-WB}$, W·m$^{-2}$) across all 15 cases–including isolated convective and radiative heat transfer, isolated convective mass transfer, and combined heat and mass transfer. $\Delta T_{air-sk}$ (°C) is the air-to-skin temperature difference. Data points are averages of three replicates with 95% confidence intervals (combined Type A and B uncertainty). A detailed breakdown of the whole-body results across 12 anatomical segments is provided in the **Supplementary Materials.**

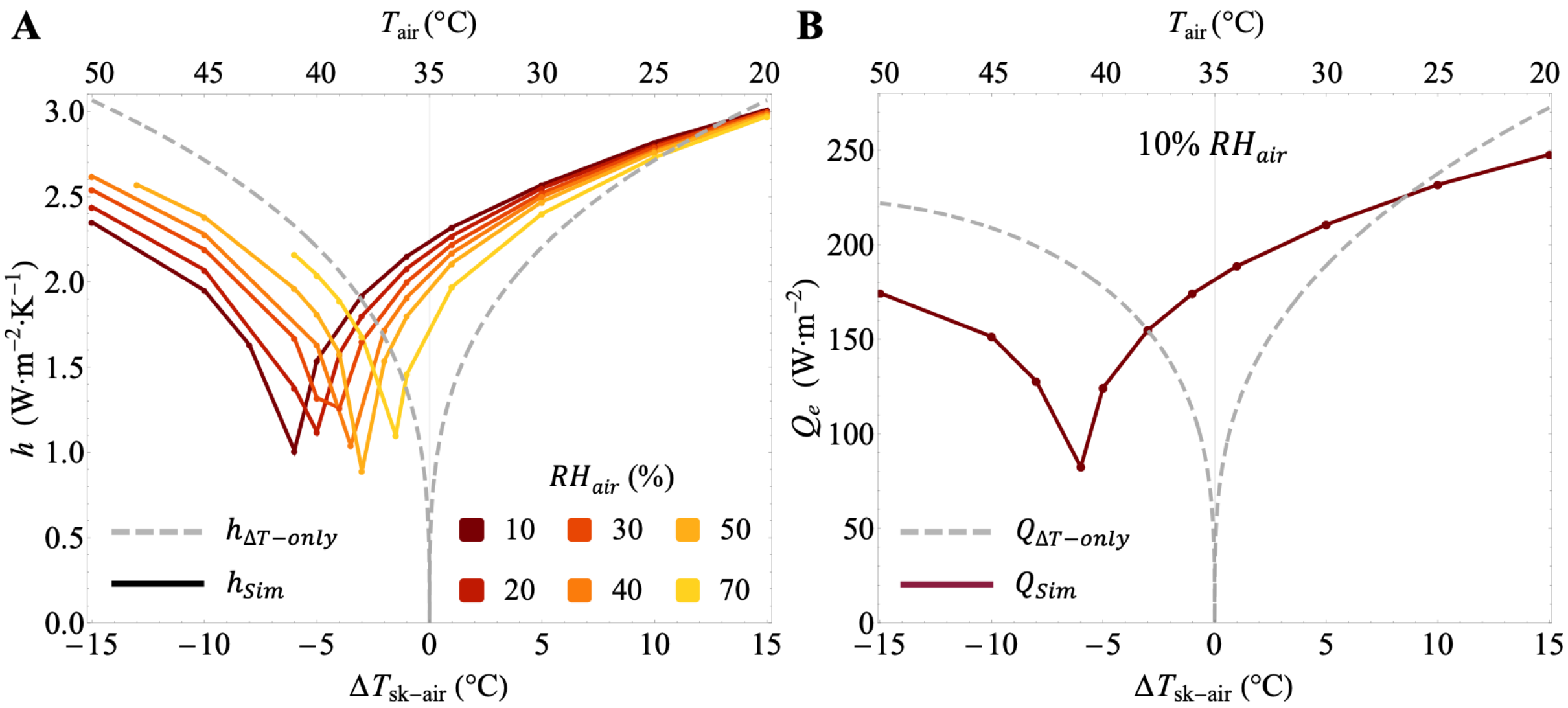

**Fig. 2. Coupled buoyancy effects on free convective heat transfer coefficient ($h$, W·m⁻²·K⁻¹) and sweat evaporation flux ($Q_e$, W·m⁻²).** Simulated whole-body **(A)** free convective heat transfer coefficients ($h_{Sim}$, W·m$^{-2}$·K$^{-1}$) and **(B)** evaporative fluxes ($Q_{Sim}$, W·m$^{-2}$) are compared with temperature-difference based empirical correlations across a range of air temperatures ($T_{air}$, °C) and selected relative humidities ($RH_{air}$). The empirical correlations are $h_{\Delta T-only} = 1.36\,|\Delta T_{sk-air}|^{0.3}$ (W·m$^{-2}$·K$^{-1}$) for free convection (*34*), and $Q_{\Delta T-only} = 16.5\, h_{\Delta T-only} \Delta p_{sk-air}$ (W·m$^{-2}$) for evaporative heat flux (*7*), where $\Delta T_{sk-air}$ (°C) and $\Delta p_{sk-air}$ (kPa) are the skin-to-air temperature and vapor pressure differences, respectively.

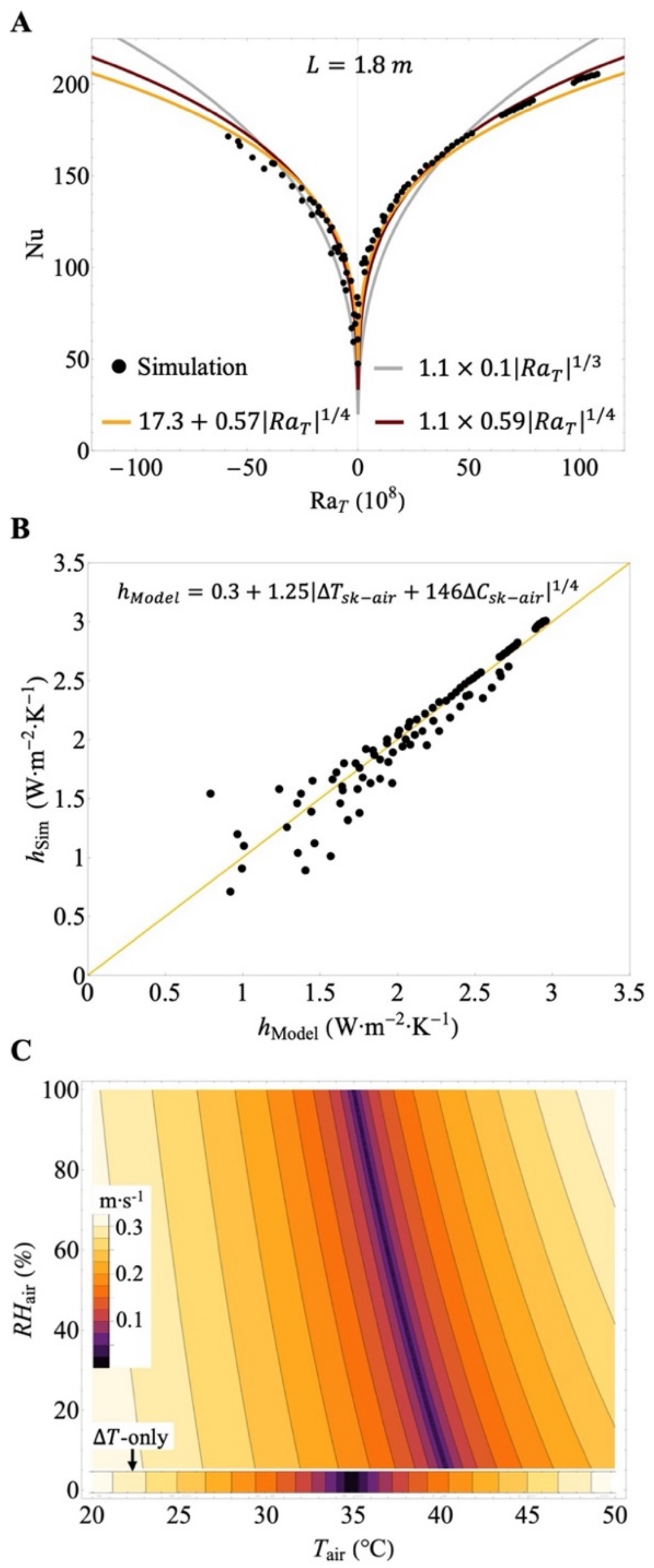

**Fig. 3. Comparison of the non-dimensional and dimensional free convection coefficients against classical correlations and model predictions, and threshold air speed for forced convection surpassing mixed convection (A)** The non-dimensional simulated whole-body free convection coefficients ($h_{Sim}$, W·m$^{-2}$·K$^{-1}$), expressed as a Nusselt number ($Nu$), follows the fitted relation $Nu = 17.3 + 0.57|Ra_T|^{1/4}$ (**Eq. 5**), where $Ra_T$ is the total Rayleigh number. The fitted trend closely aligns with the classical laminar vertical-plate correlations $Nu = 1.1 \times 0.59|Ra_T|^{1/4}$ (**Eq. 3)** than with turbulent flow correlations $Nu = 1.1 \times 0.1|Ra_T|^{1/3}$ (**Eq. 4**). Characteristic length ($L$) of 1.8 m was used (manikin height) while fitting. **(B)** Simulated whole-body free convective coefficients ($h_{Sim}$, W·m$^{-2}$·K$^{-1}$) strongly agree with the physics-informed model predictions ($h_{Model}$, W·m$^{-2}$·K$^{-1}$, **Eq. 8**), across all 100 simulated ambient conditions. $\Delta T_{sk-air}$ (°C) and $\Delta C_{sk-air}$ (kg·m$^{-3}$) are the skin-to-air temperature and concentration differences, respectively. (**C**) Threshold air speed for forced convection ($h_F$) surpassing mixed convection ($h_{Mix}$) calculated using the criterion of $h_{Mix} \leq 1.05\ h_F$ with free convection being driven by temperature-only (contour line at the bottom of the plot) and additionally by humidity (the 2D contour plot).

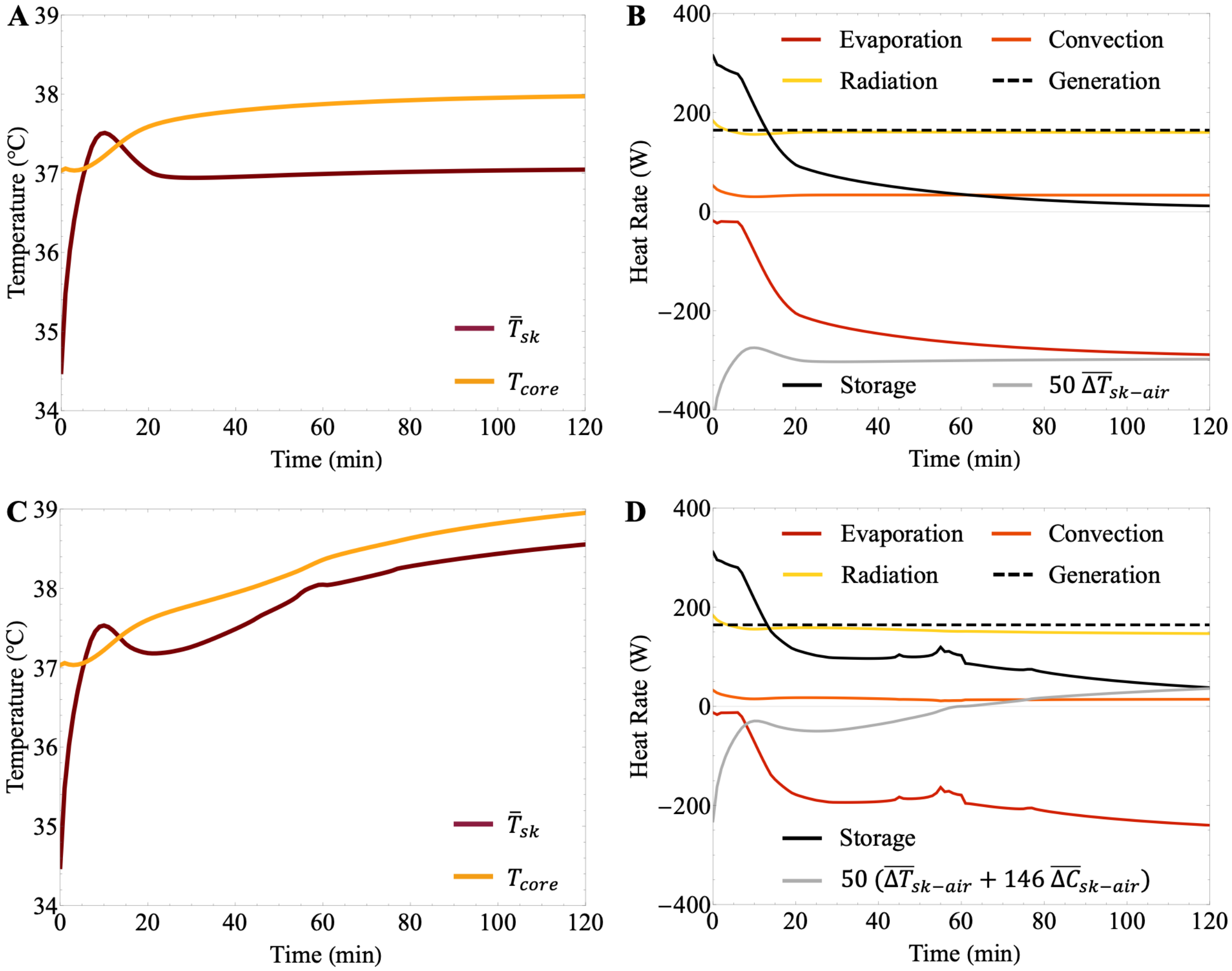

**Fig. 4. Thermophysiological responses predicted by the classical and coupled buoyancy effects included in updated Stolwijk thermoregulation model.** Simulations were performed for a lightly dressed (0.25 clo) young male with minimal activity (1.5 Met) inside a semi-outdoor canopy-covered tent with air at 43 °C and 20% $RH_{air}$, a mean radiant temperature of 55 °C, and fully stagnant air for 2 hours. Evolution of **(A)** mean skin temperature ($\overline{T}_{sk}$, °C), core temperature ($T_{core}$, °C), and **(B)** associated heat rates (W) (free convective, radiative, evaporative, metabolic heat generation and heat storage) predicted by the temperature difference based classical Stolwijk model (*10*). **(C, D)** Corresponding predictions from the updated physics-informed Stolwijk model incorporating the combined temperature- and humidity-difference driven buoyancy effects. **(B,D)** For visualization, the buoyancy-driving terms for each model ($\overline{\Delta T}_{sk-air}$ vs $\overline{\Delta T}_{sk-air} + 146\,\overline{\Delta C}_{sk-air}$) are scaled by a factor of 50. $\Delta T_{sk-air}$ (°C) and $\Delta C_{sk-air}$ (kg·m$^{-3}$) are the skin-to-air temperature and concentration differences, respectively.

**Table 1. Summary of empirical coefficients for the free convection correlations (Eq.7 with 146 coefficients for $\Delta C_{sk-air}$) fitted for all body segments under positive and negative buoyancy regimes.** $h_0$ (diffusion-only term, W·m$^{-2}$·K$^{-1}$) and B are fitted parameters, $\sigma_{h_0}$ and $\sigma_B$ are their standard errors, respectively, and $R^2$ is the coefficient of determination. $\Delta T_{sk-air}$ (°C) and $\Delta C_{sk-air}$ (kg·m$^{-3}$) are the skin-to-air temperature and concentration differences, respectively.

| | $\Delta T_{sk-air}$ + 146$\Delta C_{sk-air}$ > 0 | | | | | $\Delta T_{sk-air}$ + 146 $\Delta C_{sk-air}$ < 0 | | | | |
|---|---|---|---|---|---|---|---|---|---|---|
| **Segment** | $h_0$ | **B** | $\sigma_{h_0}$ | $\sigma_B$ | $R^2$ | $h_0$ | **B** | $\sigma_{h_0}$ | $\sigma_B$ | $R^2$ |
| Head | 0.567 | 1.036 | 0.050 | 0.032 | 0.997 | 0.567 | 1.398 | 0.050 | 0.029 | 0.991 |
| Chest | 0.227 | 0.958 | 0.035 | 0.022 | 0.998 | 0.227 | 0.227 | 0.227 | 0.227 | 0.227 |
| Back | 0.248 | 0.940 | 0.035 | 0.022 | 0.998 | 0.248 | 1.173 | 0.035 | 0.026 | 0.987 |
| Whole torso | 0.290 | 0.996 | 0.037 | 0.023 | 0.996 | 0.290 | 1.025 | 0.037 | 0.024 | 0.984 |
| Upper Arm | 0.854 | 1.158 | 0.049 | 0.031 | 0.998 | 0.854 | 1.187 | 0.049 | 0.033 | 0.988 |
| Forearm | 1.116 | 1.535 | 0.060 | 0.038 | 0.999 | 1.116 | 1.114 | 0.060 | 0.033 | 0.989 |
| Whole arm | 0.937 | 1.278 | 0.052 | 0.033 | 0.998 | 0.937 | 1.164 | 0.052 | 0.033 | 0.988 |
| Pelvis | 0.360 | 1.061 | 0.040 | 0.025 | 0.998 | 0.360 | 0.872 | 0.040 | 0.021 | 0.987 |
| Thigh | 0.538 | 1.166 | 0.044 | 0.028 | 0.998 | 0.538 | 0.654 | 0.044 | 0.017 | 0.990 |
| Calf | 0.608 | 1.291 | 0.052 | 0.033 | 0.998 | 0.608 | 0.596 | 0.052 | 0.015 | 0.992 |
| Whole leg | 0.575 | 1.233 | 0.048 | 0.031 | 0.998 | 0.575 | 0.623 | 0.048 | 0.016 | 0.991 |
| Foot | 1.151 | 1.673 | 0.081 | 0.051 | 0.998 | 1.151 | 0.831 | 0.081 | 0.032 | 0.986 |

**Table 2. Summary of regional free convective coefficients ($h_c$, W·m⁻²·K⁻¹) used across various modeling approaches namely Stolwijk** (*10*), **JOS-2 & JOS-3** (*46*), **Wissler** (*8*) **and the present study (Eq. 8).** $N = \Delta T_{sk-air} + 146\Delta C_{sk-air}$, where $N$ is the coupled buoyancy driving term, $h_0$ (diffusion-only term, W·m$^{-2}$·K$^{-1}$ ) and B are fitted parameters, $\Delta T_{sk-air}$ (°C) and $\Delta C_{sk-air}$ (kg·m$^{-3}$) are the skin-to-air temperature and concentration differences respectively (see **Supplementary Materials** for details).

| **Model** | **Stolwijk** (*10*, *39*) | **JOS-2 & 3** (*46*) | **Wissler** (*8*) $h_c = A(\Delta T_{sk-air})^{1/4}$ | **Present study** $h_c = h_0 + B\lvert\Delta T_{sk-air} + 146\Delta C_{sk-air}\rvert^{1/4}$ | | |
|---|---|---|---|---|---|---|
| **Zone** | $h_c$ **(W·m$^{-2}$·K$^{-1}$)** | $h_c$ **(W·m$^{-2}$·K$^{-1}$)** | $A$ | $h_0$ **(W·m$^{-2}$·K$^{-1}$)** | $B\ (N > 0)$ | $B\ (N < 0)$ |
| Head | 0.660 | 4.480 | 2.600 | 0.567 | 1.036 | 1.398 |
| Torso | 1.860 | 2.910 | 1.500 | 0.290 | 0.996 | 1.025 |
| Arms | 3.950 | 3.590 | 1.900 | 0.937 | 1.535 | 1.164 |
| Hands | 6.050 | 3.670 | 2.700 | 0.500 | 2.400 | 2.400 |
| Legs | 3.610 | 2.530 | 2.150 | 0.575 | 1.233 | 0.623 |
| Feet | 5.930 | 2.040 | 1.570 | 1.151 | 1.673 | 0.831 |

**Supporting Information for**

# Perspiration vapor lightens near-skin air—but hinders human evaporative cooling in arid heat

Shri H. Viswanathan, Ankit Joshi, Isabella DeClair, Bryce Twidwell, Muhammad Abdullah, Lyle Bartels, Faisal Abedin, Joseph Rotella, Cibin T. Jose, and Konrad Rykaczewski

Konrad Rykaczewski
Email:konradr@asu.edu

**This PDF file includes:**

Supplementary Text
Figs. S1 to S23
Tables S1 to S10
Matlab thermoregulation code
References

**Supplementary Text S1**

**1. Chilton-Colburn analogy and Lewis relationship**

During simultaneous heat (convection) and mass (sweat evaporation) transfer, the transport processes are analogous and the convective heat ($h_c$, W·m$^{-2}$·K$^{-1}$) and mass ($h_{mass}$, m·s$^{-1}$) transfer coefficients are related through the Chilton-Colburn analogy (**Eq. S1**) (for $0.6 < Pr < 60$ and $0.6 < Sc < 3000$), where the Lewis number ($Le = \alpha/D = k/(\rho c_p D)$, dimensionless, (*1*)) is defined as the ratio of thermal ($\alpha$, m$^2$·s$^{-1}$) to mass diffusivity ($D$, m$^2$·s$^{-1}$), $k$–thermal conductivity of air (W·m$^{-1}$·K$^{-1}$), $\rho$–air density (kg·m$^{-3}$), $c_p$–specific heat at constant pressure (J·kg$^{-1}$·K$^{-1}$). **Eq. S2 & S3** describe the mass flux rate of sweat evaporated ($\dot{m}$, kg·m$^{-2}$·s$^{-1}$) and evaporative heat flux ($Q_e$, W·m$^{-2}$), where $\lambda$ is the latent heat of vaporization (J·kg$^{-1}$), $p$–atmospheric pressure (Pa), $M_v$ & $M_{air}$–molar masses of water vapor and moist air mixture respectively (kg·mol$^{-1}$), $p_{sk}$ & $p_{air}$–vapor pressures at the skin surface and ambient air respectively (kPa). Substituting **Eq. S1 & S2** into **Eq. S3** gives **Eq. S4**. The terms within the square brackets are grouped together and evaluated at air temperature of 25 °C (the air properties variation has minimally impact on the calculations across the 20–50 °C range), and skin temperature of 32 °C (*2*), forming the Lewis relationship ($LR = 16.5$ K·kPa$^{-1}$) (**Eq. S5**). Finally, the evaporative transfer coefficient ($h_e$, W·m$^{-2}$·kPa$^{-1}$) is expressed as a function of the Lewis relationship (*3*) and the convective heat transfer coefficient (**Eq. S6**).

$$\frac{h_c}{h_{mass}} = \rho c_p Le^{2/3} \quad \textbf{(S1)}$$

$$\dot{m} = \frac{h_{mass}\rho}{p}\frac{M_v}{M_{air}}(p_{sk} - p_{air}) \quad \textbf{(S2)}$$

$$Q_e = \dot{m}\lambda \quad \textbf{(S3)}$$

$$Q_e = \left[\frac{\lambda}{p c_p Le^{2/3}}\frac{M_v}{M_{air}}\right] h_c(p_{sk} - p_{air}) \quad \textbf{(S4)}$$

$$Q_e = LR\, h_c(p_{sk} - p_{air}) = 16.5\, h_c(p_{sk} - p_{air}) = h_e(p_{sk} - p_{air}) \quad \textbf{(S5)}$$

$$h_e = 16.5\, h_c \quad \textbf{(S6)}$$

**Supplementary Text S2**

**2. Experimental verification of the Lewis relationship under forced convective conditions**

Controlled experiments were conducted within a climate-controlled wind enclosure using a thermal manikin (**Fig. S1A–F**) and the experimentally measured evaporative coefficients were compared against the Lewis factor (16.5 K·kPa$^{-1}$, **Eq. S6**) scaled dry forced convective coefficient.

**2.1 Wind Enclosure**

A custom wind enclosure (3.5 m length, 1.14 m width, 2.28 m height) was built within a climatic chamber (6 m length, 2.75 m width, 3 m height) (**Fig. S1C**) using T-slot aluminum extrusions and polycarbonate panels (*4*, *5*). To simplify the computational estimation of radiative heat transfer, the enclosure's inner walls, ceiling and floor were covered with highly emissive black kraft paper with a longwave emissivity of 0.99 (measured using a PerkinElmer Frontier Fourier Transform Infrared Spectrometer (FTIR) (*4*, *6*)). To ensure a horizontal homogenous airflow with low turbulence, air was pulled across the enclosure by three fans (XPOWER FD-630D, 0.61 m diameter) through a 6 cm thick polycarbonate honeycomb panel (manufactured by Plascore, Inc., 6 mm pores) at the entrance (**Fig. S1D**). The mean upstream air speed ($v_{air}$) was measured across five locations using five omnidirectional hot wire anemometers (**Fig. S1E**). Moreover, the inlet air temperature ($T_{air}$), and relative humidity ($RH_{air}$) (**Fig. S1F**) were also measured using the instruments outlined in **Table S1**.

**2.2 Thermal manikin – ANDI**

To investigate the heat and moisture transport around the human body, experiments were conducted using a custom-built thermal manikin, "Advanced Newton Dynamic Instrument" (ANDI) developed by Thermetrics, LLC (Seattle, USA). ANDI (**Fig. S1A–B**) represents a 50th percentile western male, standing 1.78 m tall with a total surface area of 1.86 m$^2$. The manikin consists of 35 independently controlled segments, each with area-distributed resistive temperature sensors embedded beneath the surface (*4*, *5*, *7*, *8*). ANDI was placed inside the enclosure and was mounted on a non-reflective mobile stand at its lower back. The actuated joints of ANDI's limbs were covered with masking tape to prevent the air flow from entering the hollow manikin shell (*4*). The manikin's surface can be operated in either constant uniform skin temperature ($T_{sk}$) or heat flux mode. The former is more replicative of the human skin in a thermally neutral environment (*3*, *9*). Moreover, this method averts the complications of inter-segment radiation which would otherwise occur during constant uniform heat flux mode (*4*) due to non-uniform skin temperature.

While legacy thermal manikins are designed solely to quantify heat loss into the surroundings, ANDI's area-distributed differential heat flux sensors, embedded beneath the surface, can measure both heat loss to and gain from the surroundings. Additionally, the internal cooling channels present under the surface absorb the excess incoming heat, preventing the manikin skin temperature ($T_{sk}$) from exceeding its setpoint (*7*, *8*). These capabilities enable its utility in both hot and cold environments. Furthermore, ANDI replicates the human sweat evaporation process through its 151-flow rate controlled sweat pores spread evenly across the body. To achieve a uniform moisture dispersion across the surface, the manikin was dressed in an artificial textile "skin" (sweat wicking polyester onesie, 0.5 mm thick). Excess water runoff from the manikin was collected in a tray placed beneath the feet. To prevent water evaporating from the tray and disrupting the manikin measurements, a thin layer of oil was applied to water surface.

**2.3 Experimental procedure**

To verify the applicability of the Lewis relationship under forced convective conditions, two complementary sets of experiments were conducted using the thermal manikin present within the wind enclosure. For each set, five different air speed settings were chosen between 0.34 and 1.71 m·s$^{-1}$. Firstly, to isolate the sensible-heat exchange and estimate the dry forced convective heat transfer coefficient ($h_{dry}$, W·m$^{-2}$·K$^{-1}$), the manikin surface was maintained dry at constant uniform temperature of 35 °C, while the inlet air was held at 22.9 ± 0.1 °C and 37.9 ± 1.2% RH.

Secondly, to eliminate sensible heat exchange and isolate the latent-heat exchange (only velocity driven mass transfer at constant $RH_{air}$ i.e., sweat evaporation), $T_{sk}$ and $T_{air}$ were maintained at 35 °C (isothermal conditions). The manikin was dressed in its water-saturated textile "skin", while $RH_{air}$ was held at 49.4 ± 5.5 % RH. The wet textile "skin" was characterized to offer a thermal resistance ($R_{txt}$, Inset of **Fig. S2**) of 0.0054 m$^2$·K·W$^{-1}$. During experimentation, all the enclosure walls were nearly in equilibrium with $T_{air}$. Once the enclosure's thermal environment stabilized, all zonal manikin parameters (e.g., skin temperature, heat flux etc.) were recorded every 5 s. All 10 cases were repeated thrice leading to 30 trials, where each trial lasted 75 minutes. During the final 30 minutes, the segmental skin temperature fluctuations remained under 0.05 °C, indicating steady state condition. We note that under the low heat fluxes occurring during free convection, the textile skin thermal resistance has negligible impacts.

**2.4 Experimental results**

The whole-body total heat flux ($Q_{Total}$, W·m$^{-2}$) was computed by segmental area weighting of the zonal-time average data collected during steady state. Specifically for the dry forced convection trials, the computationally estimated whole-body radiative heat flux (60.1 W·m$^{-2}$) was subtracted from $Q_{Total}$, and the remainder was divided by the skin-to-air temperature gradient, to yield the dry forced convective heat transfer coefficient ($h_{dry}$, W·m$^{-2}$·K$^{-1}$). To obtain its evaporative-equivalent, $h_{dry}$ was multiplied by the Lewis factor (16.5 K·kPa$^{-1}$, **Eq. S6**). On other hand, for the isothermal velocity driven mass transfer experiments, the total heat flux ($Q_{Total}$) corresponds only to the evaporative heat losses ($Q_e$) from the manikin. Accounting for the textile skin's thermal resistance, there exists a significant temperature drop between manikin surface ($T_{sk}$) and textile "skin" ($T_{txt}$). So $Q_e$ was divided by the textile-to-air vapor pressure gradient ($p_{txt|\,T_{txt}} - p_{air|\,T_{air}}$) (Inset of **Fig. S2**) to obtain the corrected evaporative heat transfer coefficient ($h_{exp}$, W·m$^{-2}$·kPa$^{-1}$).

As illustrated in **Fig. S2**, there exists a strong agreement between the evaporative equivalent of $h_{dry}$ and $h_{exp}$ across the tested air speed range, thus experimentally demonstrating the validity of the Lewis relationship under forced convective conditions. Furthermore, the evaporative equivalent of the empirical dry forced convective coefficient proposed by Joshi et al. (*7*) aligned closely with our experimental results. In contrast, the evaporative coefficients (uncorrected $h_{exp}$, W·m$^{-2}$·kPa$^{-1}$) computed without considering the textile skin's resistance ($Q_e$ directly divided by skin-to-air vapor pressure gradient ($p_{sk|\,T_{sk}} - p_{air|\,T_{air}}$)) were consistently lower than $h_{exp}$ across the entire test range, highlighting the importance of including the manikin's textile "skin" resistance in evaporative flux measurements.

**Supplementary Text S3**

**3. Experimental validation of the coupled free convection, sweat evaporation and radiation model using thermal manikin**

This section presents the experimental validation of the coupled free convection, evaporation and radiation model using the sweating thermal manikin ANDI. We conducted controlled experiments inside a sealed enclosure under a range of temperature and humidity conditions and replicated these cases using multiphysics simulations. The measured whole-body total heat fluxes closely matched the simulations, confirming the model's reliability across diverse conditions.

**3.1 Sealed enclosure**

Controlled experiments were conducted inside a sealed enclosure located within a climatic chamber (**Fig. S3A–F**) as previously described in **Supplementary Text S2**. To study free convection within the enclosure, we ensured the air speed within the enclosure was less than 0.02 m·s$^{-1}$ by completely sealing the enclosure (including the honeycomb entrance wall and fan wall) (**Fig. S3D**) with highly emissive black kraft paper. Moreover, the black kraft paper also simplified the computational estimation of radiative heat transfer. Additionally, a chimney (0.81 m length, 1.14 m width, 0.31 m height) was integrated into the enclosure to function either as an inlet for fresh air from the chamber or an outlet for the rising thermal plume, depending on the thermal conditions and the resulting buoyancy-driven flow (*4*). The instrumentation used within the enclosure to measure the air temperature ($T_{air}$), relative humidity ($RH_{air}$), air speed ($v_{air}$) and wall temperature ($T_{wall}$) (**Fig. S3A–E**) are outlined in **Table S2**.

**3.2 Experimental procedure**

We conducted a comprehensive set of experiments under controlled conditions (15 cases) as listed in **Table S3**, to generate buoyant flow near the skin driven by temperature-only, humidity-only, and combined temperature-humidity effects. Each case was repeated thrice to demonstrate reproducibility, resulting in 45 trials. Before each run, the thermal manikin (as described in **Supplementary Text S2**) was positioned beneath the chimney, and the enclosure was sealed. The air speed ($v_{air}$) (**Fig. S3B**) within the enclosure remained under 0.02 m·s$^{-1}$, indicating conditions suitable for investigating free convection.

To isolate temperature-only driven free convection (cases 1–4), the manikin surface was held dry (**Fig. S3F**). $T_{air}$ was varied between 40.3 and 44.8 °C at constant absolute humidity (12.55 g of water per kg of dry air). Here, the heat flux measured by the manikin corresponds only to free convective and radiative heat gain from the environment ($T_{air} > T_{sk}$). To isolate humidity-only driven sweat evaporation (cases 5–8), both the manikin surface and surrounding air were isothermally held at 35 °C, while $RH_{air}$ was varied between 35.6 and 81.4%. To enable a uniform moisture dispersion across the manikin surface, replicative of the human skin, the manikin was covered with the water-saturated (at 35 °C) textile "skin" (**Fig. S3F**). Finally, the combined effects of temperature and humidity were examined in cases 9-15 by maintaining $T_{air}$ 4.6 °C or 7.5 °C above $T_{sk}$, meanwhile varying $RH_{air}$.

Once the enclosure's thermal environment stabilized, the manikin was switched on, and its zonal parameters (e.g., skin temperature, heat flux etc.) were recorded every 5 s. Each run lasted 75 minutes, with segmental skin temperature fluctuations remained under 0.05 °C during the final 30 minutes, indicating steady state condition. The far-field air temperature ($T_{air}$) (**Fig. S3C**) measured within the sealed enclosure was used for analysis and was consistently 0.1 to 1.6 °C lower than the chamber setpoints ($T_c$, °C). For each trial, the whole-body total heat flux ($Q_{Exp-WB}$, W·m$^{-2}$) was computed by segmental area weighting of the zonal–time average data collected during steady state.

The Type A (experimental) uncertainty ($u_A$) and Type B (instrumental) uncertainty ($u_B$) were estimated for each zonal total heat fluxes ($Q_{Exp-Z}$) and combined to determine the total standard uncertainty ($u_C$). A coverage factor ($k$) was multiplied to $u_C$ to expand the uncertainty to form the overall uncertainty ($U$) (**Eq.

**S7**). For a two-tailed Student's t–distribution with 2 degrees of freedom, 3 repetitions, 95% confident level, $k$ was 4.303.

$$U = ku_C = k\sqrt{(u_A^2 + u_B^2)} \quad \textbf{(S7)}$$

**3.3 Coupled radiative, flow, moisture and heat transfer simulation**

A multiphysics computational model was developed in COMSOL Multiphysics 6.3 (finite element-based software) to quantify coupled heat and mass transfer around the thermal manikin and validate the whole-body and zonal heat fluxes against experimental measurements. The experimental sealed enclosure was geometrically replicated and a water-tight surface mesh of the thermal manikin, used in our prior work (*4*), was imported into the virtual enclosure as shown in **Fig. S4A**. The virtual manikin closely matched its physical counterpart with its total surface mesh area differing only by 1.7% from the actual manikin. To reduce the computational time, symmetry conditions were applied for the entire geometry along the sagittal/median plane.

The numerical model solves for various physics modules namely, "Fluid Flow" (Low Re k-ε Turbulence model with Low Re wall functions), " Moisture Transport", "Heat transfer in fluids" and "Surface-to-Surface Radiation". These modules were coupled together through "Non–Isothermal flow", "Heat transfer with surface–to–surface radiation", "Heat and moisture flow", and "Moisture flow" multiphysics modules. Moist air was chosen as the fluid medium, and gravity was enabled to account for the buoyancy effect. A weakly compressible flow was assumed to simulate the buoyancy generated due to moisture concentration and temperature gradients (density variations induced by fluid pressure gradients were neglected). For cases 1 to 4 and 9 to 15, since $T_{air} > T_{sk}$, a downward plume was established by drawing fresh air into the enclosure from the chamber through the chimney. Accordingly, the chimney's top surface served as the inlet, with a velocity of 0.02 m·s$^{-1}$ (5% turbulence intensity and geometry based turbulence length scale) and thermal conditions identical to those of the chamber ($T_c$ & $RH_c$, **Table S3**), while the entire enclosure floor served as the outlet. In contrary, for cases 5 to 8), the inlet and outlet boundaries were interchanged, due the upward plume ($T_{air} < T_{sk}$).

All the model reference values were computed at $T_{air}$ & $RH_{air}$, representative of the thermal conditions within the enclosure measured far from the manikin. The virtual manikin's surface temperature was set to $T_{sk}$ from **Table S3**. While simulating the isothermal cases 5-8 and the combined heat and mass transfer experiments (Cases 9-15), the evaporative rate parameter of the virtual manikin's surface was set to 100 m·s$^{-1}$, representative of fully wet surface. The enclosure's humidity was set to $RH_{air}$ from **Table S3**. The radiative heat exchange was computed using surface-to-surface radiation method (hemicube technique with resolution of 64). Since all the surfaces were assumed to be diffuse and gray, the surface radiative properties were set to "wavelength independent". The emissivity of the enclosure and the manikin surface were set to 0.99 and 0.98 respectively. Since $T_{air}$ was consistently cooler than $T_c$, to model the heat gained by the enclosure from the chamber, an effective heat transfer coefficient of 5.6 W·m$^{-2}$·K$^{-1}$ was applied on the enclosure walls (*4*).Unlike in **Supplementary Text S2**, where accounting for the textile skin's thermal resistance was crucial in the evaporative heat loss calculations, it was neglected and not modelled computationally in cases 5-15 because the measured $Q_{Exp-WB}$ were small, leading to a negligible temperature difference ($T_{sk} - T_{txt}$< 0.5 °C) between the manikin surface and the textile skin.

The entire enclosure was meshed using free triangular cells for the floor and prism elements for the remaining volume. Two high-density cuboidal mesh volumes (refinement zones) were added around the manikin to accurately resolve the velocity, temperature and moisture gradients near the manikin surface. **Fig. S4B** demonstrates the mesh refinement study that was conducted to determine when the free convective heat transfer coefficient and total heat flux becomes independent of the number of mesh elements. The final mesh contained 507510 elements with a maximum element size of 7.2 cm, a minimum element size of 1.6 cm, an element growth rate of 1.14, a curvature factor of 0.5, and a resolution of narrow region of 0.8. Following the COMSOL Multiphysics software's recommendations (*10*), the "distance to the cell center in viscous units" was maintained below 0.5 by applying 15 boundary layers, stretching factor of 1.2, and a total boundary layer thickness of 1 cm. The average element quality was 0.76. Following meshing, the model was initialized using a wall distance step and solved with a time-dependent study over 300 s. A quasi-steady state was reached between 240 and 300 s, with net heat flux variations under 5%.

Time-averaged free convective, radiative and evaporative heat fluxes were computed to estimate the total heat flux ($Q_{Sim-WB}$, W·m$^{-2}$) which was benchmarked against the experimental measurements.

## 3.4 Experimental validation of the multiphysics model

**Fig. S5A–D** compares whole-body and zonal total heat fluxes measured experimentally and estimated by simulations for all the 15 cases. Overall, the multiphysics model closely agrees with the experimental measurements, with simulated total heat fluxes generally within the experimental uncertainties. In cases 7, 9 & 13, the model underestimates the experimental heat fluxes by up to 23%, slightly exceeding the uncertainty bounds. To facilitate regional comparison, ANDI's 35 zones were grouped into 12 larger anatomical regions by area-weighted summation. Hands and feet were excluded from the analysis for several reasons. In particular, for cases 5–15, excess sweat from other zones accumulated on the fingers and feet and dripped into the water tray present below, which was not modeled in the simulations. In addition, only the manikin's palmar regions were heated and sweating, while the fingers were passive, hence hands were excluded.

**Fig. S6A–D** shows strong overall agreement between experimental ($Q_{Exp-Z}$, W·m$^{-2}$) and simulated ($Q_{Sim-Z}$, W·m$^{-2}$) zonal total heat fluxes. In particular, the head, back torso, upper arms, forearms, thighs, and calves exhibit good consistency, with simulated values falling within experimental uncertainties in at least 60% of cases. Minor discrepancies in the aforementioned regions are primarily attributable to simplified geometry of local zones and the partially loose fit of the manikin's textile skin (sweat wicking garment). For instance, simulated fluxes for the upper arms, forearms, thighs and calves agree with experiments in 68.3% of cases, with average deviations of only 10.1%, 14.1%, 9.5% and 13.3% respectively. Minor discrepancies arise in the remaining cases because the physical manikin has passive actuated joints around the elbows and kneecaps that are unheated and lack sweat pores, whereas the virtual model simplifies these regions as one continuous heated wet surface. For the back torso, simulations matched experimental values in 10 of 15 cases, with an average deviation of 12.9%. Discrepancies in the remaining cases likely arise from the unmodeled fixtures in the lower back—such as the mobile stand attachment, cooling water lines, and communication cables—that disrupt the local buoyant plume. This plume disruption also affected the pelvis region, where additional factors amplified the average deviations to 18.3% from experimental values. Particularly, a passive insulated joint along the sagittal plane, designed to support leg movement, was geometrically simplified in the virtual model as an actively heated, wet surface. Furthermore, the textile "skin" was loosely fitted near the crotch, creating air gaps that altered local heat fluxes.

Simulated total heat fluxes for the head closely matched experimental measurements for 10 of 15 cases, with an average deviation of 11.9%. These discrepancies were primarily due to the manikin's crown that comprises of a passive hollow shell intended for occasional attachment to the mobile stand, and partial coverage of the head and neck by the textile skin's zipper. In cases 9–12, nearly 70% of the head remained partially dry due to clogged sweat pores, reducing evaporation and total heat flux experimentally. After unclogging the pores, the head was maintained fully wet in cases 13–15, resulting in closer agreement with simulations. For the front torso, simulated total heat fluxes deviated on average by 22.1% from experimental measurements. The largest contributor to this discrepancy was the simplified armpit geometry in the model, whereas the physical manikin features a passive actuated upper arm–shoulder joint that was not represented computationally. Additional factors included the loosely fitting textile "skin" around the armpit, which occasionally produced partially dry patches in the pectoral and underarm regions, further affecting local heat fluxes.

Additionally, the wall temperature at the location shown in **Fig. S3A** and the air speed measured above the manikin's head (**Fig. S3E**) exhibited good agreement with the corresponding simulated values, as illustrated in **Fig. S7A–D**, further supporting the validity of the numerical model.

## Supplementary Text S4

## 4.1 Experimentally validated multiphysics model setup for parametric study

The parametric study was performed in a simplified virtual enclosure (2 m length, 1 m width, 3 m height) (**Fig. S8A**). A 3D watertight geometry of a western average male (50$^{th}$ percentile height (m) and BMI (kg·m$^{-}$ [2])) was positioned at the geometric center of the enclosure (**Fig. S8B**). Since the entire geometry was symmetrical, only one-half of the geometry was simulated, to reduce the computational cost. The model consisted of "Turbulent Fluid Flow", "Heat Transfer in Fluids", and "Moisture Transport" physics modules and were interlinked through "Non-Isothermal flow", "Heat and Moisture Transfer" and "Moisture flow" multiphysics couplings. A Low-Re k-ε Turbulence model (Low-Re wall functions) was used to accurately resolve the flow features near the skin surface. A weakly compressible flow of moist air was assumed, and the gravity module was enabled to capture the buoyancy-driven flow. The skin was maintained at a constant temperature of 35 °C and modeled as a saturated wet surface. To prevent numerical convergence issues, the inlet and outlet boundaries of the enclosure were dynamically assigned based on the natural flow direction. For cases with an ascending buoyant plume, the floor served as the inlet and the ceiling as the outlet and vice versa for a descending plume. The inlet $T_{air}$ and $RH_{air}$ were parametrically varied between 20–50 °C and 10–90% respectively.

The entire geometry was discretized using free triangular elements for the floor and prism elements in the remaining volume. Two high-density cuboidal mesh volumes were generated around the body to accurately capture the steep velocity, temperature and moisture gradients near the skin surface. A mesh refinement study was conducted to identify when the free convective coefficient ($h_{sim}$, W·m$^{-2}$·K$^{-1}$) became independent of mesh density (**Fig. S8C**). The final optimized mesh contained 370576 elements with a maximum and minimum element size of 7.5 and 2 cm respectively, an element growth rate of 1.13, a curvature factor of 0.5, and a narrow region resolution of 0.8. The "distance to the cell center in viscous units" was maintained below 0.5 by applying 15 boundary layers, stretching factor of 1.2, and a total boundary layer thickness of 2 cm, resulting in an average element quality of 0.75. The model was initialized with a wall distance method and solved in a transient study (50 s) to compute the free convective heat transfer coefficients ($h_{sim}$) and evaporative heat losses ($Q_{sim}$, W·m$^{-2}$). The simulated free convection coefficient ($h_{sim}$) is computed by extracting the surface-integrated convective heat flux from the COMSOL Multiphysics solver (variable: ht.ntflux) and dividing by the absolute skin-to-air temperature difference ($|\Delta T_{sk-air}|$). The variation in $h_{sim}$ decreased by less than 2% after 30 s, indicating that a quasi-steady state had been achieved.

**4.2 Buoyancy forces induced by temperature- and humidity-differences**

The individual buoyant forces arising from the temperature ($\Delta T_{sk-air} = T_{sk} - T_{air}$) and humidity or moisture concentration differences ($\Delta C_{sk-air} = C_{sk} - C_{air}$), namely, $B_{\Delta T} = g\rho\beta_T\Delta T_{sk-air}$ and $B_{\Delta C} = g\rho\beta_C\Delta C_{sk-air}$, were computed, where $g$ is the gravitational acceleration (9.81 m·s$^{-2}$), $\beta_T$ is the thermal coefficient of expansion (reciprocal of the absolute average of $T_{air}$ and $T_{sk}$), and $\beta_C$ is the coefficient of expansion due to compositional changes. $\beta_C$ can be obtained from the density of air ($\rho$), the average molecular weight of air ($MW_a$ =29 g·mol$^{-1}$), and that of water vapor ($MW_C$ =18.02 g·mol$^{-1}$) as $(MW_a/MW_C - 1)/\rho$ or $0.607/\rho$ (*11*). **Table S4** outlines the behavior of $B_{\Delta T}$, $B_{\Delta C}$ and the net buoyancy $B_{Net}$ —which is the sum of both buoyancies—across various air temperature regimes. $B_{\Delta T}$, $B_{\Delta C}$ and $B_{Net}$ are computed and plotted as function of skin-to-air temperature differences for each $RH_{air}$ as shown in **Fig. S9A–C, Fig. S10A–C**, and **Fig. S11A–C**

**4.3 Coupled temperature and humidity impact on free convection and sweat evaporation**

Using the experimentally validated multiphysics model, we systematically varied the $T_{air}$ (20–50 °C) and $RH_{air}$ (10–90%) (100 cases total using Research Computing at Arizona State University (*12*)) to investigate the coupled impact of thermal- and humidity-gradients on whole-body free convective coefficients ($h_{Sim}$, W·m$^{-2}$·K$^{-1}$) and evaporative heat fluxes ($Q_{Sim}$, W·m$^{-2}$), for an average western male (50$^{th}$ percentile BMI and height), while maintaining a water-saturated skin surface at 35 °C. **Fig. S12A–F, Fig. S13A–F**, and **Fig. S14A–F** compare the simulated values against the widely used empirical correlations that are solely a function of the skin-to-air temperature-differences.

$$h_{\Delta T-only} = 1.36\,|\Delta T_{sk-air}|^{0.3} \quad \textbf{(S8)}$$

$$Q_{\Delta T-only} = 16.5\,h_{\Delta T-only}\,\Delta p_{sk-air} \quad \textbf{(S9)}$$

Where, $h_{\Delta T-only}$ (W·m$^{-2}$·K$^{-1}$) is the empirical whole-body dry free convective heat transfer coefficient correlation derived from our prior studies ($T_{sk} > T_{air}$, (*4*)) and validation cases 1–4 as described in **Supplementary Text S3**. $Q_{\Delta T-only}$ (W·m$^{-2}$) is the evaporative heat flux obtained as a function of the empirical dry free convective coefficient (**Eq. S8**), Lewis factor (16.5 K·kPa$^{-1}$, **Eq. S6**) and the skin-to-air vapor pressure differences ($\Delta p_{sk-air} = p_{sk} - p_{air}$). Complete simulation results for all cases are provided in **Table S8**, **Table S9** and **Table S10**.

**Supplementary Text S5**

**5. Physics-based model for free convective coefficients accounting for humidity-induced buoyancy**

$$h_c = h_0 + B|\Delta T_{sk-air} + 146\Delta C_{sk-air}|^{1/4} \quad \textbf{(S10)}$$

Here, $h_0$ is the diffusion-only value, $B$ is the coefficient specifying geometrical, flow, and fluid property impacts, and $\Delta T_{sk-air} + 146C_{sk-air}$ is the combined temperature and concentration buoyancy driving term. Fitting this physics-informed model (**Eq. S10**) to individual body segments yields excellent agreement with simulation results (see Manuscript, **Table 1** and **Fig. S15A–F**, **Fig. S16A–F** & **Fig. S17A–F**).

**Supplementary Text S6**

**6.1 Local free convective correlations for hands and feet**

Simulated local convective coefficients ($h_c$, W·m$^{-2}$·K$^{-1}$) for hands were fitted using temperature-difference scaling (**Eq. S11**) and combined temperature- and concentration-difference scaling (**Eq. S12**) approaches and compared with average Quintela et al. correlations (*13*) (**Eq. S13–S14**) for hands across skin-to-air temperature differences for dry skin (i.e., $\Delta C_{sk-air} = 0$) (**Fig. S18A**). Similarly for feet, the fitted correlations using both approaches (**Eq. S15–S16**) were compared in **Fig. S18B**.

$$h_c = 2.7\Delta T_{sk-air}{}^{1/4} \quad \textbf{(S11)}$$
$$h_c = 0.5 + 2.4\Delta T_{sk-air}{}^{1/4} \quad \textbf{(S12)}$$
$$h_c = 1.58\,\Delta T_{sk-air}{}^{0.51} \text{ (Left hand)} \quad \textbf{(S13)}$$
$$h_c = 3.55\,\Delta T_{sk-air}{}^{0.11} \text{ (Right hand)} \quad \textbf{(S14)}$$
$$h_c = 1.57\Delta T_{sk-air}{}^{1/4} \quad \textbf{(S15)}$$
$$h_c = 0.969 + 0.99\Delta T_{sk-air}{}^{1/4} \quad \textbf{(S16)}$$

**6.2 Posture influence on free convection**

Apart from the standing posture used in this study, equivalent simulations were performed for seated and supine postures. The standing manikin in the validated numerical model was replaced with manikins in seated and supine postured manikins, and air temperature was varied between 20–50 °C with skin temperature fixed at 35 °C. Whole-body heat transfer coefficients were extracted and fitted yielding $h_{seated} = 0.48 + 1.19\,|\Delta T_{sk-air}|^{1/4}$ and $h_{supine} = 0.34 + 1.46\,|\Delta T_{sk-air}|^{1/4}$ (**Fig. S18C**).

**Supplementary Text S7**

**7. Domain-size sensitivity analysis for diffusion baseline ($\mathrm{Nu_0}$)**

Pure heat diffusion simulations were performed in a rectangular enclosure (**Fig.S20C**) by simultaneously varying the width (0.8–3 m) and length (1–6 m), while holding the height constant at 3 m. Ambient temperature was varied from 20 to 50 °C in steps of 5°C. The manikin geometry and all other settings were identical to the main parametric model. Enclosures smaller than the original domain (2 m × 1 m × 3 m) yielded slightly elevated $\mathrm{Nu_0}$ values due to artificial wall proximity effects. From the original domain size onwards, the domain-size had negligible impact on $\mathrm{Nu_0}$, as shown in **Fig.S20A**. The domain-averaged $\mathrm{Nu_0}$ computed from stabilized cases (original domain size and above) was 23.44 ± 2.59.

To confirm that $\mathrm{Nu_0}$ is not sensitive to enclosure geometry, the diffusion simulations were repeated in a cylindrical enclosure (**Fig.S20D**) with a radius varied from 0.8 to 3.2 m, height fixed at 3 m. A consistent trend was observed — slightly elevated $\mathrm{Nu_0}$ at the smallest radius (0.8 m), with negligible impact with larger diameter, see **Fig.S20B**. The domain-averaged $\mathrm{Nu_0}$ from stabilized cases was 23.14 ± 2.58, in close agreement with the rectangular enclosure results.

**Supplementary Text S8**

**8. Air properties used in calculations**

The thermal conductivity, kinematic viscosity, and thermal diffusivity were interpolated in Wolfram Mathematica using Interpolate[] function from values obtained from Bergman et al. textbook (*14*). The values of the diffusion coefficient were calculated using Marrero and Mason formula (*15*) of $D(T) = 0.187 \times 10^{-9}(T + 273.15)^{2.072}$, where temperature is in degrees Celsius. The air density and coefficient of expansion due to compositional changes were calculated from in-built thermodynamic data within Wolfram Mathematica. The properties are specified for the range of 20 to 50 °C and 10 to 100% relative humidity in **Tables S5 and S6**.

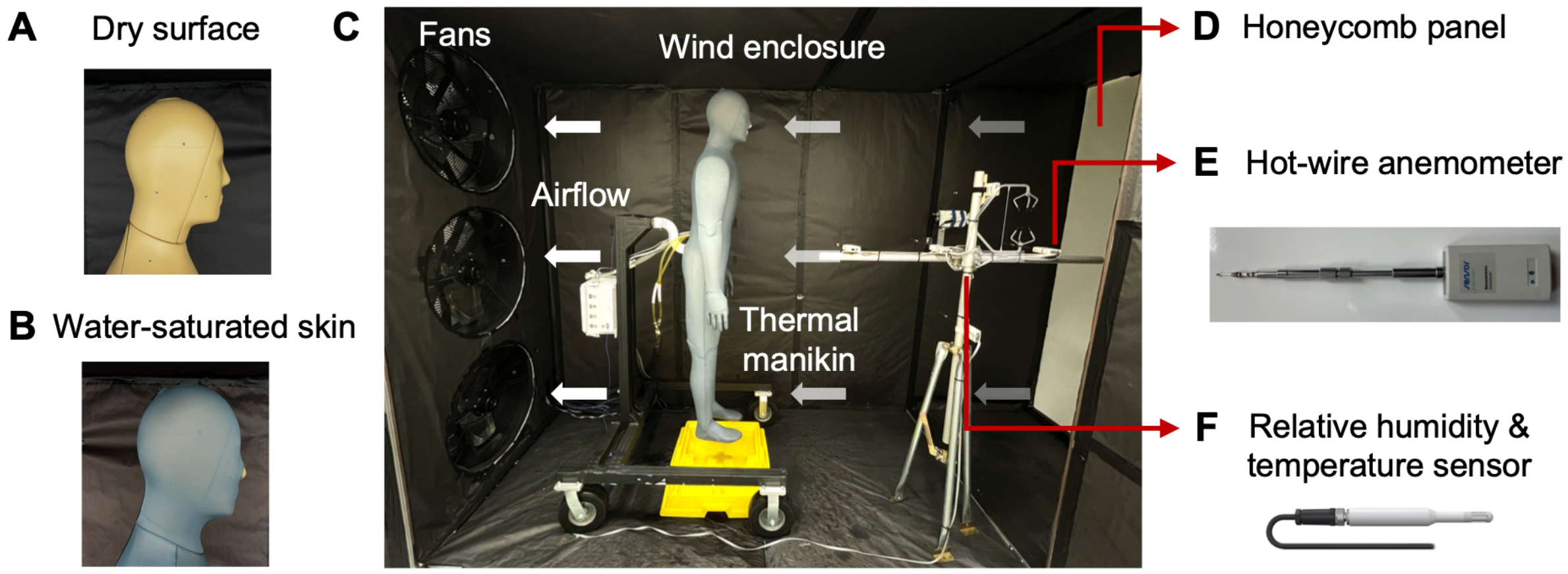

**Fig. S1. Forced convection experimental setup with the thermal manikin inside the wind enclosure.** **(A)** Dry manikin surface during sensible-heat exchange experiments, **(B)** Water-saturated textile skin worn by the manikin during latent-heat exchange experiments, **(C–D)** Airflow drawn through the honeycomb panel by the fans, **(E)** Five hot-wire anemometers (SensoAnemo 5100 LSF) and **(F)** Relative humidity and temperature sensor (EE181), as listed in **Table S1.**

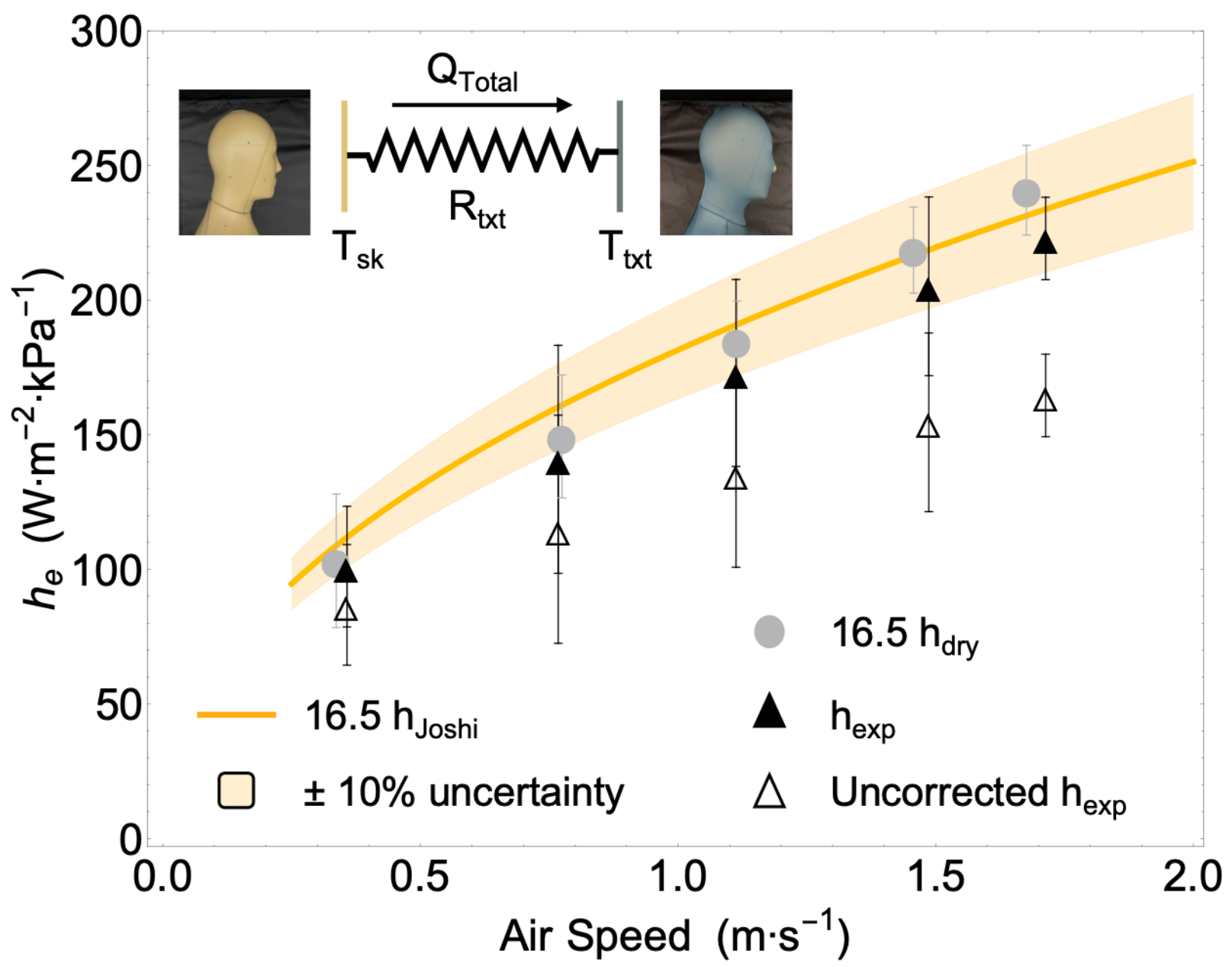

**Fig. S2. Comparison of measured and derived evaporative transfer coefficients and validating the applicability of Lewis relationship under forced convection.** $h_e$ (W·m$^{-2}$·kPa$^{-1}$) – evaporative heat transfer coefficient**.** The whole-body dry forced convective coefficients, $h_{dry}$ (W·m$^{-2}$K$^{-1}$) and $h_{Joshi}$ (W·m$^{-2}$K$^{-1}$) were obtained from (i) sensible-heat exchange experiments conducted inside the wind enclosure and (ii) empirical correlations of Joshi et al. (7)**,** respectively. These dry coefficients were multiplied by the Lewis factor (16.5 K·kPa$^{-1}$) to estimate their evaporative equivalents. The experimentally measured whole-body evaporative heat transfer coefficients, $h_{exp}$ (W·m$^{-2}$·kPa$^{-1}$), were determined from isothermal latent-heat exchange experiments after accounting for the manikin's water-saturated textile skin's thermal resistance (0.0054 m$^2$·K·W$^{-1}$, see inset). The uncorrected $h_{exp}$ (W·m$^{-2}$·kPa$^{-1}$) computed without including textile's thermal resistance, are also shown for comparison. The strong agreement between the evaporative equivalents of $h_{dry}$, $h_{Joshi}$ and $h_{exp}$ across air speeds experimentally demonstrates the validity of the Lewis relationship under forced convective conditions. Experimental data points are averages of three replicates with 95% confidence intervals (combined Type A and B uncertainty).

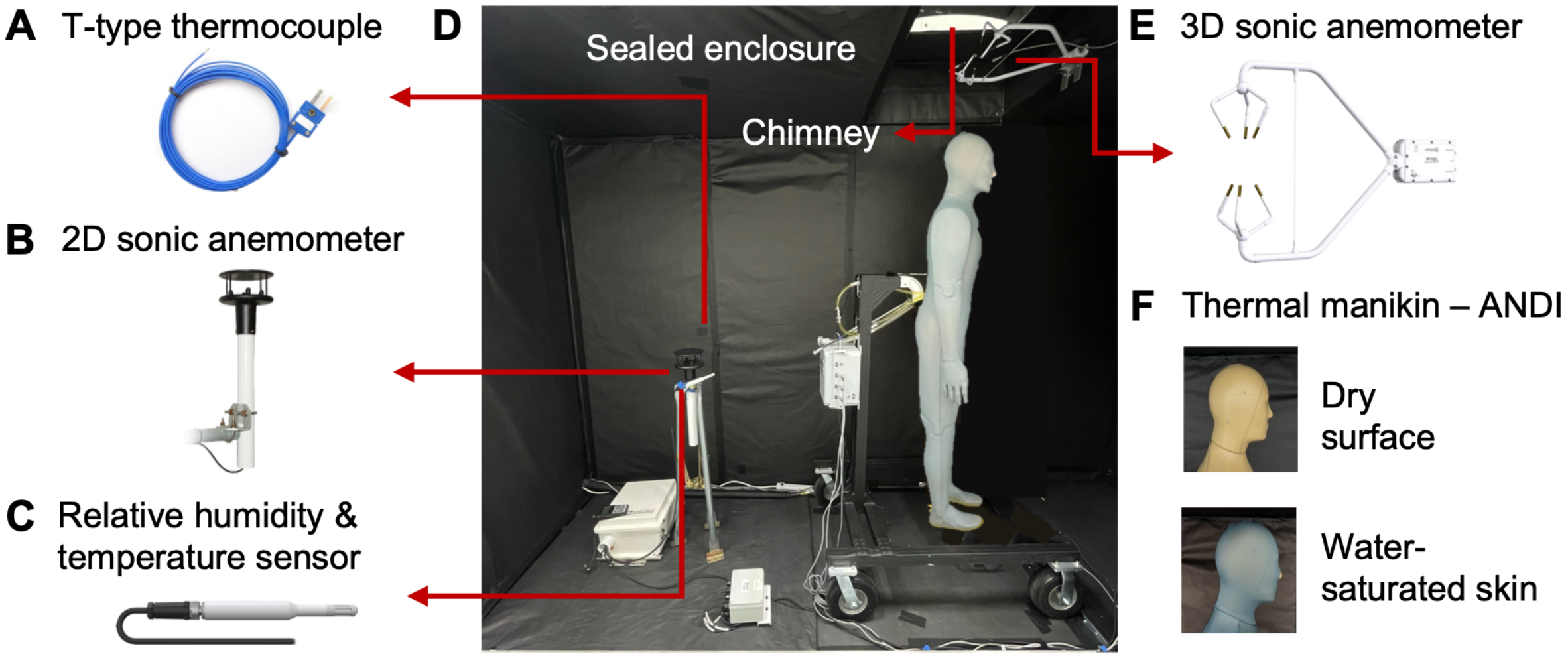

**Fig. S3. Free convection experimental setup with the thermal manikin inside the sealed enclosure.** **(A)** T-type thermocouple measured the enclosure wall temperature, **(B)** 2D sonic anemometer (WINDSONIC1-L) and **(C)** Relative humidity and temperature sensor (EE181) measured the air speed, relative humidity and temperature within the **(D)** sealed enclosure respectively, **(E)** 3D sonic anemometer measured the air speed above the manikin's head. Instrumentation details are provided in **Table S2**. **(F)** Dry manikin surface used for temperature-only driven heat transfer experiments and water-saturated textile skin worn by the manikin for experiments involving sweat evaporation.

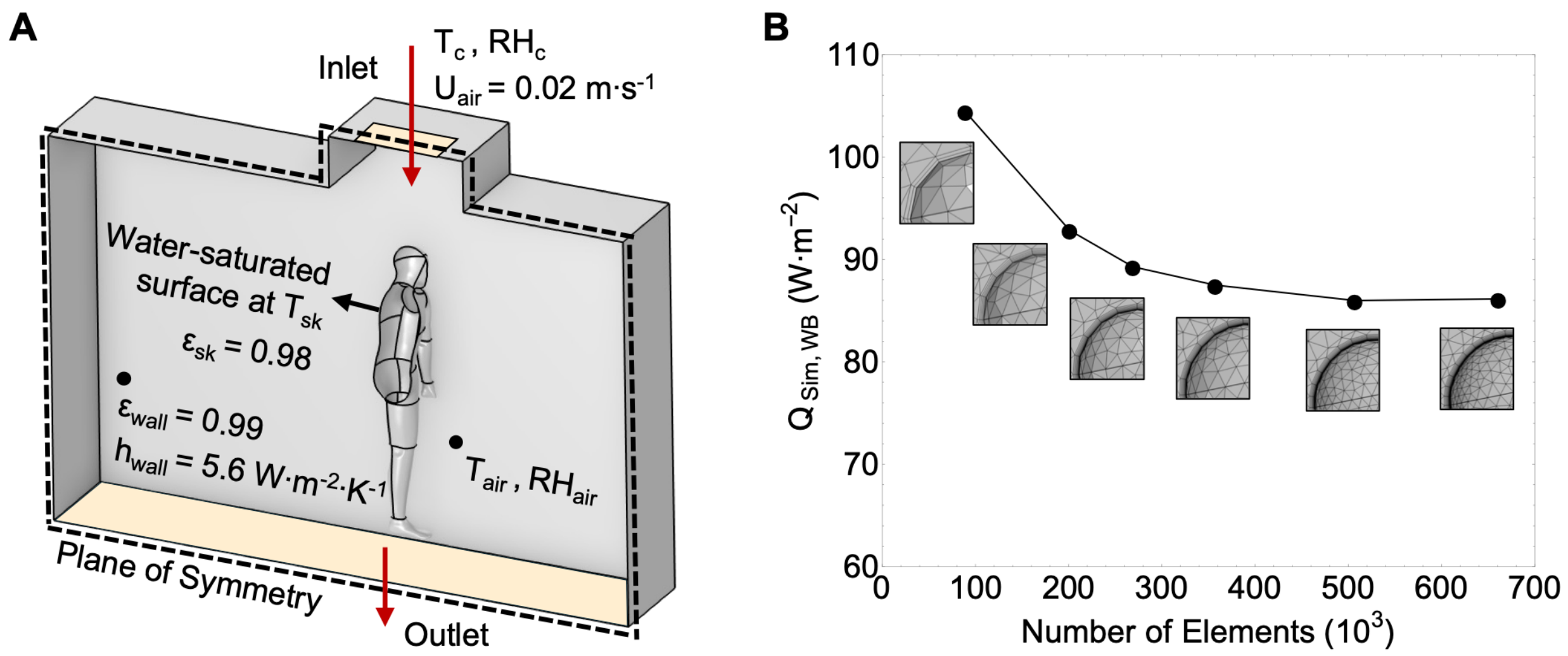

**Fig. S4. (A) Schematic and (B) Mesh refinement study of the coupled multiphysics model used for validation.** $T_{sk}$ (°C) and $\varepsilon_{sk}$ – manikin's surface temperature and emissivity respectively, $h_{wall}$ (W·m$^{-2}$·K$^{-1}$) – equivalent heat transfer coefficient of the enclosure walls accounting for wall resistance and external convection, $\varepsilon_{wall}$ – wall emissivity, $T_{air}$ (°C) and $RH_{air}$ (%) – enclosure air temperature and relative humidity respectively, $T_c$ (°C) and $RH_c$ (%) – climatic chamber air temperature and relative humidity respectively, $U_{air}$ (m·s$^{-1}$) – inlet air speed, $Q_{Sim,WB}$ (W·m$^{-2}$) – whole-body total heat flux computed from the simulations.

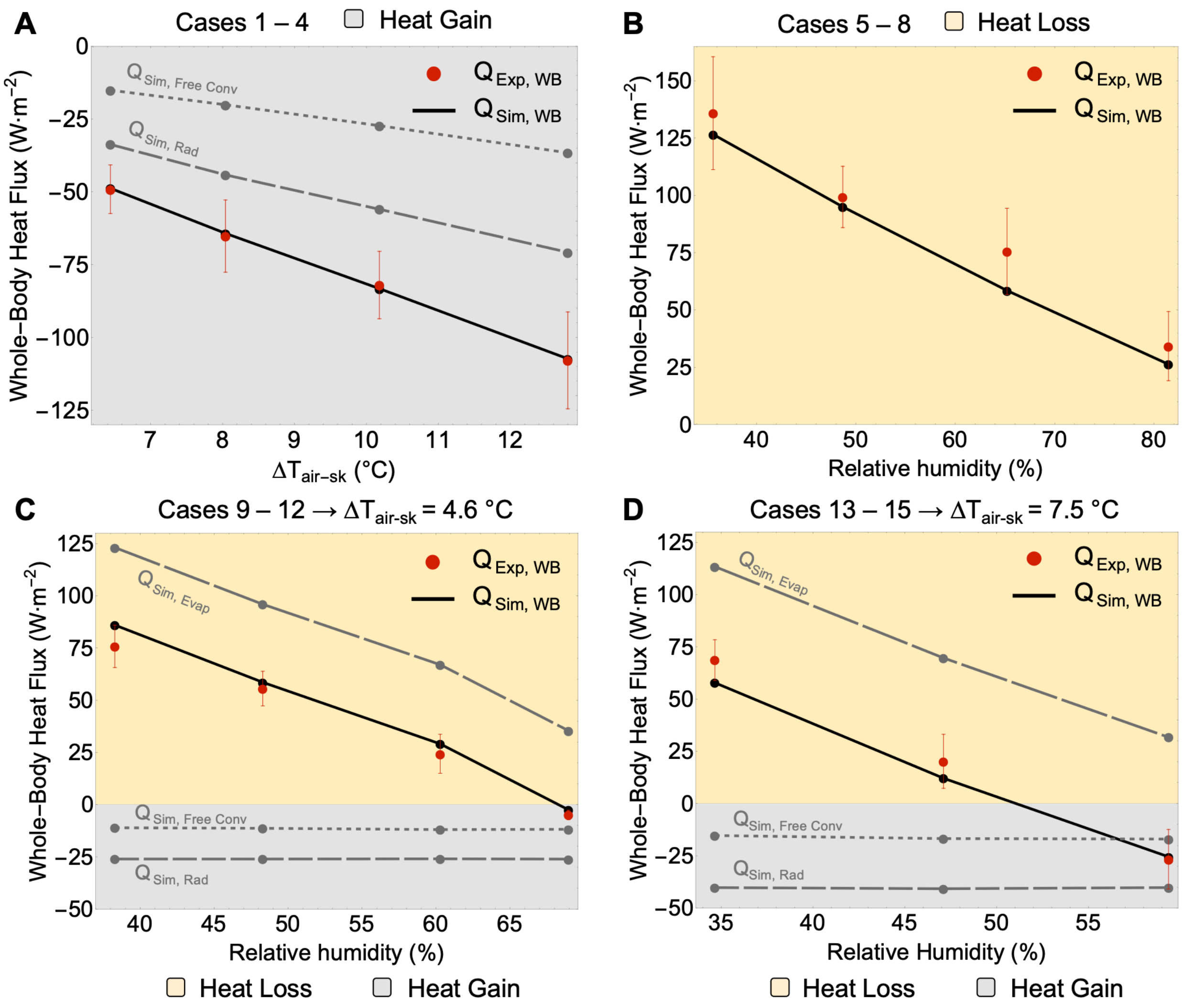

**Fig. S5. Comparison of whole-body heat fluxes between experiments and simulations.** $Q_{Exp,WB}$ (W·m$^{-2}$) and $Q_{Sim,WB}$ (W·m$^{-2}$) – whole-body total heat fluxes from experiments and simulations respectively. **(A)** Temperature-only driven free convection (Cases 1–4), **(B)** Humidity-only driven sweat evaporation (Cases 5–8), **(C–D)** Combined temperature and humidity-driven heat transfer (Cases 9–15). Strong agreement between experimental and simulated results at the whole-body level confirms the reliability of the coupled multiphysics model. Data points are averages of three replicates with 95% confidence intervals (combined Type A and B uncertainty).

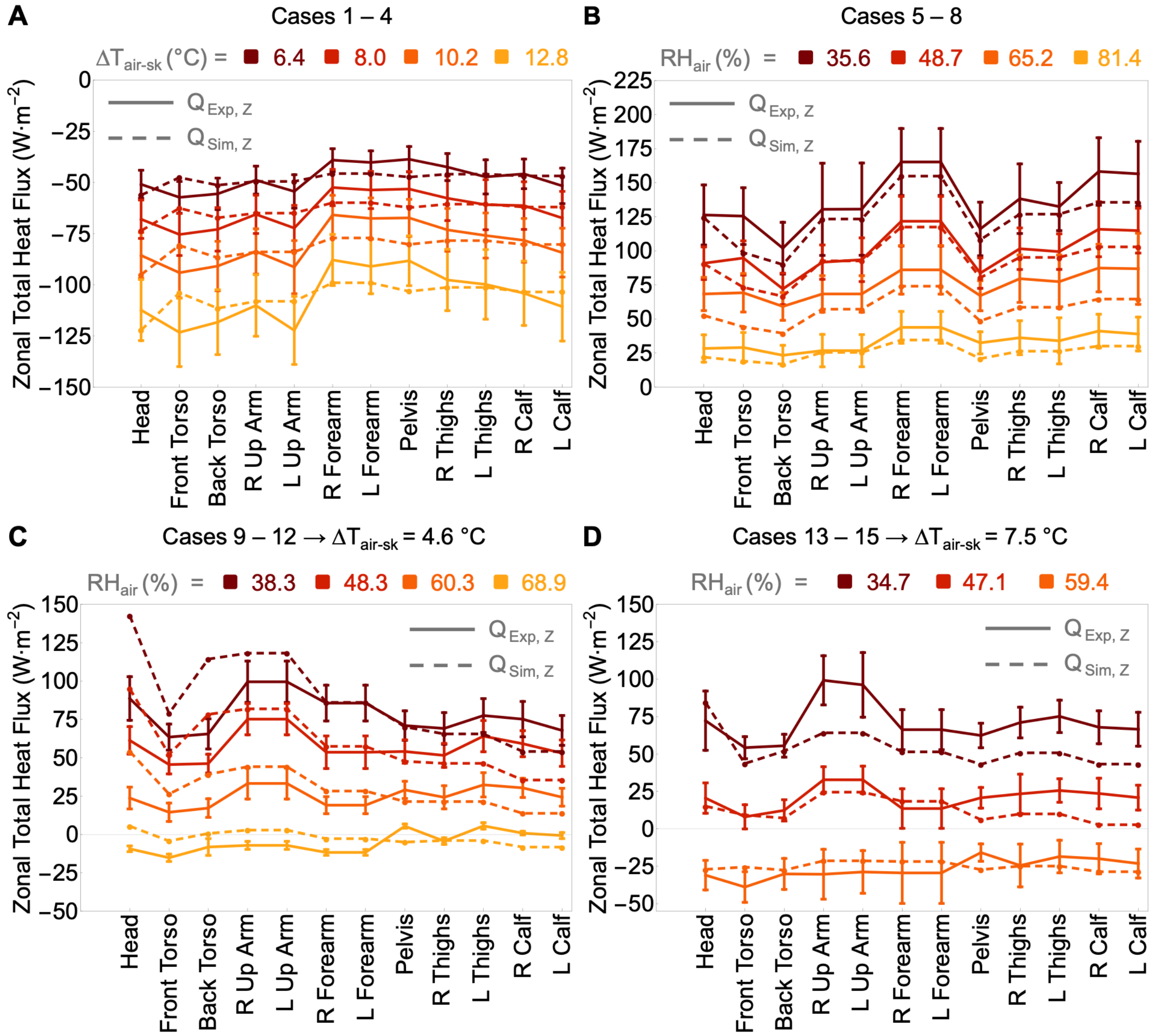

**Fig. S6. Comparison of zonal total heat fluxes between experiments and simulations.** $Q_{Exp,Z}$ (W·m$^{-2}$) – Zonal measured total heat flux, $Q_{Sim,Z}$ (W·m$^{-2}$) – Zonal simulated total heat flux **(A)** Temperature-only driven free convection (Cases 1–4), **(B)** Humidity-only driven sweat evaporation (Cases 5–8), **(C–D)** Combined temperature and humidity-driven heat transfer (Cases 9–15). Overall strong agreement, with minor discrepancies at specific zones, confirms the reliability of the coupled multiphysics model. Data points are averages of three replicates with 95% confidence intervals (combined Type A and B uncertainty).

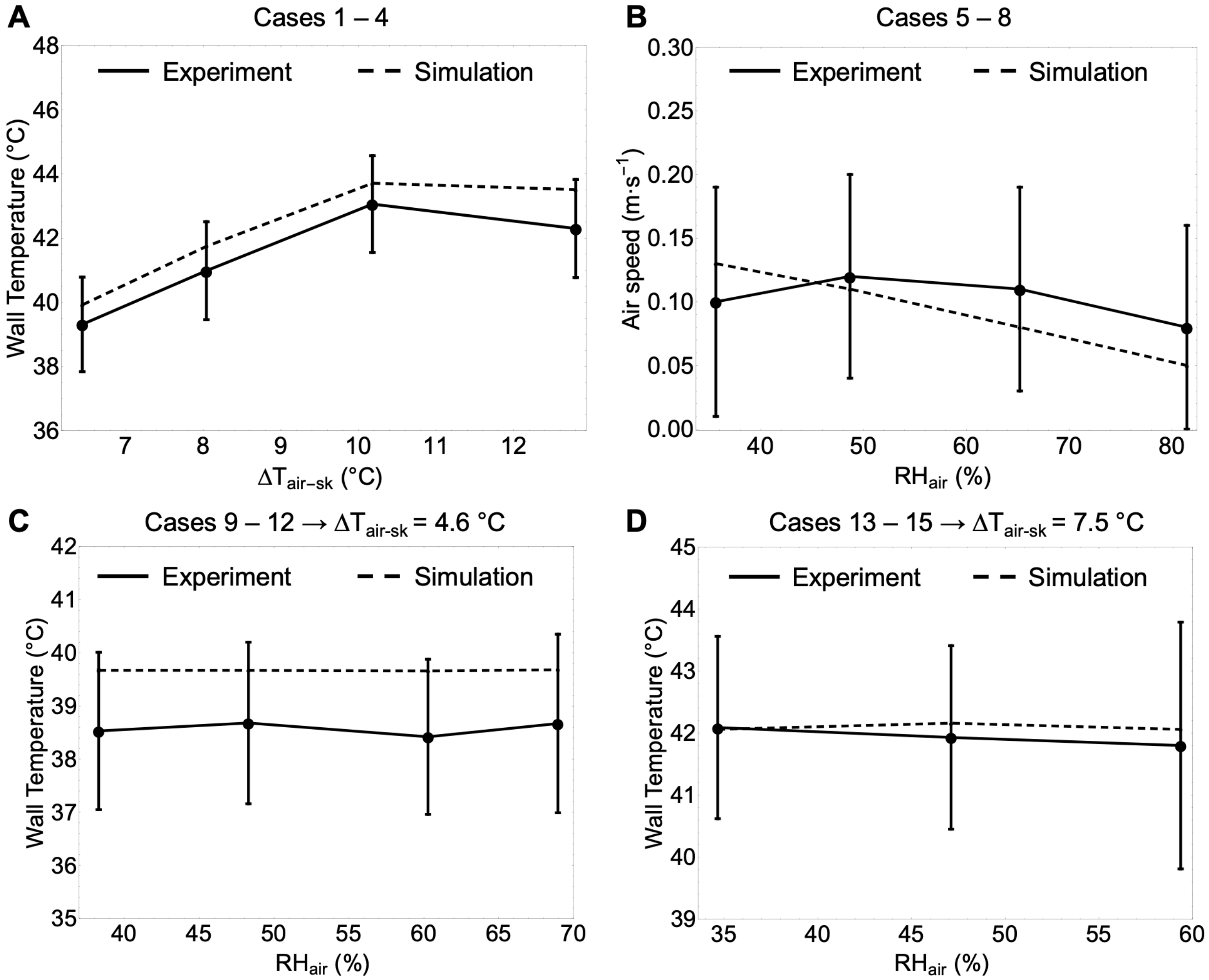

**Fig. S7. Comparison of key environmental parameters between experiments and simulations. (A, C, D)** Enclosure wall temperature (**Fig. S3A**) for temperature-only driven free convection (Cases 1–4), and combined temperature and humidity-driven heat transfer (Cases 9–15). **(B)** Air speed above the manikin's head (**Fig. S3E**) for humidity-only driven sweat evaporation (Cases 5–8). All comparisons show good agreement, further confirming the reliability of the coupled multiphysics model. Data points are averages of three replicates with ±1 SD (standard deviation) (combined Type A and B uncertainty).

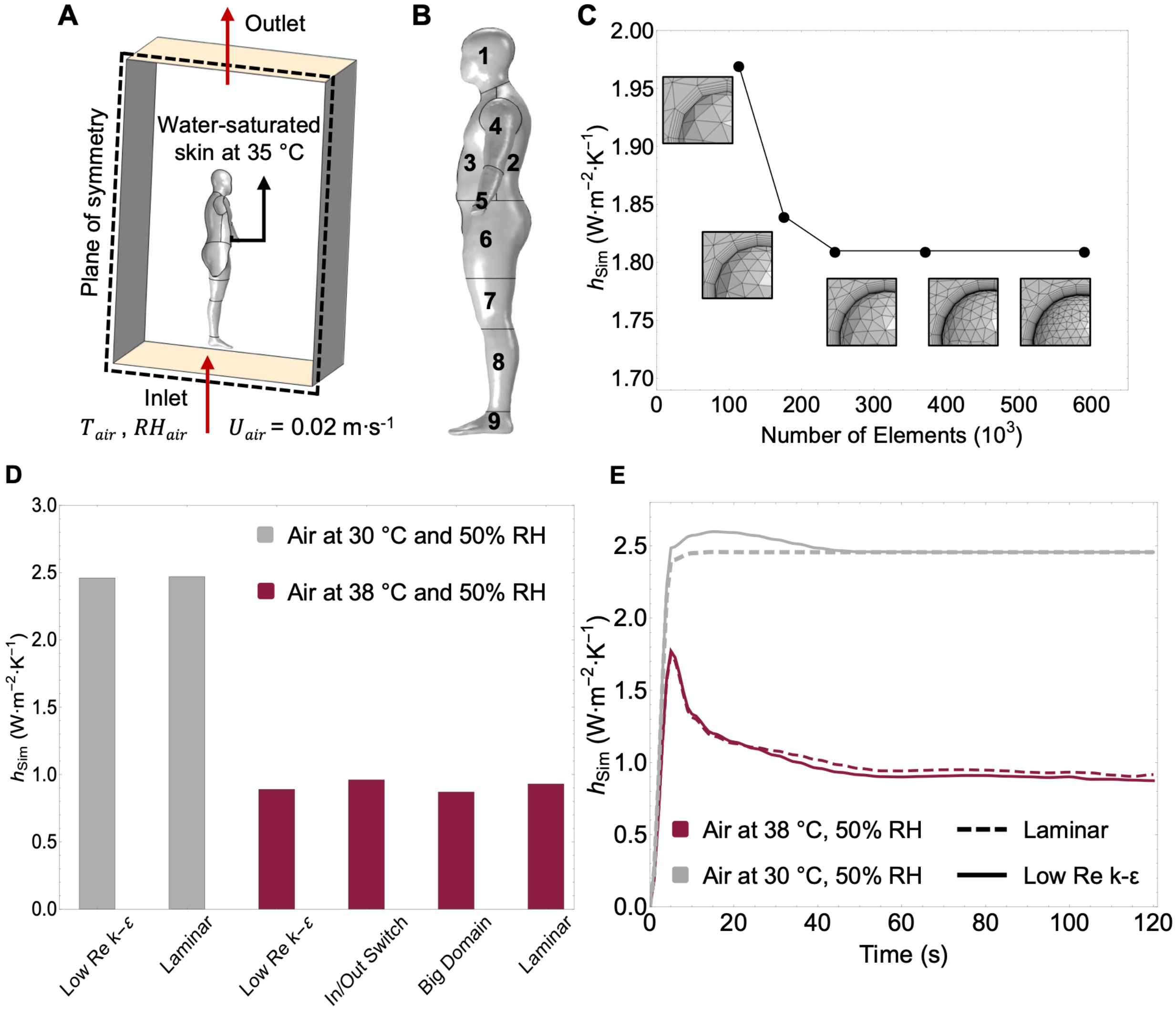

**Fig. S8. Development of multiphysics model for parametric study. (A)** Schematic of the digital twin, $T_{air}$ (°C) and $RH_{air}$ (%) – air temperature and relative humidity respectively, $U_{air}$ ($m{\cdot}s^{-1}$) – inlet air speed. **(B)** Cross-sectional side view of an average western young male (50th percentile height (m) and BMI ($kg{\cdot}m^{-2}$)) with 9 body segments. **(C)** Mesh refinement study of the validated coupled multiphysics model used for parametric study. $h_{sim}$ ($W{\cdot}m^{-2}{\cdot}K^{-1}$), – simulated whole-body free convective coefficient. Inset: Zoomed in pictures of how the boundary layer on the manikin's head grows denser with finer mesh sizes. **(D)** Steady-state $h_{Sim}$ values computed across Low-Re k-ε, laminar, in/out dynamic switch, and larger domain configurations for two thermal conditions (air at 30 and 38°C, 50% RH). Results remain consistent across all sensitivity configurations, confirming computational robustness. **(E)** Transient evolution of whole-body free convective heat transfer coefficient $h_{Sim}$ ($W{\cdot}m^{-2}{\cdot}K^{-1}$) over 120 s for stagnation (38°C, 50% RH) and buoyancy-enhanced (30°C, 50% RH) cases comparing Low-Re k-ε and laminar formulations. Simulations stabilize by 30–40s, validating the original 50 s window.

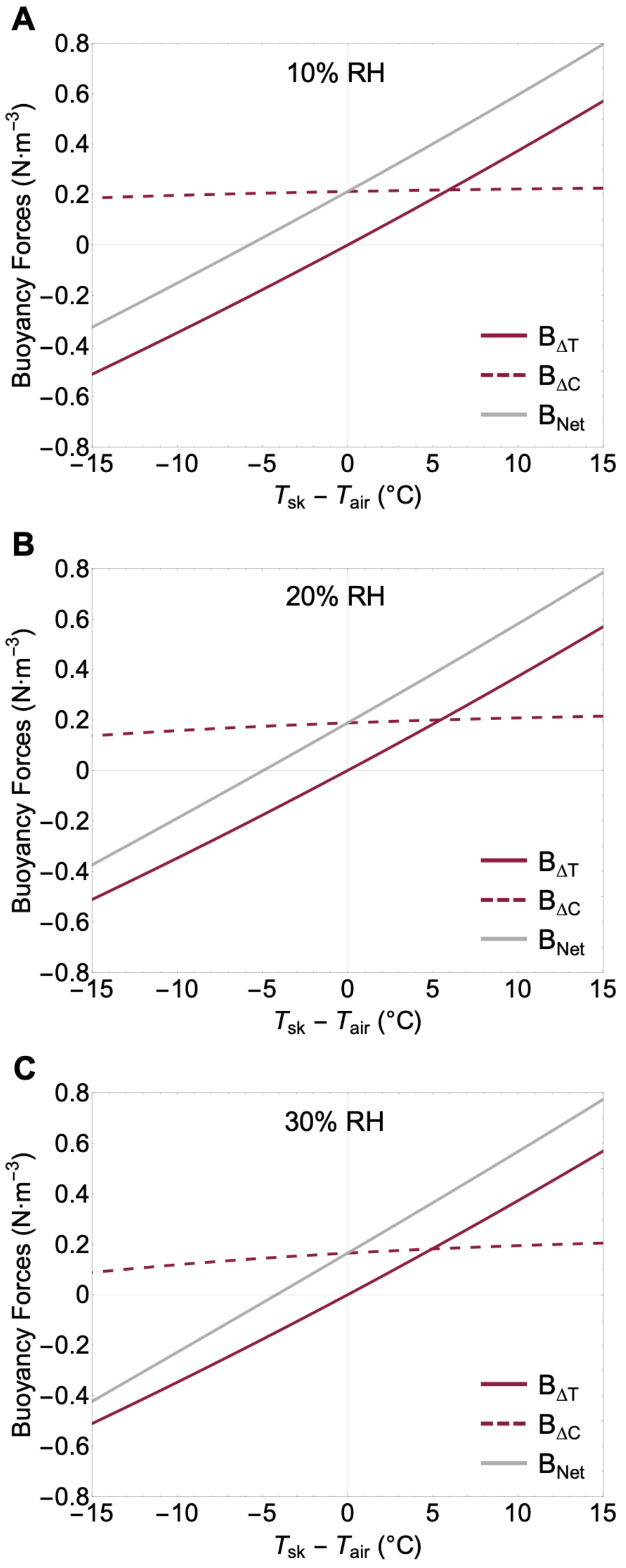

**Fig. S9. Comparison of different buoyancy forces across varying ambient conditions.** Individual buoyancy forces generated by temperature difference-driven ($B_{\Delta T}$, N·m$^{-3}$), humidity or moisture concentration differences-driven ($B_{\Delta T}$, N·m$^{-3}$) and their sum ($B_{Net}$, N·m$^{-3}$) are compared across the entire air temperature range at **(A)** 10% RH, **(B)** 20% RH, and **(C)** 30% RH. $T_{sk}$– skin temperature (at 35 °C), $T_{air}$– air temperature (20–50 °C).

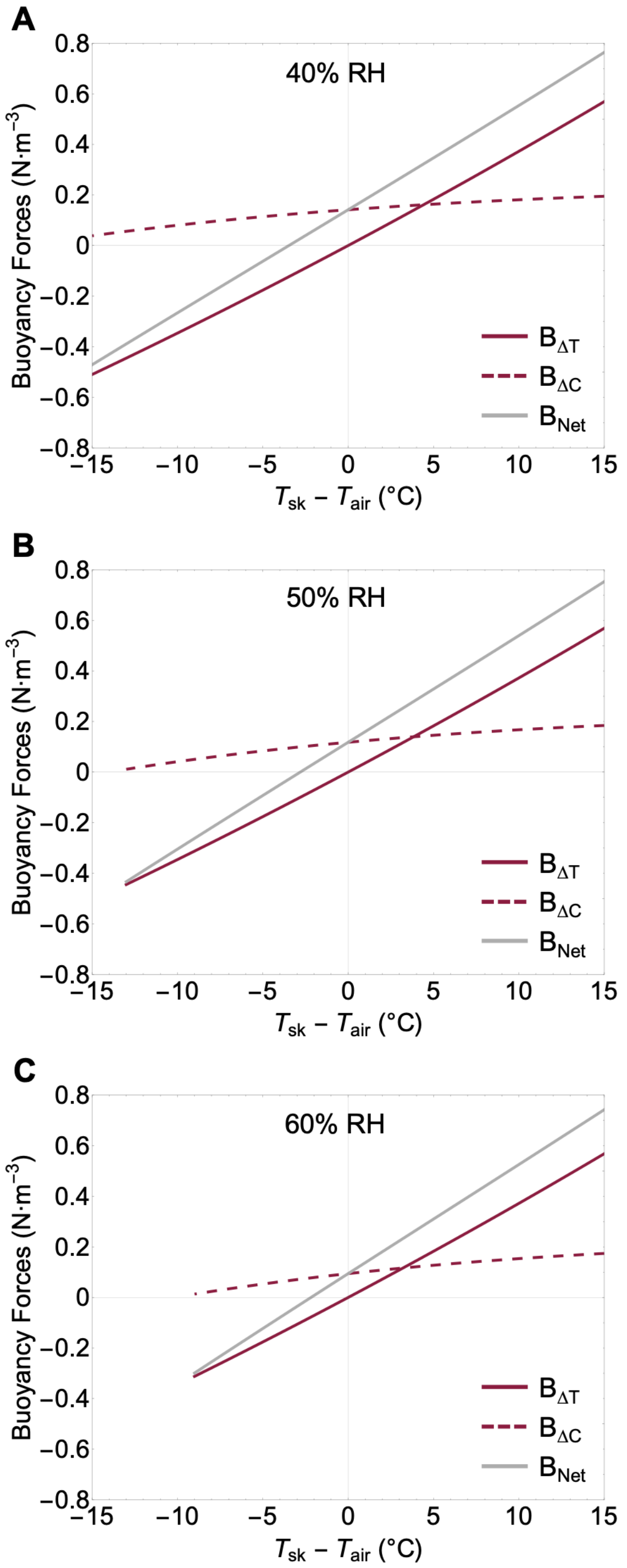

**Fig. S10. Comparison of different buoyancy forces across varying ambient conditions.** Individual buoyancy forces generated by temperature difference-driven ($B_{\Delta T}$, N·m$^{-3}$), humidity or moisture concentration differences-driven ($B_{\Delta T}$, N·m$^{-3}$) and their sum ($B_{Net}$, N·m$^{-3}$) are compared across the entire air temperature range at **(A)** 40% RH, **(B)** 50% RH, and **(C)** 60% RH. $T_{sk}$– skin temperature (at 35 °C), $T_{air}$– air temperature (°C). Only cases with a positive skin-to-air moisture concentration difference ($C_{sk} > C_{air}$) was computed; scenarios with $C_{sk} < C_{air}$—which would cause condensation on the skin—were omitted.

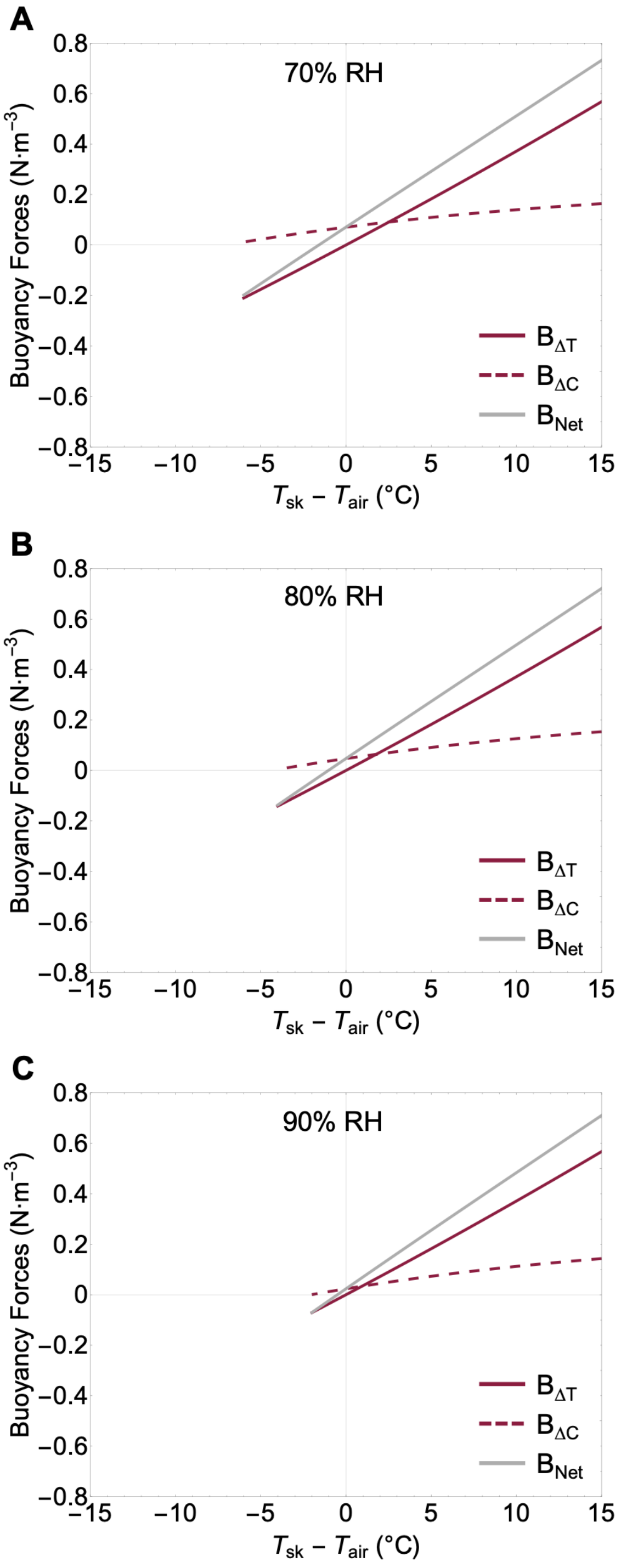

**Fig. S11. Comparison of different buoyancy forces across varying ambient conditions.** Individual buoyancy forces generated by temperature difference-driven ($B_{\Delta T}$, N·m$^{-3}$), humidity or moisture concentration differences-driven ($B_{\Delta T}$, N·m$^{-3}$) and their sum ($B_{Net}$, N·m$^{-3}$) are compared across the entire air temperature range at **(A)** 40% RH, **(B)** 50% RH, and **(C)** 60% RH. $T_{sk}$– skin temperature (at 35 °C), $T_{air}$– air temperature (°C). Only cases with a positive skin-to-air moisture concentration difference ($C_{sk} > C_{air}$) was computed; scenarios with $C_{sk} < C_{air}$—which would cause condensation on the skin—were omitted.

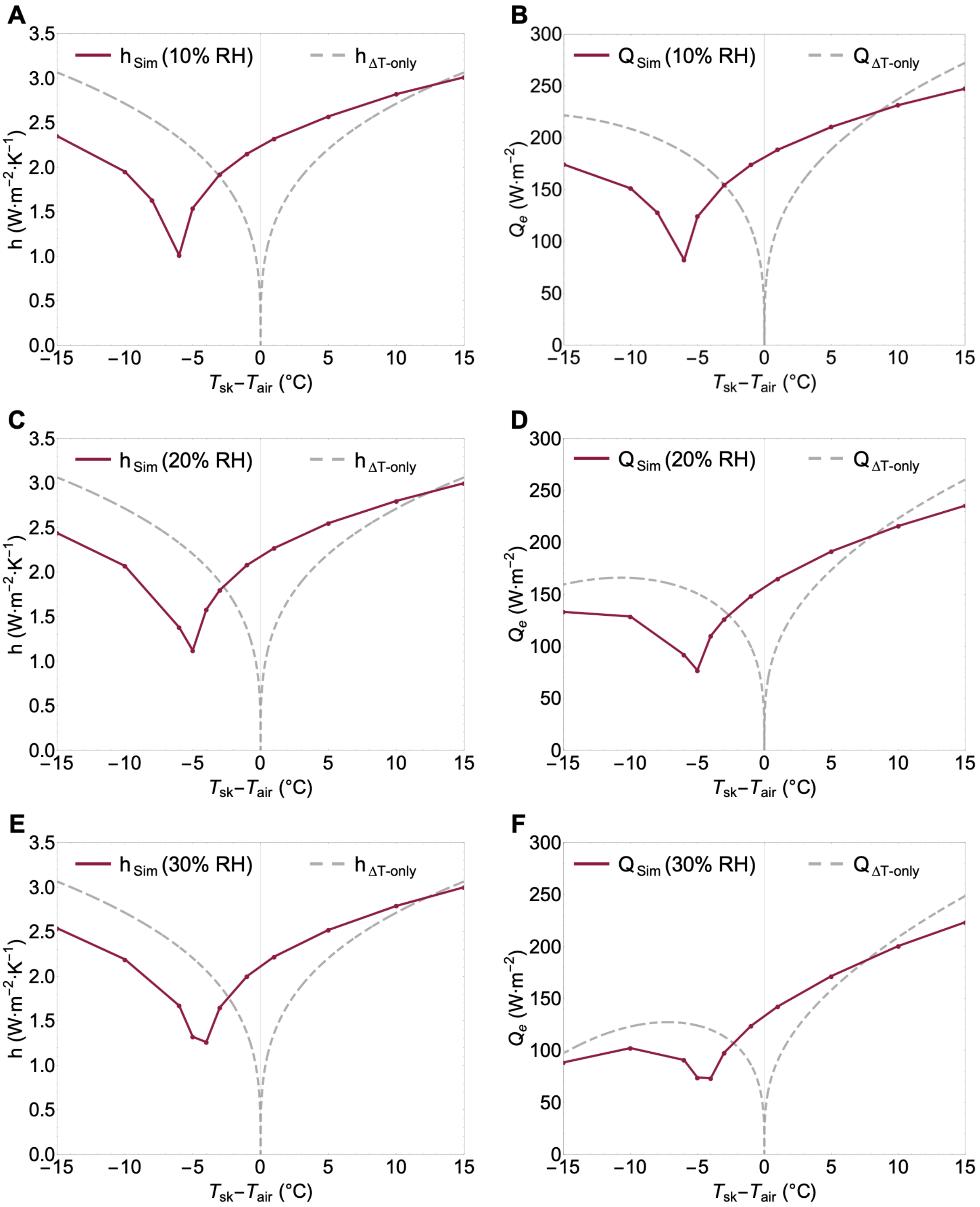

**Fig. S12. Comparison of simulated whole-body free convective coefficients and evaporative fluxes accounting for temperature- and humidity-differences against only temperature-difference only empirical correlations.** $h_{Sim}$ (W·m$^{-2}$·K$^{-1}$) and $Q_{Sim}$ (W·m$^{-2}$) – Simulated whole-body free convective coefficients and evaporative fluxes respectively, $h_{\Delta T-only}$ (W·m$^{-2}$·K$^{-1}$) and $Q_{\Delta T-only}$ (W·m$^{-2}$) – thermal-gradient only empirical correlations (**Eq. S8**–**S9**) respectively, **(A–B)** 10% $RH_{air}$, **(C–D)** 20% $RH_{air}$, **(E–F)** 30% $RH_{air}$.

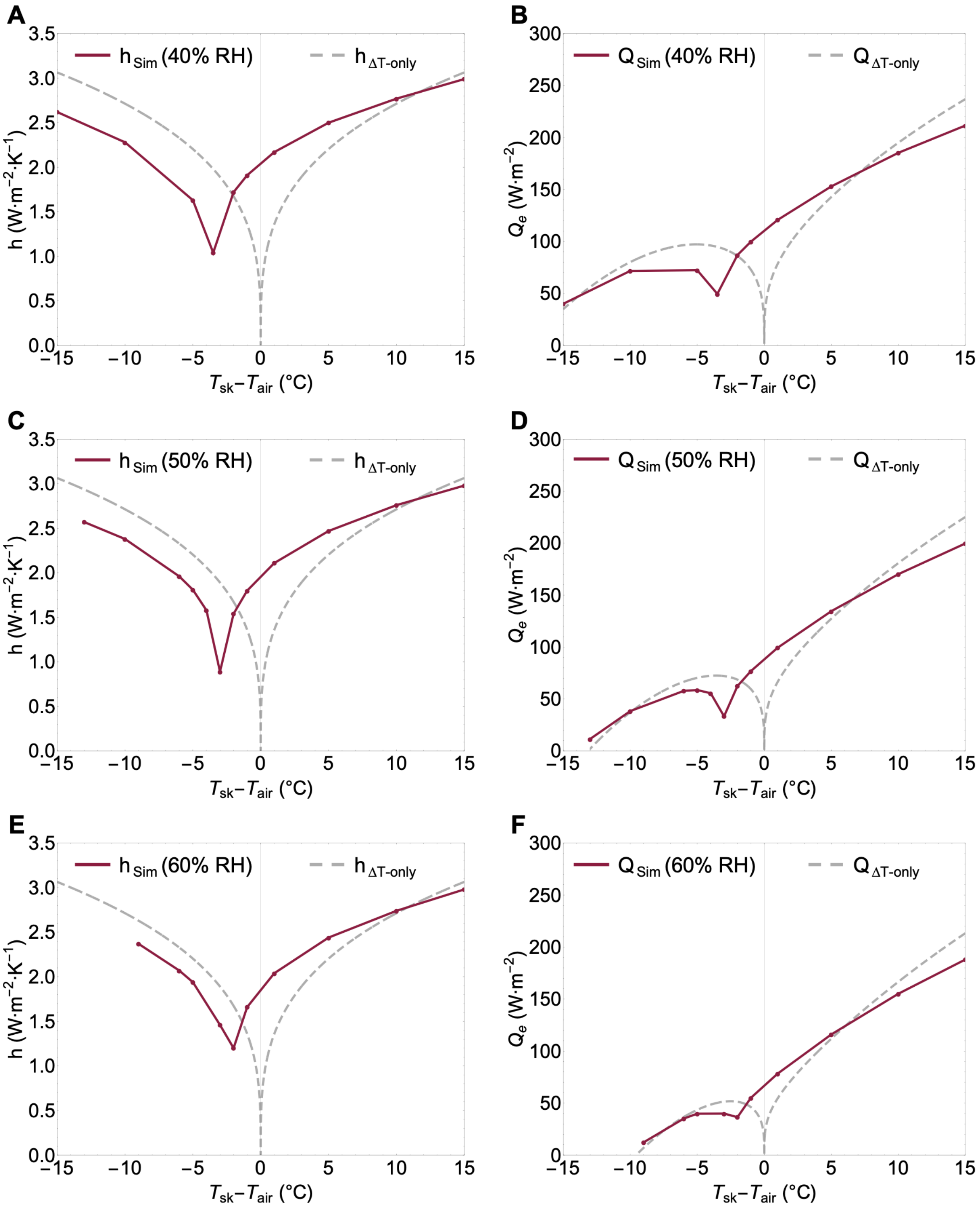

**Fig. S13. Comparison of simulated whole-body free convective coefficients and evaporative fluxes accounting for temperature- and humidity-differences against only temperature-difference only empirical correlations.** $h_{Sim}$ (W·m$^{-2}$·K$^{-1}$) and $Q_{Sim}$ (W·m$^{-2}$) – Simulated whole-body free convective coefficients and evaporative fluxes respectively, $h_{\Delta T-only}$ (W·m$^{-2}$·K$^{-1}$) and $Q_{\Delta T-only}$ (W·m$^{-2}$) – thermal-gradient only empirical correlations (**Eq. S8**–**S9**) respectively, **(A–B)** 40% $RH_{air}$, **(C–D)** 50% $RH_{air}$, **(E–F)** 60% $RH_{air}$.

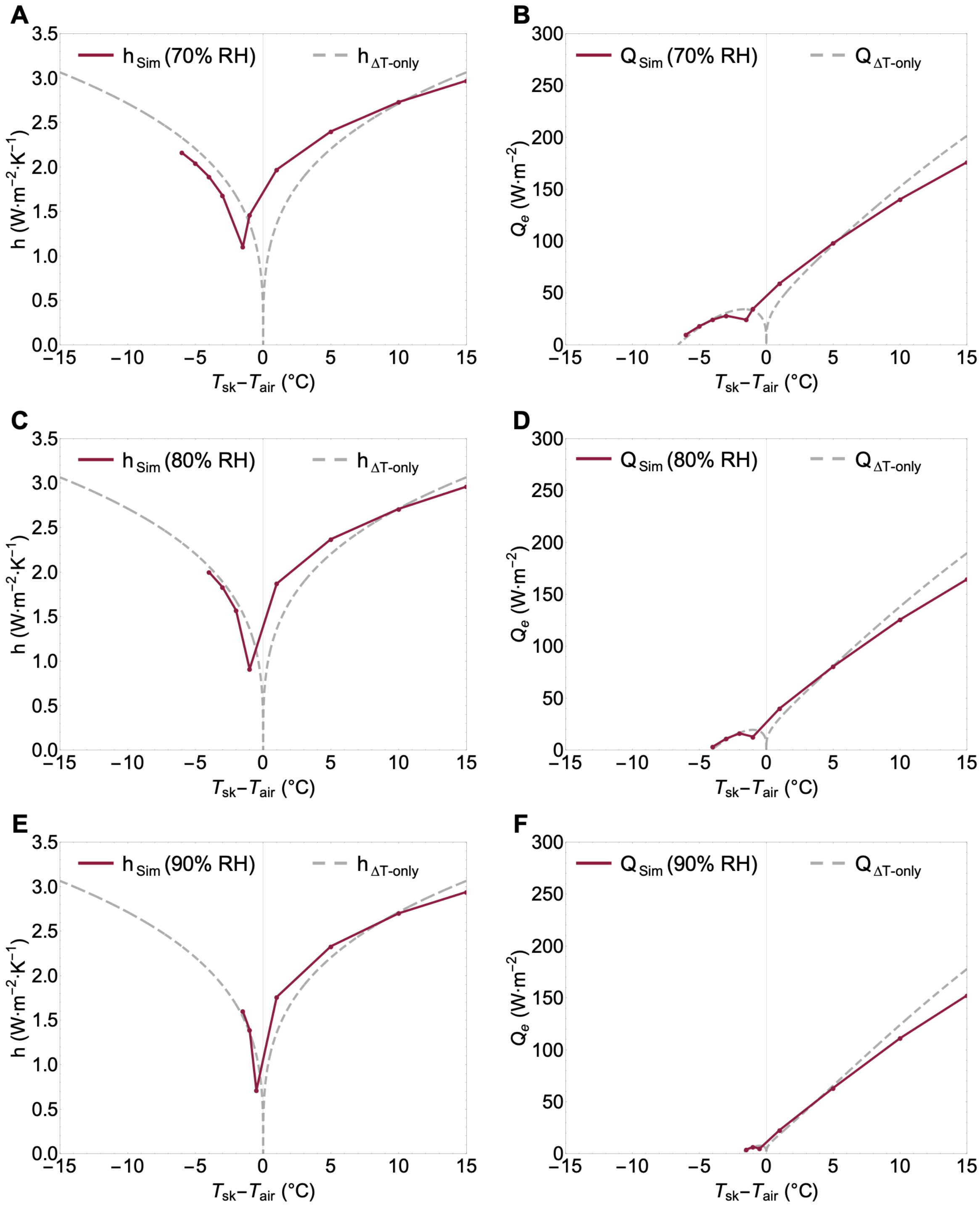

**Fig. S14. Comparison of simulated whole-body free convective coefficients and evaporative fluxes accounting for temperature- and humidity-differences against only temperature-difference only empirical correlations.** $h_{Sim}$ (W·m$^{-2}$·K$^{-1}$) and $Q_{Sim}$ (W·m$^{-2}$) – Simulated whole-body free convective coefficients and evaporative fluxes respectively, $h_{\Delta T-only}$ (W·m$^{-2}$·K$^{-1}$) and $Q_{\Delta T-only}$ (W·m$^{-2}$) – thermal-gradient only empirical correlations (**Eq. S8**–**S9**) respectively, **(A–B)** 70% $RH_{air}$, **(C–D)** 80% $RH_{air}$, **(E–F)** 90% $RH_{air}$.

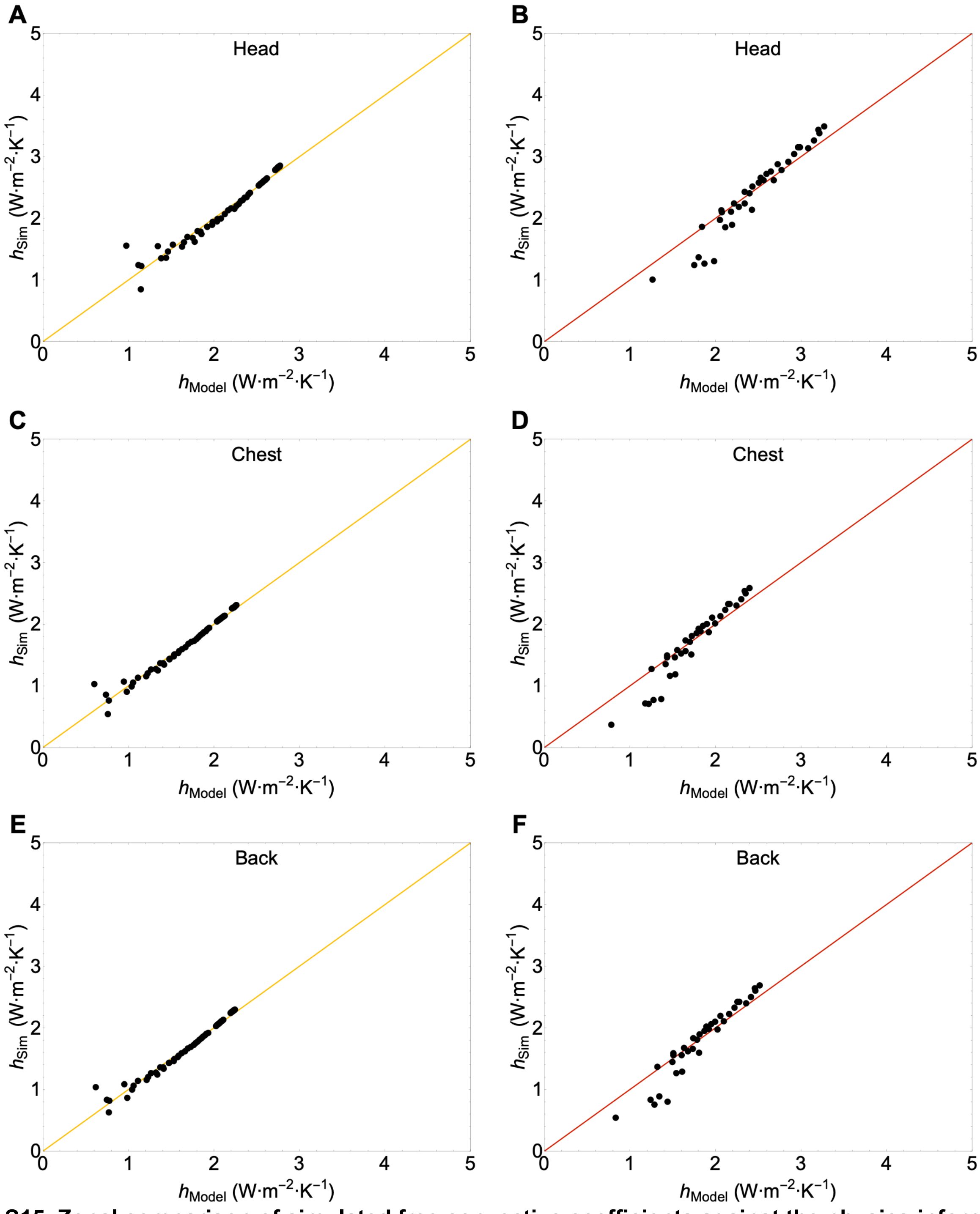

**Fig. S15. Zonal comparison of simulated free convective coefficients against the physics-informed model.** Simulated free convective coefficients ($h_{Sim}$, W·m$^{-2}$·K$^{-1}$) are compared with the model predictions ($h_{Model}$, W·m$^{-2}$·K$^{-1}$) from **Eq. S10** across **(A–B)** Head, **(C–D)** Chest, **(E–F)** Back. For each zone, left panels correspond to positive (supporting) combined temperature and concentration buoyancy ($\Delta T_{sk-air} + 146C_{sk-air}$), and right panels correspond to negative (opposing) buoyancy. The strong agreement demonstrates that the physics-informed model robustly captures zonal free convection across both buoyancy regimes.

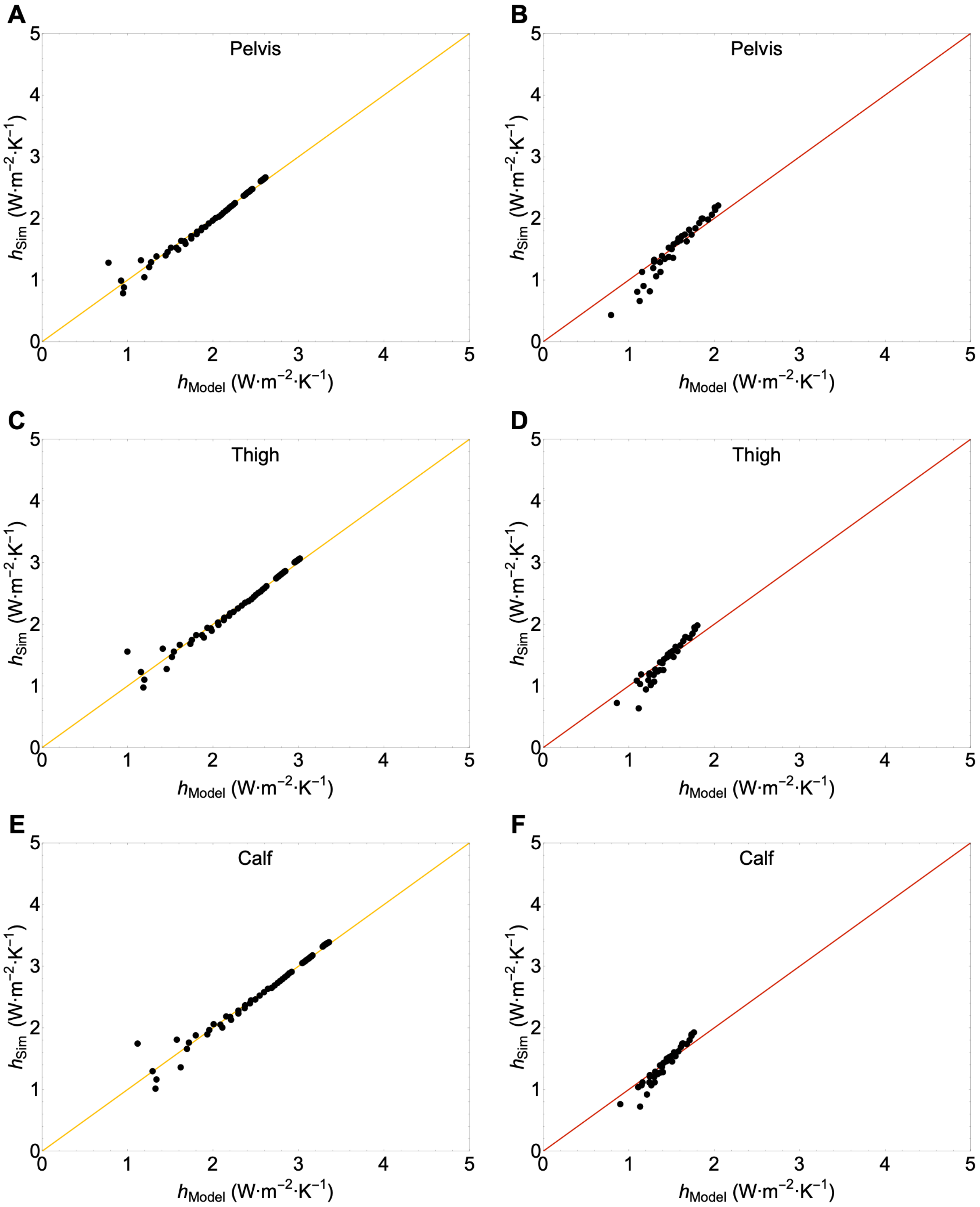

**Fig. S16. Zonal comparison of simulated free convective coefficients against the physics-informed model.** Simulated free convective coefficients ($h_{Sim}$, W·m$^{-2}$·K$^{-1}$) are compared with the model predictions ($h_{Model}$, W·m$^{-2}$·K$^{-1}$) from **Eq. S10** across **(A–B)** Pelvis, **(C–D)** Thigh, **(E–F)** Calf. For each zone, left panels correspond to positive (supporting) combined temperature and concentration buoyancy ($\Delta T_{sk-air} + 146C_{sk-air}$), and right panels correspond to negative (opposing) buoyancy. The strong agreement demonstrates that the physics-informed model robustly captures zonal free convection across both buoyancy regimes.

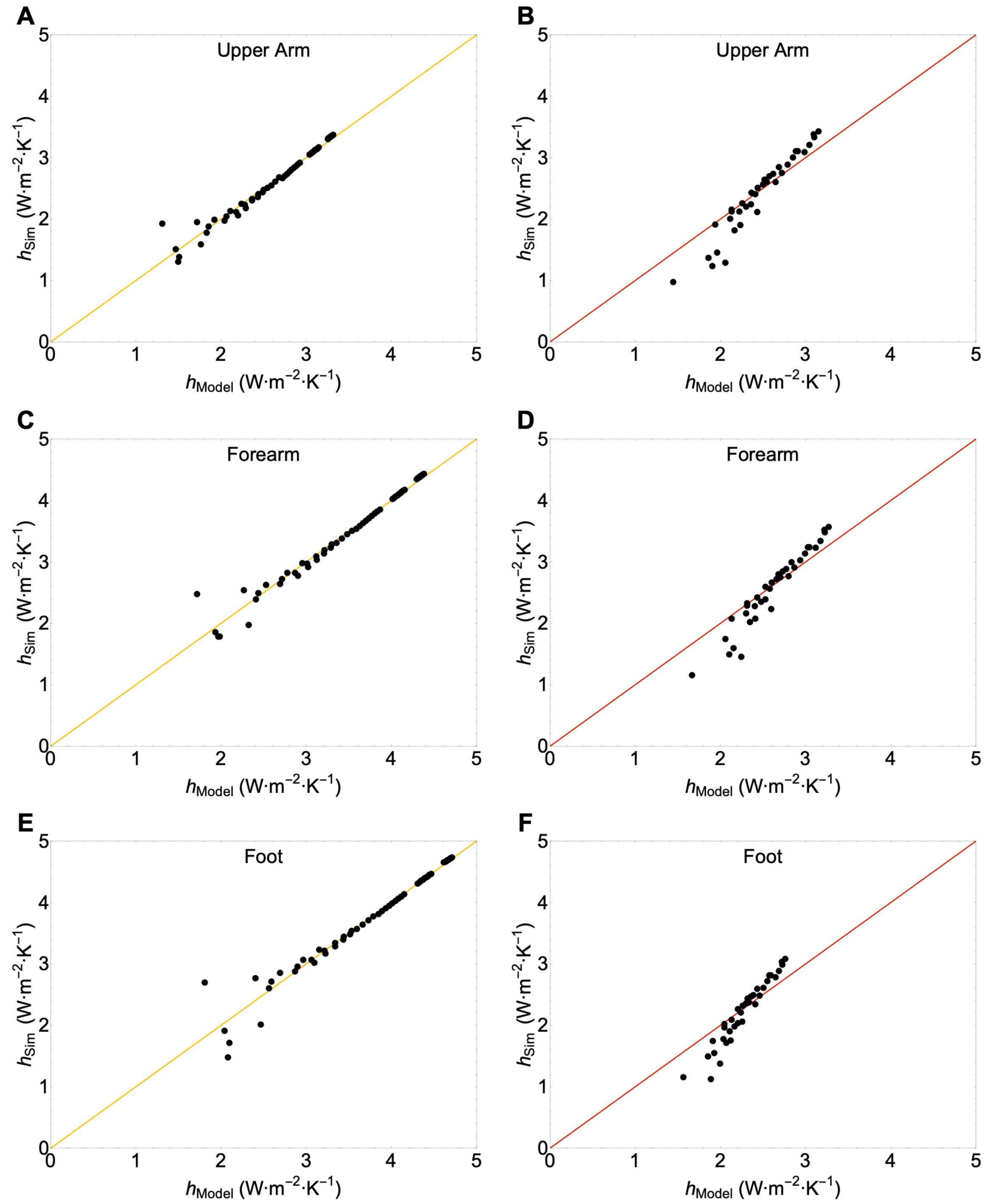

**Fig. S17. Zonal comparison of simulated free convective coefficients against the physics-informed model.** Simulated free convective coefficients ($h_{Sim}$, W·m$^{-2}$·K$^{-1}$) are compared with the model predictions ($h_{Model}$, W·m$^{-2}$·K$^{-1}$) from **Eq. S10** across **(A–B)** Upper arm, **(C–D)** Forearm, **(E–F)** Foot. For each zone, left panels correspond to positive (supporting) combined temperature and concentration buoyancy ($\Delta T_{sk-air} + 146C_{sk-air}$), and right panels correspond to negative (opposing) buoyancy. The strong agreement demonstrates that the physics-informed model robustly captures zonal free convection across both buoyancy regimes.

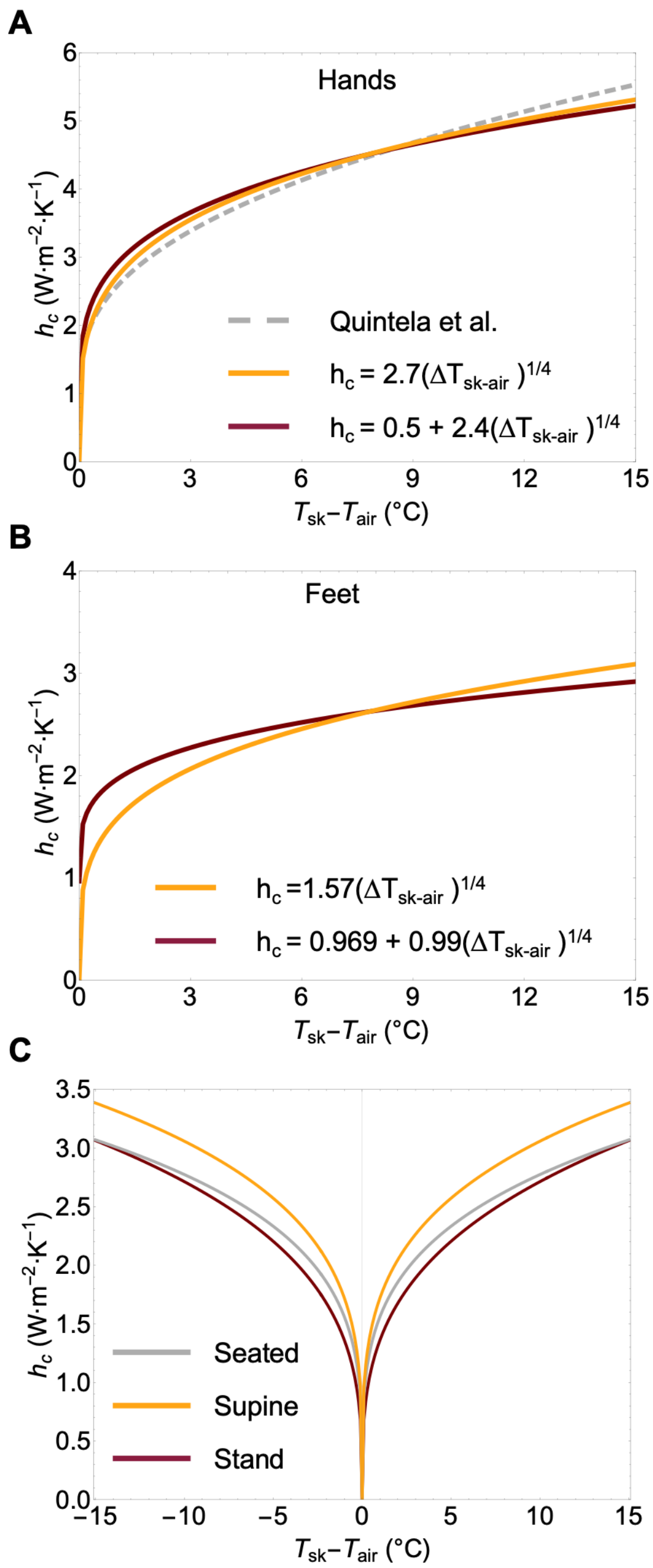

**Fig. S18. Comparison of fitted free convective correlations for hands and feet. (A)** Hand: Simulated free convective coefficients ($h_c$, W·m$^{-2}$·K$^{-1}$) for hands were fitted using temperature-difference scaling (**Eq. S11**) and combined temperature- and concentration-difference scaling (**Eq. S12**) and compared with average correlations from Quintela et al. (*13*) (**Eq. S13–S14**). **(B)** Feet: $h_c$ for feet fitted with temperature-difference scaling (**Eq. S15**) compared with combined temperature- and concentration- difference scaling (**Eq. S16**). **(C)** Whole-body free convective heat transfer coefficient $h_c$ as a function of absolute skin-to-air temperature difference |ΔT| for standing, seated, and supine positions.

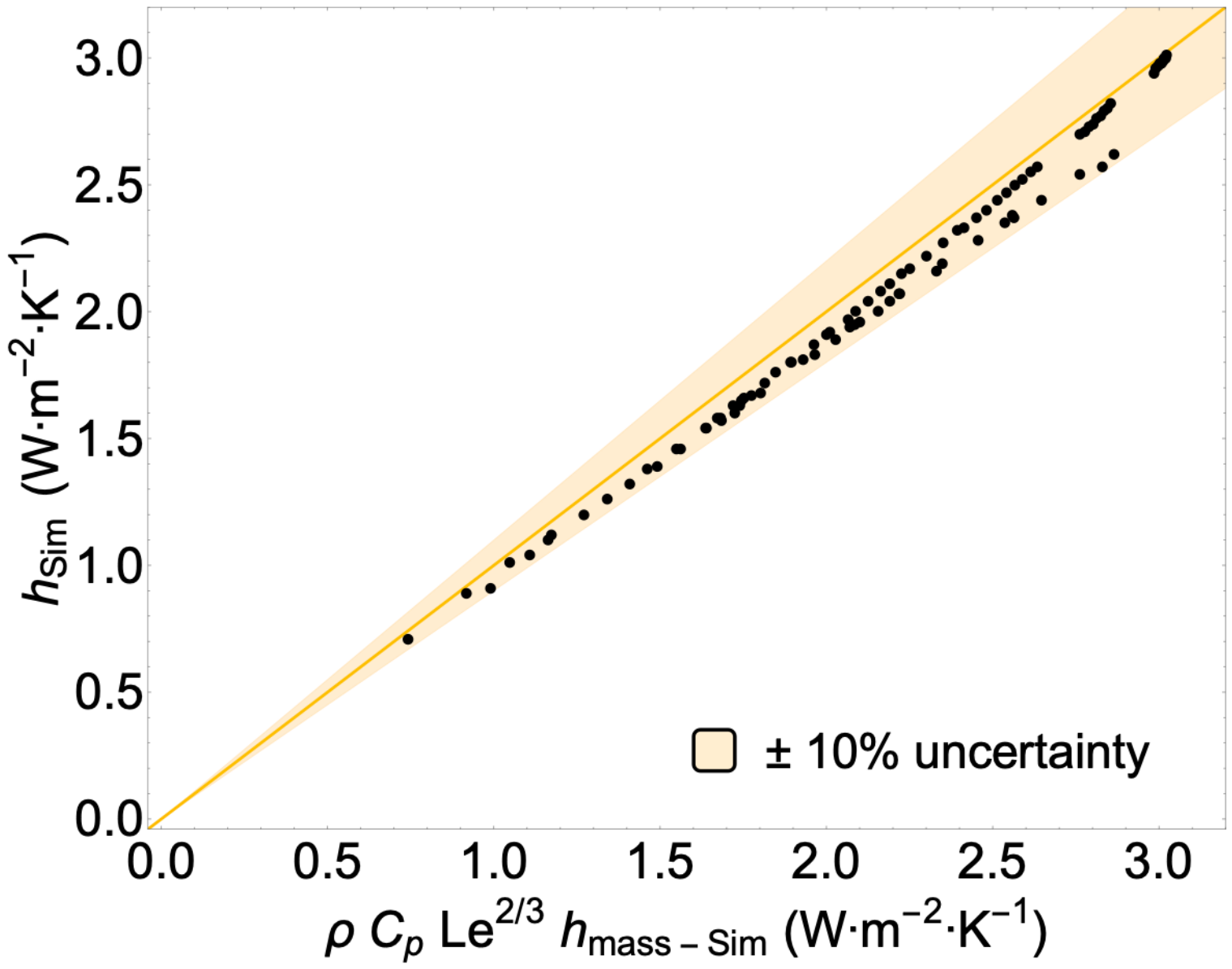

**Fig. S19. Comparison of simulated whole-body free convective coefficients and mass transfer-derived equivalents and validating the Chilton-Colburn analogy.** Simulated whole-body mass transfer coefficients ($h_{mass-Sim}$, m·s$^{-1}$) were multiplied by moist air density (ρ, kg·m$^{-3}$), specific heat ($C_p$, J·kg$^{-1}$·K$^{-1}$), and Lewis number ($Le$, dimensionless) (**Eq. S1**) to estimate the mass transfer-derived equivalents of $h_{Sim}$ (W·m$^{-2}$·K$^{-1}$). A strong agreement shows that Chilton-Colburn analogy holds under free convective conditions only when $h_{Sim}$ accounts for coupled temperature and humidity-differences.

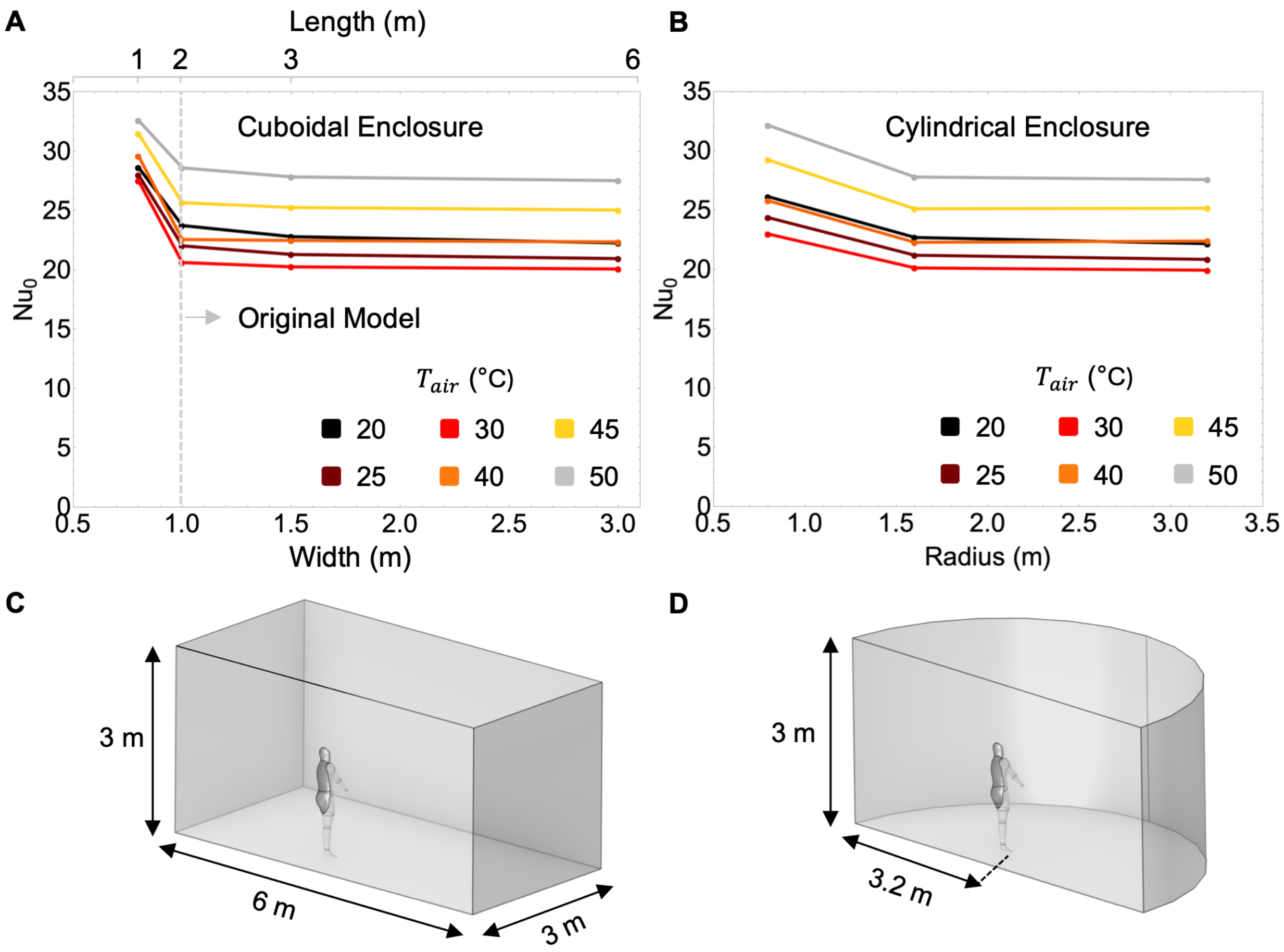

**Fig. S20. Domain-size sensitivity analysis for diffusion baseline ($\mathrm{Nu}_0$).** Simulated whole-body heat transfer coefficients for heat diffusion only simulations expressed in dimensionless form (L=1.8 m and thermal conductivity calculated at film temperature) for manikin skin at 35 °C, but varying air temperature and size of **(A)** rectangular and **(B)** cylindrical domains. **(C)** and **(D)** illustrate the biggest cuboidal and cylindrical domains tested for these "diffusion-only" simulations

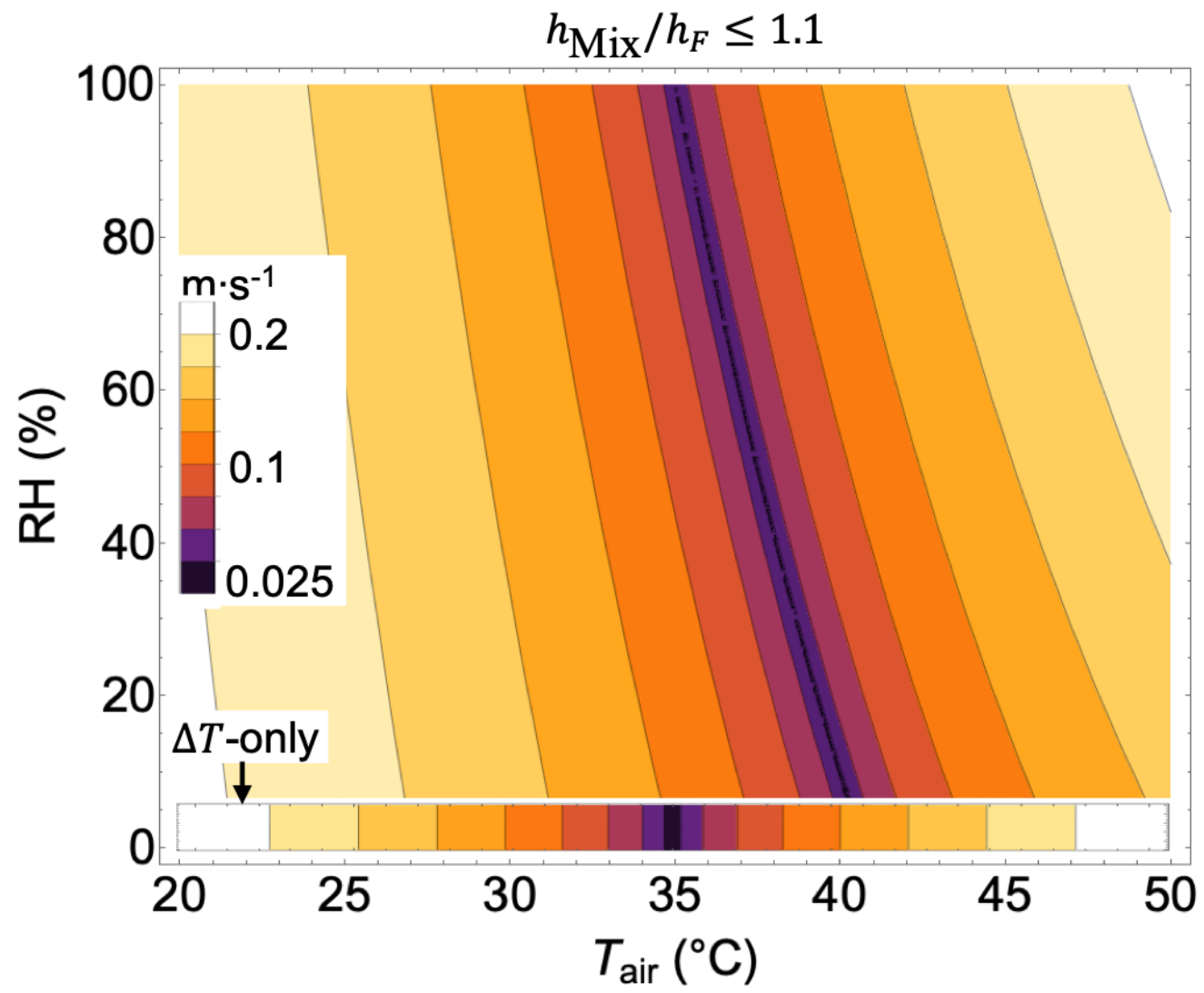

**Fig. S21. Threshold air speed for forced convection ($h_F$) surpassing mixed convection ($h_{Mix}$)** calculated using criterion of $h_{Mix} \leq 1.1\, h_F$ free convection being driven by temperature-only (line at the bottom of the plot) and additionally by humidity.

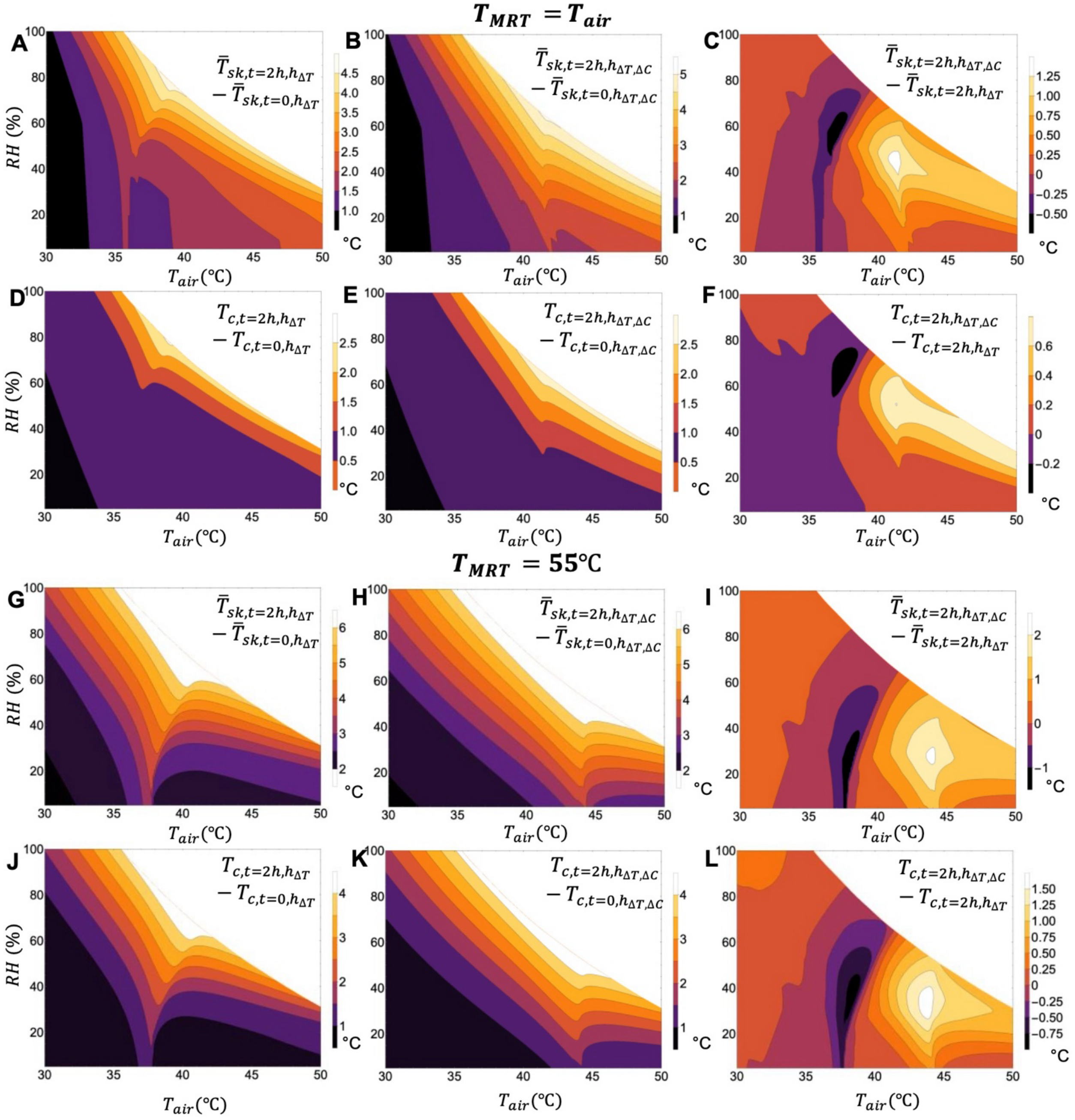

**Fig. S22. Impact of free convection model on mean skin ($\overline{\Delta T}_{sk}$) and core ($T_c$) temperature predictions after resting (1.5 MET) for two hours (subscript *t=2h*) in fully shady (A-F, mean radiant temperature, $T_{mrt} = T_{air}$) and partially-shaded (G-L, $T_{mrt} = 55$℃) fully stagnant conditions.** 1st column (**A**, **D**, **G**, and **J**): buoyancy driven by temperature only (subscript $h_{\Delta T}$), 2nd column (**B**, **E**, **H**, and **K**): buoyancy driven by humidity and temperature (subscript $h_{\Delta T,\Delta C}$), and 3rd column (**C**, **F**, **I**, and **L**) the difference between second and first column. The core and mean skin temperatures are reduced by initial values of 37.03 °C and 34.48 °C at *t=0*, respectively. The values are cutoff at survivability limits determined by Vanos et al.(*16*).

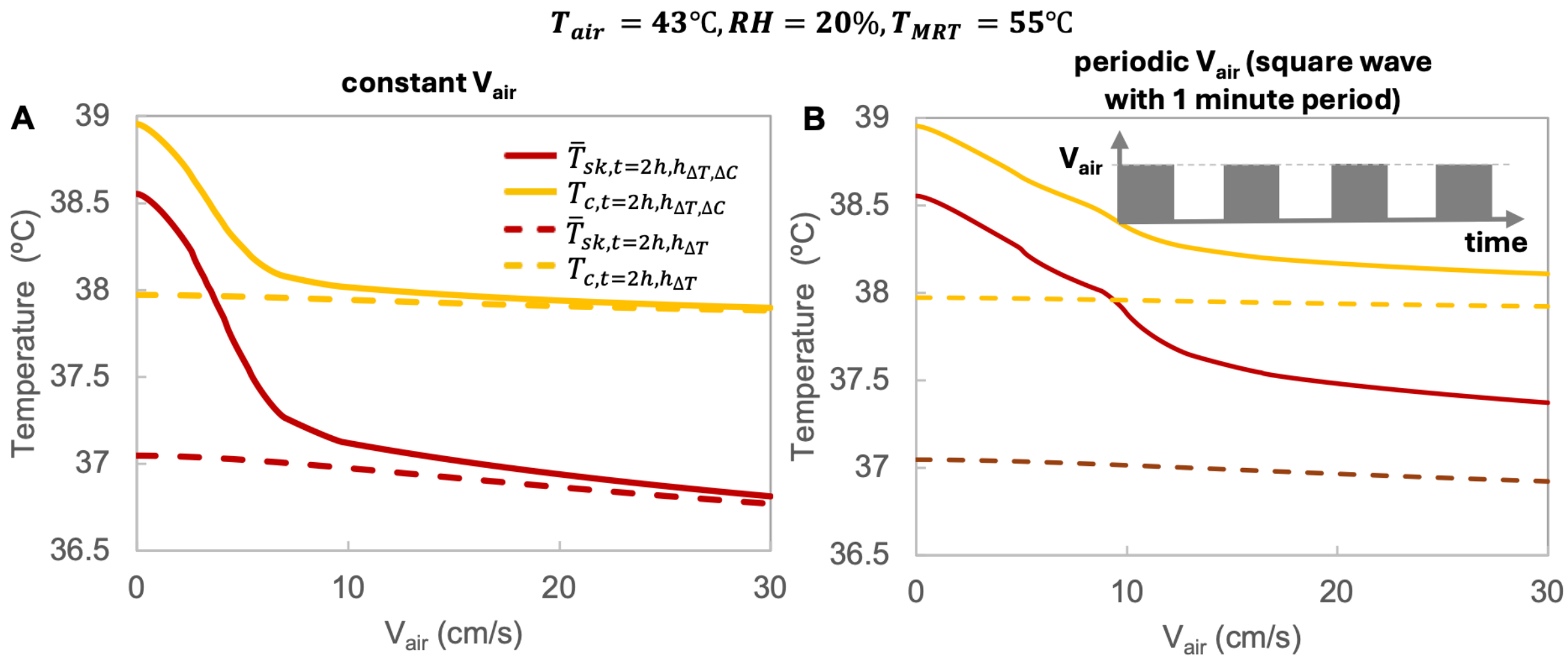

**Fig. S23. Impact mixed convection and air speed ($V_{air}$)** on mean skin ($\overline{\Delta T}_{sk}$) and core ($T_c$) temperature predictions after resting (1.5 MET) for two hours (subscript *t=2h*) in partially-shaded ($T_{mrt} = 55℃$) conditions: **(A)** constant $V_{air}$ and **(B)** periodic $V_{air}$ (a square wave altering between 0 and $V_{air}$ with period of 1 minute).

**Table S1. Detailed specifications of the instruments used to measure the inlet air temperature ($T_{air}$), relative humidity ($RH_{air}$), and air speed ($v_{air}$) within the enclosure.**

| Sensor | Manufacturer | Variable (s) Measured | Quantity | Uncertainty | Frequency (Hz) |
|---|---|---|---|---|---|
| EE181 | Campbell Scientific, Inc. | $T_{air}$ & $RH_{air}$ | 1 | ± 0.2 °C<br>± (1.4 + 0.01 RH%) | 1 |
| Senso Anemo5100 LSF Transducer | Sensor Electronic | $v_{air}$ | 5 | ± 0.02 $m \cdot s^{-1}$ | 8 |

**Table S2. Detailed specifications of the sensors used to measure air temperature ($T_{air}$), wall temperatures ($T_{wall}$), relative humidity ($RH_{air}$), and air velocity ($v_{air}$) inside the enclosure. The manufacturer specified the uncertainties of the 3D sonic anemometer in absolute terms and the 2D sonic wind sensor in relative terms.**

| **Sensor** | **Manufacturer** | **Variable (s) Measured** | **Quantity** | **Uncertainty** | **Frequency (Hz)** |
|---|---|---|---|---|---|
| CSAT3B | Campbell Scientific, Inc. | $v_{air}$ | 1 | ± 0.08 m·s$^{-1}$ ($U_x$, $U_y$)<br>± 0.04 m·s$^{-1}$ ($U_z$) | 1 |
| WINDSONIC1-L | Campbell Scientific, Inc. | $v_{air}$ | 1 | ± 2% | 1 |
| EE181 | Campbell Scientific, Inc. | $T_{air}$ & $RH_{air}$ | 1 | ± 0.2 °C<br>± (1.4 + 0.01 RH) % | 0.33 |
| Type-T Thermocouple | Minnesota Measurement Instruments LLC | $T_{wall}$ | 2 | ± 0.5 °C | 1 |

**Table S3. Summary of enclosure's environmental and manikin skin conditions for each experimental case. $T_c$ & $RH_c$ – air temperature and relative humidity setpoints of the climatic chamber, $T_{air}$ & $RH_{air}$ – air temperature and relative humidity measured within the enclosure (Fig. S1.B–C), $T_{sk}$ – manikin skin temperature, $\Delta T_{air-sk}$ – temperature difference between $T_{air}$ & $T_{sk}$**

| Cases | $T_c$ (°C) | $RH_c$ (%) | $T_{air}$ (°C) | $RH_{air}$ (%) | $T_{sk}$ (°C) | Skin wettedness | $\Delta T_{air-sk}$ (°C) |
|---|---|---|---|---|---|---|---|
| **Temperature-only driven free convection (Dry free convection and radiation only)** | | | | | | | |
| 1 | 40 | 24.6 | 40.4 ± 0.2 | 27.6 ± 1.9 | 34 | 0 | 6.4 ± 0.2 |
| 2 | 42 | 22.2 | 42.0 ± 0.2 | 25.4 ± 2.3 | 34 | 0 | 8.0 ± 0.2 |
| 3 | 44 | 20 | 44.2 ± 0.2 | 23.5 ± 1.6 | 34 | 0 | 10.2 ± 0.2 |
| 4 | 44 | 20 | 43.8 ± 0.2 | 23.8 ± 1.6 | 31 | 0 | 12.8 ± 0.2 |
| **Humidity-only driven sweat evaporation only** | | | | | | | |
| 5 | 35 | 27 | 35.3 ± 0.2 | 35.6 ± 1.5 | 35.0 ± 0.2 | 1 | 0.3 ± 0.1 |
| 6 | 35 | 44 | 35.3 ± 0.2 | 48.7 ± 1.5 | 35.0 ± 0.1 | 1 | 0.3 ± 0.1 |
| 7 | 35 | 60 | 35.5 ± 0.2 | 65.2 ± 1.5 | 35.4 ± 0.1 | 1 | 0.1 ± 0.1 |
| 8 | 35 | 80 | 35.5 ± 0.2 | 81.4 ± 1.5 | 35.4 ± 0.1 | 1 | 0.1 ± 0.1 |
| **Combined temperature and humidity effects (Free convection, radiation and sweat evaporation)** | | | | | | | |
| 9 | 40 | 19 | 39.6 ± 0.2 | 38.3 ± 1.6 | 35.0 | 1 | 4.6 ± 0.2 |
| 10 | 40 | 34 | 39.6 ± 0.2 | 48.3 ± 1.5 | 35.0 | 1 | 4.6 ± 0.2 |
| 11 | 40 | 45 | 39.5 ± 0.2 | 60.3 ± 1.9 | 35.0 | 1 | 4.5 ± 0.2 |
| 12 | 40 | 61 | 39.6 ± 0.2 | 68.9 ± 1.6 | 35.0 | 1 | 4.6 ± 0.2 |
| 13 | 44 | 22 | 42.4 ± 0.2 | 34.7 ± 1.7 | 35.0 | 1 | 7.4 ± 0.2 |
| 14 | 44 | 34 | 42.5 ± 0.2 | 47.1 ± 2.0 | 35.0 | 1 | 7.5 ± 0.2 |
| 15 | 44 | 49 | 42.4 ± 0.2 | 59.4 ± 2.0 | 35.0 | 1 | 7.4 ± 0.2 |

**Table S4. Summary of the behavior of temperature- and humidity-induced buoyancies across air temperature regimes.** $T_{sk}$ **(°C) and** $T_{air}$ **(°C) are the skin and air temperatures respectively,** $C_{sk}$ **(kg·m$^{-3}$) and** $C_{air}$ **(kg·m$^{-3}$) are the corresponding water-vapor concentrations,** $\Delta T_{sk-air}$ **(°C) and** $\Delta C_{sk-air}$ **(kg·m$^{-3}$) are the skin-to-air temperature and concentration differences, respectively.** $B_{\Delta T} = g\rho\beta_T\Delta T_{sk-air}$ **and** $B_{\Delta C} = g\rho\beta_C\Delta C_{sk-air}$**, are the thermal- and concentration-driven buoyancies respectively, where** $g$ **is the gravitational acceleration (9.81 m·s$^{-2}$),** $\beta_T$ **is the thermal coefficient of expansion (reciprocal of the absolute average of** $T_{air}$ **and** $T_{sk}$**), and** $\beta_C$ **is the coefficient of expansion due to compositional changes. Net buoyancy** $B_{net}$ **is the sum of** $B_{\Delta C}$ **and** $B_{\Delta T}$**. Stagnation temperature** $T_{stag}$ **is the air temperature at which** $B_{\Delta C}$ **and** $B_{\Delta T}$ **cancel out, producing a near stagnant flow for a particular** $RH_{air}$

| Regime | $\Delta T_{sk-air}$ | $\Delta C_{sk-air}$ | $B_{\Delta T}$ | $B_{\Delta C}$ | Dominant force | $B_{net}$ | Plume Direction |
|---|---|---|---|---|---|---|---|
| $T_{air} < T_{sk}$ | $> 0$ | $> 0$ | $> 0$ | $> 0$ | $B_{\Delta T}$ | $B_{\Delta C} + B_{\Delta T}$ | Up |
| $T_{sk} < T_{air} < T_{stag}$ | $< 0$ | $> 0$ | $< 0$ | $> 0$ | $B_{\Delta C}$ | $B_{\Delta C} - B_{\Delta T}$ | Up |
| $T_{sk} < T_{air} = T_{stag}$ | $< 0$ | $> 0$ | $< 0$ | $> 0$ | $B_{\Delta T} \approx B_{\Delta C}$ | $\sim 0$ | Nearly stagnant |
| $T_{air} > T_{stag} > T_{sk}$ | $< 0$ | $> 0$ | $< 0$ | $> 0$ | $B_{\Delta T}$ | $B_{\Delta C} - B_{\Delta T}$ | Down |

**Table S5. Air properties as function of temperature ($k$-thermal conductivity, $\upsilon$-kinematic viscosity, $\alpha$-thermal diffusivity, $\beta_{\Delta T}$-volumetric coefficient of thermal expansion, and $D$-diffusion coefficient)**

| $T_{air}$ (°C) | $k$(W·m$^{-1}$·K$^{-1}$) | $\upsilon$ (m$^2$·s$^{-1}$) | $\alpha$(m$^2$·s$^{-1}$) | $\beta_{\Delta T}$ (1/K) | $D$ (m$^2$·s$^{-1}$) |
|---|---|---|---|---|---|
| 20 | 0.0257787 | 1.5243E-05 | 2.1553E-05 | 0.00341122 | 2.4191E-05 |
| 25 | 0.0261598 | 1.5714E-05 | 2.2243E-05 | 0.00335402 | 2.5054E-05 |
| 30 | 0.0265378 | 1.6191E-05 | 2.2941E-05 | 0.0032987 | 2.5932E-05 |
| 35 | 0.026913 | 1.6674E-05 | 2.3646E-05 | 0.00324517 | 2.6826E-05 |
| 40 | 0.0272859 | 1.7162E-05 | 2.4361E-05 | 0.00319336 | 2.7736E-05 |
| 45 | 0.0276568 | 1.7655E-05 | 2.5083E-05 | 0.00314317 | 2.8662E-05 |
| 50 | 0.0280261 | 1.8154E-05 | 2.5815E-05 | 0.00309454 | 2.9603E-05 |

**Table S6. Air density (kg·m$^{-3}$) as function of temperature and relative humidity.**

| T (°C) | RH (%) | | | | | | | | | | |
|---|---|---|---|---|---|---|---|---|---|---|---|
| | **0** | **10** | **20** | **30** | **40** | **50** | **60** | **70** | **80** | **90** | **100** |
| **20** | 1.20 | 1.20 | 1.20 | 1.20 | 1.20 | 1.20 | 1.20 | 1.20 | 1.20 | 1.19 | 1.19 |
| **25** | 1.18 | 1.18 | 1.18 | 1.18 | 1.17 | 1.17 | 1.17 | 1.17 | 1.17 | 1.17 | 1.17 |
| **30** | 1.16 | 1.15 | 1.15 | 1.15 | 1.15 | 1.15 | 1.15 | 1.15 | 1.15 | 1.15 | 1.15 |
| **35** | 1.13 | 1.13 | 1.13 | 1.13 | 1.13 | 1.13 | 1.13 | 1.13 | 1.12 | 1.12 | 1.12 |
| **40** | 1.11 | 1.11 | 1.11 | 1.11 | 1.11 | 1.10 | 1.10 | 1.10 | 1.10 | 1.10 | 1.10 |
| **45** | 1.09 | 1.09 | 1.09 | 1.08 | 1.08 | 1.08 | 1.08 | 1.08 | 1.07 | 1.07 | 1.07 |
| **50** | 1.07 | 1.06 | 1.06 | 1.06 | 1.06 | 1.05 | 1.05 | 1.05 | 1.05 | 1.05 | 1.04 |

**Table S7. Air coefficient of expansion due to compositional changes (m$^3$·kg$^{-3}$) a function of temperature and relative humidity.**

| T (°C) | RH (%) | | | | | | | | | | |
|---|---|---|---|---|---|---|---|---|---|---|---|
| | **0** | **10** | **20** | **30** | **40** | **50** | **60** | **70** | **80** | **90** | **100** |
| **20** | 0.50 | 0.50 | 0.50 | 0.51 | 0.51 | 0.51 | 0.51 | 0.51 | 0.51 | 0.51 | 0.51 |
| **25** | 0.51 | 0.51 | 0.51 | 0.51 | 0.52 | 0.52 | 0.52 | 0.52 | 0.52 | 0.52 | 0.52 |
| **30** | 0.52 | 0.52 | 0.52 | 0.52 | 0.53 | 0.53 | 0.53 | 0.53 | 0.53 | 0.53 | 0.53 |
| **35** | 0.53 | 0.53 | 0.53 | 0.54 | 0.54 | 0.54 | 0.54 | 0.54 | 0.54 | 0.54 | 0.54 |
| **40** | 0.54 | 0.54 | 0.55 | 0.55 | 0.55 | 0.55 | 0.55 | 0.55 | 0.55 | 0.55 | 0.55 |
| **45** | 0.56 | 0.56 | 0.56 | 0.56 | 0.56 | 0.56 | 0.56 | 0.56 | 0.56 | 0.56 | 0.57 |
| **50** | 0.57 | 0.57 | 0.57 | 0.57 | 0.57 | 0.57 | 0.57 | 0.58 | 0.58 | 0.58 | 0.58 |

**Table S8. Simulation results for whole-body heat transfer coefficients ($h_{Sim}$, W·m$^{-2}$·K$^{-1}$) and evaporative heat loss ($Q_{Sim}$, W·m$^{-2}$) across varying air temperature and relative humidity conditions, with skin temperature fixed at 35°C. "Temperature-difference only" free convective coefficients ($h_{\Delta T-only}$, W·m$^{-2}$·K$^{-1}$) and corresponding evaporative heat fluxes ($Q_{\Delta T-only}$, W·m$^{-2}$) ($Q_{\Delta T-only} = 16.5 \times h_{\Delta T-only} \times \Delta p_{sk-air}$).**

| Cases | $T_{air}$ (°C) | $RH_{air}$ (%) | $\Delta T_{sk-air}$ (°C) | $h_{Sim}$ (W·m$^{-2}$·K$^{-1}$) | $Q_{Sim}$ (W·m$^{-2}$) | $h_{\Delta T-only}$ (W·m$^{-2}$·K$^{-1}$) | $Q_{\Delta T-only}$ (W·m$^{-2}$) |
|---|---|---|---|---|---|---|---|
| 1 | 20 | 10 | 15.0 | 3.01 | 247.6 | 3.06 | 272.4 |
| 2 | 25 | 10 | 10.0 | 2.82 | 231.7 | 2.71 | 237.5 |
| 3 | 30 | 10 | 5.0 | 2.57 | 210.6 | 2.2 | 189 |
| 4 | 34 | 10 | 1.0 | 2.32 | 188.7 | 1.36 | 114.2 |
| 5 | 36 | 10 | -1.0 | 2.15 | 174.1 | 1.36 | 112.8 |
| 6 | 38 | 10 | -3.0 | 1.92 | 154.7 | 1.87 | 153 |
| 7 | 40 | 10 | -5.0 | 1.54 | 124.3 | 2.17 | 174.8 |
| 8 | 41 | 10 | -6.0 | 1.01 | 82.4 | 2.29 | 182.8 |
| 9 | 43 | 10 | -8.0 | 1.63 | 127.9 | 2.49 | 195.1 |
| 10 | 45 | 10 | -10.0 | 1.95 | 151.4 | 2.65 | 204 |
| 11 | 50 | 10 | -15.0 | 2.35 | 174.4 | 2.98 | 215.9 |
| 12 | 20 | 20 | 15.0 | 3 | 235.5 | 3.06 | 260.6 |
| 13 | 25 | 20 | 10.0 | 2.8 | 215.9 | 2.71 | 223.3 |
| 14 | 30 | 20 | 5.0 | 2.55 | 191.4 | 2.2 | 173.6 |
| 15 | 34 | 20 | 1.0 | 2.27 | 165.2 | 1.36 | 102.3 |
| 16 | 36 | 20 | -1.0 | 2.08 | 148.3 | 1.36 | 99.5 |
| 17 | 38 | 20 | -3.0 | 1.8 | 126 | 1.87 | 132.6 |
| 18 | 39 | 20 | -4.0 | 1.58 | 110.1 | 2.03 | 141.7 |
| 19 | 40 | 20 | -5.0 | 1.12 | 77 | 2.17 | 148.4 |
| 20 | 41 | 20 | -6.0 | 1.38 | 92 | 2.29 | 153.4 |
| 21 | 45 | 20 | -10.0 | 2.07 | 128.8 | 2.65 | 162.1 |
| 22 | 50 | 20 | -15.0 | 2.44 | 133.2 | 2.98 | 155.2 |
| 23 | 20 | 30 | 15.0 | 3 | 223.5 | 3.06 | 248.8 |
| 24 | 25 | 30 | 10.0 | 2.79 | 200.6 | 2.71 | 209.1 |
| 25 | 30 | 30 | 5.0 | 2.52 | 171.6 | 2.2 | 158.1 |
| 26 | 34 | 30 | 1.0 | 2.22 | 142.4 | 1.36 | 90.3 |
| 27 | 36 | 30 | -1.0 | 2 | 123.6 | 1.36 | 86.2 |
| 28 | 38 | 30 | -3.0 | 1.65 | 97.9 | 1.87 | 112.2 |
| 29 | 39 | 30 | -4.0 | 1.26 | 73.2 | 2.03 | 118.2 |
| 30 | 40 | 30 | -5.0 | 1.32 | 74 | 2.17 | 122 |
| 31 | 41 | 30 | -6.0 | 1.67 | 90.8 | 2.29 | 124.1 |
| 32 | 45 | 30 | -10.0 | 2.19 | 102.3 | 2.65 | 120.2 |
| 33 | 50 | 30 | -15.0 | 2.54 | 88.4 | 2.98 | 94.4 |
| 34 | 20 | 40 | 15.0 | 2.99 | 211.6 | 3.06 | 237 |
| 35 | 25 | 40 | 10.0 | 2.77 | 185.3 | 2.71 | 195 |
| 36 | 30 | 40 | 5.0 | 2.5 | 153 | 2.2 | 142.7 |
| 37 | 34 | 40 | 1.0 | 2.17 | 120.7 | 1.36 | 78.4 |
| 38 | 36 | 40 | -1.0 | 1.91 | 99.8 | 1.36 | 72.8 |
| 39 | 37 | 40 | -2.0 | 1.72 | 86.9 | 1.66 | 85.4 |

**Table S9. Simulation results for whole-body heat transfer coefficients ($h_{Sim}$, W·m$^{-2}$·K$^{-1}$) and evaporative heat loss ($Q_{Sim}$, W·m$^{-2}$) across varying air temperature and relative humidity conditions, with skin temperature fixed at 35°C. "Temperature-difference only" free convective coefficients ($h_{\Delta T-only}$, W·m$^{-2}$·K$^{-1}$) and corresponding evaporative heat fluxes ($Q_{\Delta T-only}$, W·m$^{-2}$) ($Q_{\Delta T-only} = 16.5 \times h_{\Delta T-only} \times \Delta p_{sk-air}$).**

| Cases | $T_{air}$ (°C) | $RH_{air}$ (%) | $\Delta T_{sk-air}$ (°C) | $h_{Sim}$ (W·m$^{-2}$·K$^{-1}$) | $Q_{Sim}$ (W·m$^{-2}$) | $h_{\Delta T-only}$ (W·m$^{-2}$·K$^{-1}$) | $Q_{\Delta T-only}$ (W·m$^{-2}$) |
|---|---|---|---|---|---|---|---|
| 40 | 38.5 | 40 | -3.5 | 1.04 | 49.5 | 1.96 | 93.6 |
| 41 | 40 | 40 | -5.0 | 1.63 | 72.3 | 2.17 | 95.6 |
| 42 | 45 | 40 | -10.0 | 2.28 | 71.7 | 2.65 | 78.3 |
| 43 | 50 | 40 | -15.0 | 2.62 | 40 | 2.98 | 33.6 |
| 44 | 20 | 50 | 15.0 | 2.98 | 199.7 | 3.06 | 225.1 |
| 45 | 25 | 50 | 10.0 | 2.76 | 170.1 | 2.71 | 180.8 |
| 46 | 30 | 50 | 5.0 | 2.47 | 134.5 | 2.2 | 127.3 |
| 47 | 34 | 50 | 1.0 | 2.11 | 99.4 | 1.36 | 66.5 |
| 48 | 36 | 50 | -1.0 | 1.8 | 76.8 | 1.36 | 59.5 |
| 49 | 37 | 50 | -2.0 | 1.54 | 62.5 | 1.66 | 68.2 |
| 50 | 38 | 50 | -3.0 | 0.89 | 33.3 | 1.87 | 71.3 |
| 51 | 39 | 50 | -4.0 | 1.58 | 55.6 | 2.03 | 71.3 |
| 52 | 40 | 50 | -5.0 | 1.81 | 58.6 | 2.17 | 69.2 |
| 53 | 41 | 50 | -6.0 | 1.96 | 57.9 | 2.29 | 65.4 |
| 54 | 45 | 50 | -10.0 | 2.38 | 38.3 | 2.65 | 36.4 |
| 55 | 48 | 50 | -13.0 | 2.57 | 11.5 | 2.86 | 1.9 |
| 56 | 20 | 60 | 15.0 | 2.98 | 188 | 3.06 | 213.3 |
| 57 | 25 | 60 | 10.0 | 2.74 | 155.1 | 2.71 | 166.6 |
| 58 | 30 | 60 | 5.0 | 2.44 | 116 | 2.2 | 111.9 |
| 59 | 34 | 60 | 1.0 | 2.04 | 78.4 | 1.36 | 54.5 |
| 60 | 36 | 60 | -1.0 | 1.66 | 55.1 | 1.36 | 46.2 |
| 61 | 37 | 60 | -2.0 | 1.2 | 36.5 | 1.66 | 51 |
| 62 | 38 | 60 | -3.0 | 1.46 | 40 | 1.87 | 50.8 |
| 63 | 40 | 60 | -5.0 | 1.94 | 39.9 | 2.17 | 42.8 |
| 64 | 41 | 60 | -6.0 | 2.07 | 35.2 | 2.29 | 36.1 |
| 65 | 44 | 60 | -9.0 | 2.37 | 12.3 | 2.57 | 6.9 |
| 66 | 20 | 70 | 15.0 | 2.97 | 176.1 | 3.06 | 201.5 |
| 67 | 25 | 70 | 10.0 | 2.73 | 140.3 | 2.71 | 152.4 |
| 68 | 30 | 70 | 5.0 | 2.4 | 98 | 2.2 | 96.4 |
| 69 | 34 | 70 | 1.0 | 1.97 | 59.2 | 1.36 | 42.6 |
| 70 | 36 | 70 | -1.0 | 1.46 | 34.6 | 1.36 | 32.8 |
| 71 | 36.5 | 70 | -1.5 | 1.1 | 24.1 | 1.53 | 34 |
| 72 | 38 | 70 | -3.0 | 1.68 | 28.1 | 1.87 | 30.4 |
| 73 | 39 | 70 | -4.0 | 1.89 | 24.3 | 2.03 | 24.4 |
| 74 | 40 | 70 | -5.0 | 2.04 | 18 | 2.17 | 16.5 |
| 75 | 41 | 70 | -6.0 | 2.16 | 9.9 | 2.29 | 6.7 |
| 76 | 20 | 80 | 15.0 | 2.96 | 164.5 | 3.06 | 189.7 |
| 77 | 25 | 80 | 10.0 | 2.71 | 125.7 | 2.71 | 138.2 |
| 78 | 30 | 80 | 5.0 | 2.37 | 80.5 | 2.2 | 81 |

**Table S10. Simulation results for whole-body heat transfer coefficients ($h_{Sim}$, W·m$^{-2}$·K$^{-1}$) and evaporative heat loss ($Q_{Sim}$, W·m$^{-2}$) across varying air temperature and relative humidity conditions, with skin temperature fixed at 35°C. "Temperature-difference only" free convective coefficients ($h_{\Delta T-only}$, W·m$^{-2}$·K$^{-1}$) and corresponding evaporative heat fluxes ($Q_{\Delta T-only}$, W·m$^{-2}$) ($Q_{\Delta T-only} = 16.5 \times h_{\Delta T-only} \times \Delta p_{sk-air}$).**

| Cases | $T_{air}$ (°C) | $RH_{air}$ (%) | $\Delta T_{sk-air}$ (°C) | $h_{Sim}$ (W·m$^{-2}$·K$^{-1}$) | $Q_{Sim}$ (W·m$^{-2}$) | $h_{\Delta T-only}$ (W·m$^{-2}$·K$^{-1}$) | $Q_{\Delta T-only}$ (W·m$^{-2}$) |
|---|---|---|---|---|---|---|---|
| 79 | 34 | 80 | 1.0 | 1.87 | 40.2 | 1.36 | 30.7 |
| 80 | 36 | 80 | -1.0 | 0.91 | 12.8 | 1.36 | 19.5 |
| 81 | 37 | 80 | -2.0 | 1.57 | 16.1 | 1.66 | 16.5 |
| 82 | 38 | 80 | -3.0 | 1.83 | 11.1 | 1.87 | 10 |
| 83 | 39 | 80 | -4.0 | 2 | 3.3 | 2.03 | 1 |
| 84 | 20 | 90 | 15.0 | 2.94 | 152.4 | 3.06 | 177.8 |
| 85 | 25 | 90 | 10.0 | 2.7 | 111.2 | 2.71 | 124.1 |
| 86 | 30 | 90 | 5.0 | 2.33 | 63.3 | 2.2 | 65.6 |
| 87 | 34 | 90 | 1.0 | 1.76 | 22.8 | 1.36 | 18.7 |
| 88 | 35.5 | 90 | -0.5 | 0.71 | 4.9 | 1.11 | 7.7 |
| 89 | 36 | 90 | -1.0 | 1.39 | 6.6 | 1.36 | 6.2 |
| 90 | 36.5 | 90 | -1.5 | 1.6 | 4 | 1.53 | 3.2 |

**Mathworks Matlab thermoregulation code with setting replicating Fig.4C&D**

```
%%---------------2026 version -------------------
%%This thermoregulation model is based on J.A.J Stolwijk's model published
%%in 1971. The changes made to the original model are described by Joshi, A.,
Twidwell, B., Park, M., & Rykaczewski, K. (2025).
% Comparative analysis of thermoregulation models to assess heat strain in moderate to
extreme heat. Journal of Thermal Biology, 127, 104035.
% Additional changes including humidity induced buoyancy and mixed
% convection model are described in the main manuscript by Viswanathan et
% al. 2026
%% Material properties and other parameters were same as per stolwijk's-1971 model but
following things has been modified
%%(1) Implementation of clothing including non-uniform distribution
%%(2) 2026 Updated heat transfer coefficient values as per de Dear 1997 and
%%Rykaczewski research group findings
%%(3) Updated set-point temperatures as per Takahashi-2021 (JOS3 model)
%%(4) Updated control coefficients as per Takahashi-2021 (JOS3 model)
%%(5) Updated the blood flow

%%KR--10-19-2026--corrections added (removal of double Kelvin and
%%index variable in the Csat calculations pointed out by Prof. Yihang Lu

clc %clear the command lines
clear %clear the variables

%%---------------------------------------------------
%%--------------Input parameters----------------------
%%---------------------------------------------------

%Environmental and other parameter defined within the time/transient loop
dTime = 1;%Time step size [s] (smaller the better)
start_time = 0;%Start time in [S]
end_time = 120*60;%End time in [s] or total duration in [s]
outputInterval =60;%Output of data at given interval [s]

weight = 74.43;%kgL_c
height = 1.81;%m
age = 30;%years
%%---------------------------------------------------
%%--------------Clothing input----------------------
%%---------------------------------------------------
Rc_cl = [0, 0.054, 0, 0, 0.054, 0.054];%0.26 clo or 0.04 m2K/2 concentrated into
trunk, legs and feet (*1.34)
Re_cl = [0, 9.4, 0, 0, 9.4, 9.4];%clothing evaporative resistance for individual
segment [m2Pa/W]:Head, Trunk, Arms, Hands, Legs, Feet

%%---------------------------------------------------
%%--------------Other parametrs----------------------
%%---------------------------------------------------
P_atm = 101325;%Atmospheric pressure [Pa]
L_c = 16.5/(P_atm/101325);%Lewis coeff [K/kPa]

%%---------------------------------------------------
%%----------Thermal properties----------------------
%%---------------------------------------------------
%Variable =  [  Head_core,      Head muscle,    Head fat,   Head skin,  Trunk core,
Trunk muscle,  Trunk fat,  Trunk skin,
%               Arms_core,      Arms_muscle,    Arms_fat,   Arms_skin,  Hands_core,
Hands_muscle,  Muscle_fat,  Muscle_skin,
%               Legs_core,      Legs_muscle,    Legs_fat,   Legs_skin,  Feet_core,
Feet_muscle,    Feet_fat,   Feet_skin
```

```
%               Central blood   ];
%Heat Cacitance [J K^-1]
C = [9288.48,     1380.72,    920.48,     1004.16, ...%Head :core, muscle,fat,skin
heat Cacitance [J K^-1]
    41086.88,    67571.6,    17782,      5062.64, ...%Trunk:core, muscle,fat,skin heat
Cacitance [J K^-1]
    5899.44,     12719.36,   2426.72,    1799.12, ...%Arms :core, muscle,fat,skin heat
Cacitance [J K^-1]
    585.76,      251.04,     376.56,     711.28,  ...%Hand :core, muscle,fat,skin heat
Cacitance [J K^-1]
    17740.16,    38367.28,   5983.12,    4518.72, ...%Legs :core, muscle,fat,skin heat
Cacitance [J K^-1]
    962.32,      251.04,     543.92,     920.48,  ...%Feet :core, muscle,fat,skin heat
Cacitance [J K^-1]
    9414.00];%cental blood pool heat Cacitance [J K^-1]
%Basal metabolic heat generation [W]
Q_bm = [14.93, 0.12, 0.13, 0.09, ... %Head : core, muscle, fat, skin basal metabolic
heat generation [W]
    52.78, 5.82, 2.48, 0.47, ... %Trunk: core, muscle, fat, skin basal metabolic heat
generation [W]
    0.81,  1.10, 0.20, 0.15, ... %Arms : core, muscle, fat, skin basal metabolic heat
generation [W]
    0.09,  0.23, 0.03, 0.06, ... %Hand : core, muscle, fat, skin basal metabolic heat
generation [W]
    2.59,  3.33, 0.50, 0.37, ... %Legs : core, muscle, fat, skin basal metabolic heat
generation [W]
    0.15,  0.02, 0.05, 0.08, ... %Feet : core, muscle, fat, skin basal metabolic heat
generation [W]
    0.00]; %central blood pool basal metabolic heat generation [W]
%Basal effective blood flow [l h^-1]
BFB = [45.00,  0.12, 0.13, 1.44, ... %Head : core, muscle, fat, skin basal effective
blood flow [l h^-1]
    210.00, 6.00, 2.56, 2.10, ... %Trunk: core, muscle, fat, skin basal effective
blood flow [l h^-1]
    0.84,   1.14, 0.20, 0.50, ... %Arm  : core, muscle, fat, skin basal effective
blood flow [l h^-1]
    0.10,   0.24, 0.04, 2.00, ... %Hand : core, muscle, fat, skin basal effective
blood flow [l h^-1]
    2.69,   3.43, 0.52, 2.85, ... %Leg  : core, muscle, fat, skin basal effective
blood flow [l h^-1]
    0.16,   0.02, 0.05, 3.00, ... %Feet : core, muscle, fat, skin basal effective
blood flow [l h^-1]
    0.00]; %central blood pool basal effective blood flow [l h^-1]
%Thermal conductance between j and j+1 [W K^-1]
TC = [1.605,  13.258, 16.049, 0.000, ... %Head : core, muscle, fat, skin thermal
conductance between j and j+1 [W K^-1]
    1.593,  5.524,  23.027, 0.000, ... %Trunk: core, muscle, fat, skin thermal
conductance between j and j+1 [W K^-1]
    1.396,  10.351, 30.471, 0.000, ... %Arm  : core, muscle, fat, skin thermal
conductance between j and j+1 [W K^-1]
    6.397,  11.223, 11.514, 0.000, ... %Hand : core, muscle, fat, skin thermal
conductance between j and j+1 [W K^-1]
    10.467, 14.421, 74.432, 0.000, ... %Leg  : core, muscle, fat, skin thermal
conductance between j and j+1 [W K^-1]
    16.282, 20.585, 16.398, 0.000, ... %Feet : core, muscle, fat, skin thermal
conductance between j and j+1 [W K^-1]
    0.000]; %central blood pool thermal conductance between j and j+1 [W K^-1]
%T_set as per JOS3
T_set = [310.61, 308.69, 308.58, 308.35,... %Head :core, muscle,fat,skin set point
temperature of body segments [K]
```

```
         310.18, 309.57, 308.87, 307.93,... %Trunk:core, muscle,fat,skin set point
temperature of body segments [K]
         309.29, 307.86, 307.69, 307.34,... %Arms :core, muscle,fat,skin set point
temperature of body segments [K]
         308.20, 308.17, 307.67, 307.59,... %Hands:core, muscle,fat,skin set point
temperature of body segments [K]
         309.81, 309.29, 308.46, 307.25,... %Legs :core, muscle,fat,skin set point
temperature of body segments [K]
         307.89, 307.78, 307.46, 307.39,... %Feet :core, muscle,fat,skin set point
temperature of body segments [K]
         310.49]; %central blood pool set point temperature of body segments [K]

%%---------------------------------------------------
%%------Model constants & distribution factor------
%%---------------------------------------------------
%variable = [Head, Trunk, Arms, Hands, Legs, Feet];
SA      = [0.133,   0.680, 0.254, 0.095, 0.597, 0.130]; %Surface area of body segments
[m^2]: Head, Trunk, Arms, Hands, Legs, Feet
DF_work = [0.000,   0.300, 0.080, 0.010, 0.600, 0.010]; %Distributuion factor of total
work done by muscles in given segment [-]:Head, Trunk, Arms, Hands, Legs, Feet
DF_TR   = [0.0695,  0.4935,      0.0686,      0.1845,      0.1505,
       0.0334];%Distributuion factor of skin thermoreceptor in given segment [-]:Head,
Trunk, Arms, Hands, Legs, Feet
DF_VD   = [0.1320,  0.3220,      0.0950,      0.1210,      0.2300,
       0.1000];%Distributuion factor of VD in given segment [-]:Head, Trunk, Arms,
Hands, Legs, Feet
DF_VC   = [0.0500,  0.1500,      0.0500,      0.3500,      0.0500,
       0.3500];%Distributuion factor of VC in given segment [-]:Head, Trunk, Arms,
Hands, Legs, Feet
DF_SW   = [0.0810,  0.4810,      0.1540,      0.0310,      0.2180,
       0.0350];%Distributuion factor of SW in given segment [-]:Head, Trunk, Arms,
Hands, Legs, Feet
DF_SH   = [0.0200,  0.8500,      0.0500,      0.0000,      0.0700,
       0.0000];%Distributuion factor of shivering heat generation by muscles in given
segment [-]:Head, Trunk, Arms, Hands, Legs, Feet

%%---------------------------------------------------
%%------------Initialization-----------
%%---------------------------------------------------
numTimeSteps = round((end_time - start_time) / dTime) + 1;%number of time steps
resultIndex = 1;% Counter for result storageH
T=T_set;%starts at set point temperature [degC converted to K for calculations]
timeArray = zeros(1, numTimeSteps);%time [s]

%%---------------------------------------------------
%%--------------------Time loop--------------------
%%---------------------------------------------------
for stepIndex = 1:numTimeSteps
    currentTime = start_time + (stepIndex - 1) * dTime;%Time step for the current loop
[s]
    disp(['Current Time: ', num2str(currentTime), ' s']);%Display the time in console
[s]

     if currentTime >= 0 && currentTime <= end_time
        T_air = 43;  % Air temperature [degC]
        MRT = 55;    % Mean radiant temperature [degC]
        v_air = 0;   % Air velocity [m s^-1]
        RH_air = 0.2; % Air humidity [-]
        met = 1.5;     % Work/metabolic rate [met]
     end
```

```
    %%-------Processing environmental parameters------------------------

    T_air = T_air+273.15;
    MRT = MRT+273.15;

    Psat_air = exp(77.3450 + 0.0057 * (T_air) - 7235 / (T_air)) /
(T_air)^8.2;%Saturation pressure of water vapor in air [Pa]
    Csat_air=Psat_air*18*10^-3/((T_air)*8.314);%saturation concentration of water
vapor [kg/m3]
    P_air = RH_air*Psat_air;%Actual pressure of water vapor in air[Pa]
    C_air=Csat_air*RH_air;%water vapor concentration in air [kg/m3]
    P_atm = 101325;%Atmospheric pressure [Pa]
    L_c = (16.5/(P_atm/101325))/1000;%Lewis coeff [-]
    SA_total=sum(SA);%Total surface area in [m^2]

    %%-------Identifying the temperature at skin node------------------------
    for i = 1:6
        k = 4 * i;%To identify the skin nodes [-]
        T_sk(i) = T(k);%Skin temperature [deg K]
    end


    %%-------------------------------------------------
    %%----------Heat transfer to environment-----------
    %%-------------------------------------------------

    %%------------Heat transfer coefficients-----------
        % Preallocate h_conv for better performance
        h_conv = zeros(1, 6);%mixed or total convection
        h_convf = zeros(1, 6);%forced convection
        h_convn = zeros(1, 6);%free convection
    %%------------Heat transfer coefficients-----------
        % Forced convection
        hc_a = [3.20, 8.1, 11.09, 14.40, 11.10, 12.00]; %[Head, Trunk, Arms, Hands,
Legs, Feet]%%De Dear averages where needed
        hc_b = [0.97,  0.62,  0.59,  0.55, 0.51,   0.49]; %[Head, Trunk, Arms, Hands,
Legs, Feet]
        h_rad = [4.89, 4.26, 4.49, 4.21, 5.08, 6.14]; %Radiation heat transfer
coefficients for given body segments [W m^-2 K^-1]:[Head, Trunk, Arms, Hands, Legs,
Feet]
        for i = 1:length(hc_a)
            h_convf(i) = hc_a(i) * (v_air ^ hc_b(i)); %forced convection heat transfer
coefficient [W m^-2 K^-1]
        end

        %% dT+146dC>0
        hc_d = [0.567, 0.29, 0.937, 0.5, 0.575,1.151]; %[Head, Trunk, Arms, Hands,
Legs, Feet]
        hc_e = [1.036, 0.996, 1.278, 2.4, 1.233, 1.673]; %[Head, Trunk, Arms, Hands,
Legs, Feet]
        %% dT+146dC<0
        hc_f =  [0.567, 0.29, 0.937, 0.5, 0.575,1.151];%[Head, Trunk, Arms, Hands,
Legs, Feet]
        hc_g = [1.4, 1.03, 1.164, 2.4, 0.623, 0.831]; %[Head, Trunk, Arms, Hands,
Legs, Feet]

        Psat_sk2 = zeros(1, length(hc_d));
        Csat_sk2 = zeros(1, length(hc_d));
        for i = 1:length(hc_d)
            Psat_sk2(i) = exp(77.3450 + 0.0057 * (T(i*4)) - 7235    / (T(i*4))) /
(T(i*4))^8.2;%Saturation pressure of water vapor at skin[Pa]
```

```
            Csat_sk2(i) = Psat_sk2(i)*18*10^-3/((T(i*4))*8.314);%saturation
concentration of water vapor at skin[kg/m3]

            %calculate mean dT+146*dC from segments
                Psat_sk3=0;
                Csat_sk3=0;
                dCsat_sk3=0;
                dTsk3=0;
                dT146dC_mean=0;
               for j = 1:length(hc_d)
                     Psat_sk3 = exp(77.3450 + 0.0057 * (T(j*4)) - 7235    / (T(j*4)))
/ (T(j*4))^8.2;%Saturation pressure of water vapor at skin[Pa]
                     Csat_sk3 = Psat_sk3*18*10^-3/((T(j*4))*8.314);
                     dCsat_sk3 = Csat_sk3-C_air;%saturation concentration of water
vapor at skin[kg/m3]
                     dTsk3 =T(j * 4)-T_air;
                     dT146dC_mean=dT146dC_mean+(dTsk3+146*dCsat_sk3)*(1/6);%mean of
Dt+146dC for all 6 segments--use to decide whether flow is up or down
               end

            if dT146dC_mean > 0
            h_convn(i) = hc_d(i) + hc_e(i)*(abs((T(i * 4) - T_air)+146*(Csat_sk2(i)-
C_air))^0.25); %natural convection heat transfer coefficient [W m^-2 K^-1]

            else
            h_convn(i) = hc_f(i) + hc_g(i)*(abs((T(i * 4) - T_air)+146*(Csat_sk2(i)-
C_air))^0.25); %natural convection heat transfer coefficient [W m^-2 K^-1]
            end

            %%addition for mixed convection
            h_conv(i)=(h_convn(i)^3+h_convf(i)^3)^(1/3);
        end

 % To simulate only temperature-induced free convection replace the above
 % with:
 %        hc_a = [3.20, 8.1, 11.09, 14.40, 11.10, 12.00]; %[Head, Trunk, Arms, Hands,
Legs, Feet]%%De Dear 1997 averages where needed
 %        hc_b = [0.97, 0.62, 0.59, 0.55, 0.51, 0.49]; %[Head, Trunk, Arms, Hands,
Legs, Feet]]%%De Dear 1997 averages where needed
 %        h_rad = [4.89, 4.26, 4.49, 4.21, 5.08, 6.14]; %Radiation heat transfer
coefficients for given body segments [W m^-2 K^-1]:[Head, Trunk, Arms, Hands, Legs,
Feet]
 %        hc_aW = [2.6, 1.5, 1.9, 2.7, 2.15, 1.57]; %[Head, Trunk, Arms, Hands, Legs,
Feet] free convection Wissler table 9.16: area weighted averages, hands: Quintella et
al. 2004 average, feet: Adt^0.25 fit to Viswanathan et al. 2026
 %
 %        for i = 1:length(hc_a)
 %            h_convf(i) = hc_a(i) * (v_air ^ hc_b(i)); %forced convection heat
transfer coefficient [W m^-2 K^-1]
 %            h_convn(i) = hc_aW(i) * (abs(T(i * 4) - T_air) ^ 0.25); %natural
convection heat transfer coefficient [W m^-2 K^-1]
 %             %mixed convection 5-4-26
 %            h_conv(i)=(h_convn(i)^3+h_convf(i)^3)^(1/3);
 %        end



    %%------------Heat transfer through clothing-----------
    for i = 1:6
        f_cl(i) = 1.00+1.81*Rc_cl(i);
```

```
        h_total(i) = (Rc_cl(i)+1/(f_cl(i)*(h_conv(i)+h_rad(i))))^-1; % total heat
transfer coefficient [W m^-2 K^-1]
        T_o(i) = (h_conv(i)*T_air+h_rad(i)*MRT)/(h_conv(i)+h_rad(i));%Operative
temperature [deg K]

    end


    %%---------------------------------------------------
    %%----------Control system of human thermoreg.-----
    %%---------------------------------------------------

    %%-----------------ERR signals------------------------------
    WRM = zeros(25, 1); % Initialize warm array [-]
    CLD = zeros(25, 1); % Initialize cold array [-]
    for j = 1:25
        ERR(j) = T(j) - T_set(j);%ERR signal offset from the setpoint temperature
[degC]
        WRM(j) = max(ERR(j), 0);%Warm signal[-]
        CLD(j) = abs(min(ERR(j), 0));%Cold signal[-]
    end
    %%-----------------Integration of skin-thermoreceptor signals---------------------
--------
    for i = 1:6
        k = 4 * i;%To identify the skin nodes [-]
        WRM_skin(i) =  WRM(k) * DF_TR(i);%Weighing based on skin receptors[-]
        CLD_skin(i) =  CLD(k) *(DF_TR(i));%Weighing based on skin receptors[-]
    end
    WRMS = sum(WRM_skin);%Integration of skignal from all the zones [-]
    CLDS = sum(CLD_skin);%Integration of skignal from all the zones [-]

    %%-----------------Skin blood flow-----------------------------
    VD = max((100.5 * ERR(1)) + (6.4 * (WRMS - CLDS)),0);%VD signal JOS3
    VC =max((-10.8 * ERR(1)) + (-10.8 * (WRMS - CLDS)),0);%VC signal JOS3

    for i = 1:6
        k = 4 * i;%To identify the skin nodes [-]
        BF_skin(i) = ((BFB(k) + (DF_VD(i) *  VD)) / (1 + DF_VC(i) * VC))*  (2^ (ERR(k)
/ 6));%Skin blood flow rate [l h^-1]
        BF_skin(i) = min(BF_skin(i), 90);


    end

    %%-----------------Shivering------------------------------
    Shivering = 24.36*-ERR(1)*CLDS;%Shivering signal
    for i = 1:6
        Q_shiv(i) = DF_SH(i)*Shivering;%Heat generation caused by shivering [W]
    end

    %%-----------------Sweating------------------------------
    for i = 1:6
        k = 4 * i;%To identify the skin nodes [-]
        Psat_sk(i) = exp(77.3450 + 0.0057 * (T(k)) - 7235    / (T(k))) /
(T(k))^8.2;%Saturation pressure of water vapor at skin[Pa]
    end
    for i = 1:6
        h_evap(i) = (Re_cl(i)+1/((f_cl(i)*h_conv(i) * L_c)))^-1;%Evaporative mass
transfer coefficient [W m^-2 Pa^-1]
        Q_sweat_max(i) = (h_evap(i)*(Psat_sk(i)-P_air))*SA(i);%Maximum possible
evaporative cooling at skin [W]
```

```
    end
    SW = max((371.2 * ERR(1)) + (33.64 * (WRMS -CLDS)),0);%SW signal [W] JOS3

    for i = 1:6
        k = 4 * i;
        Q_sweat(i) =(DF_SW(i) * SW  * 2^(ERR(k) / 10));%Evaporative cooling at skin 
provided by SW [W]
        Q_sweat(i) = min(Q_sweat(i), Q_sweat_max(i));;%Maximum evaporative cooling at 
skin provided by SW [W]
        Skin_wettedness(i) = 0.06 + (0.94 * (Q_sweat(i) / Q_sweat_max(i)));%Skin 
wettedness [-]
    end

    for i = 1:6
        Q_sweat(i) = Skin_wettedness(i)*h_evap(i)*(Psat_sk(i)-
P_air)*SA(i);%Evaporative cooling at skin provided by SW [W]
    end

    %%------------------------------------------------
    %%----------Heat generation------------------------
    %%------------------------------------------------

    %%-----------------Effective heat generation by physical activity-----------------
------------
    %Q_work is the total heat production as one would determine from the oxygen uptake 
minus the basal metabolic rate and the caloric equivalent of the external work 
performed.
    Q_total = met*58.15*SA_total;%converting the metabolic rate into [W] from [met]
    if Q_total - 86.45 > 0
        Q_work = (Q_total - sum(Q_bm)) * 0.78;%[W]
    else
        Q_work = 0.0;%[W]
    end
    for i = 1:6
        Q_work_seg(i) = Q_work*DF_work(i);%[W]
    end

    %%-----------------Effective heat generation and blood flow at each segment-------
----------------------
    for i = 1:6
        k = (4 * i) - 3;

        Q(k) = Q_bm(k); %Basal metabolic heat of core node of each segment [W]
        BF(k) = BFB(k); %Total effective blood flow in core node of each segment [l 
h^-1]

        Q(k+1) = Q_bm(k+1) + Q_shiv(i) + Q_work_seg(i);%Basal, work and shivering heat 
generation at muscle node of each segment [W]
        BF(k+1) = BFB(k+1) + (Q(k+1) - Q_bm(k+1))/1.163; %Total effective blood flow 
in muscle node of each segment [l h^-1]

        Q(k+2) = Q_bm(k+2);%Basal metabolic heat of fat node of each segment [W]
        BF(k+2) = BFB(k+2);%Total effective blood flow in muscle node of each segment 
[l h^-1]

        Q(k+3) = Q_bm(k+3);%Basal metabolic heat of skin node of each segment [W]
        BF(k+3) = BF_skin(i);%Total effective blood flow in skin node of each segment 
[l h^-1]%%CORRECTED

        Q_TC(k)     = TC(k) * (T(k) - T(k+1));%Conductive heat transfer from core to 
muscle [W]
```

```
        Q_TC(k+1)   = TC(k+1) * (T(k+1) - T(k+2));%Conductive heat transfer from
muscle to fat [W]
        Q_TC(k+2)   = TC(k+2) * (T(k+2) - T(k+3));%Conductive heat transfer from fat
to skin [W]
        Q_TC(k+3)   = 0;
    end

    %%------------------Effective heat transfer between central blood node and each
segment (through artery and veins)-----------------------------
        for j = 1:24
            Q_bf(j) = 1.163*BF(j)*(T(j)-T(25));%Heat transfer by blood flow [W]
        end

    %%-----------------Net heat at each node-----------------------------
    for i = 1:6
        k=(4*i)-3;
        Q(k) =  Q(k) - Q_bf(k)  - Q_TC(k);%Net heat gain/loss at core [W]
        Q(k+1) =  Q(k+1) - Q_bf(k+1) + Q_TC(k) - Q_TC(k+1);%Net heat gain/loss at
muscle [W]
        Q(k+2) =  Q(k+2) - Q_bf(k+2) + Q_TC(k+1) - Q_TC(k+2);%Net heat gain/loss at
fat [W]
        Q(k+3) =  Q(k+3) - Q_bf(k+3) + Q_TC(k+2) - Q_sweat(i)-((h_total(i)*(T(k+3)-
T_o(i)))*SA(i));%Net heat gain/loss at skin [W]
    end
    %%-----------------Sensible and latent heat loss from skin (for post processing)---
--------------------------
    for i = 1:6
        k=(4*i);
        SHL_sk(i) = ((h_total(i)*(T(k)-T_o(i)))*SA(i));

        %%%KR ADDITION%%%%
        SHCONV_sk(i) = ((h_conv(i)*(T(k)-T_air))*SA(i));
        SHRAD_sk(i) = ((h_rad(i)*(T(k)-MRT))*SA(i));
        %%%%%%%%%%%

        LHL_sk(i) = Q_sweat(i);
        THL_sk(i) = SHL_sk(i)+LHL_sk(i);

    end

    res_sh = 0.0014  * ((34+273.15) - T_air);%Respiratory sensible heat loss [W]
    res_lh = 0.0173  * (5.87 - (P_air/1000));%Respiratory latent heat loss [W]
    RES = (res_sh+res_lh)*Q_total;%Total respiratory heat loss [W]
    Q(5) = Q(5)-RES;%Net heat at core node of trunk [W]
    Q(25) = sum(Q_bf);%Net heat at central blood pool [W] - JOS3

    %%-----------------Updated temperature based on net heat balance and Cacitance-----
------------------------
    for j = 1:25
        T_new(j) = T(j)+((Q(j)/C(j))*dTime);%Updated temperature based on net heat
balance and Cacitance [degC]
    end

    %%-----------------Post processing to calculate mean skin temp, skin wettedness,
and total evaporative coolong-----------------------------
    for i = 1:6
        k = i*4;
        T_ms(i) = (T_new(k)*(SA(i)/SA_total));%mean skin temperature [deg K]
        To_mean(i) = (T_o(i)*(SA(i)/SA_total));%mean operating temperature [deg K]
        w_sk(i) = (Skin_wettedness(i)*(SA(i)/SA_total));%mean skin wettedness [-]
    end
```

```
    T_ms = sum(T_ms);%%mean skin temperature [deg K]
    To_mean = sum(To_mean);%mean operating temperature [deg K]
    w_sk = sum(w_sk);%mean skin wettedness [-]

    Qsk_latent_total = sum(LHL_sk);%Total latent heat loss from skin [W]
    Qsk_sensible_total = sum(SHL_sk);%Total sensible heat loss from skin [W]

    Qconv=sum(SHCONV_sk);%W
    Qrad=sum(SHRAD_sk);%W

    CO = sum(BF);%Cardiac output [l/hr]
    Q_storage = (Q_total-0.22*(Q_total-sum(Q_bm)))-sum(LHL_sk)-sum(SHL_sk)-RES;%Heat
storage [W]

    %%-----------------Data output setting------------------------------
    if mod(currentTime, outputInterval) == 0
        % Store results
        TimeStamp(resultIndex) = currentTime;%[s]

        %%-----------------Physiological parameters------------------------------
        T_output(resultIndex, :) = T_new-273.15;%Temperature at each node [degC]
        w_sk_output(resultIndex, :) = w_sk;%Mean skin wettedness [-]
        Q_storage_output(resultIndex, :) =Q_storage;%Heat storage [W]
        T_ms_output(resultIndex, :) = T_ms-273.15;%Mean skin temperature[degC]
        CO_output(resultIndex, :) = CO;%Cardiac output [l/hr]
        Skin_wettedness_output(resultIndex, :) = Skin_wettedness;%Skin wettedness of
individual segment [-]
        BF_output(resultIndex, :) = BF;%blodd flow to each node [l/hr]


        %%-----------------Heat exchange with environment------------------------------
        Qsk_latent_total_output(resultIndex, :) = Qsk_latent_total;%Total latent heat
loss[W]
        Qsk_sensible_total_output(resultIndex, :) = Qsk_sensible_total;%Total sensible
heat loss[W]
        SHL_sk_output(resultIndex, :) = SHL_sk;%Sensible heat loss for individual body
segment [W]
        LHL_sk_output(resultIndex, :) = LHL_sk;%Latent heat loss for individual body
segment [W]
        THL_sk_output(resultIndex, :) = THL_sk;%Total heat loss for individual body
segment [W]
        RES_output(resultIndex, :) = RES;%Respiratory heat loss[W]

         %%%%%%%%%%%%%%%%KR OUTPUT%%%%%%%%%%%%%
         Qconv_output(resultIndex,:) = Qconv;
         Qrad_output(resultIndex,:) = Qrad;
         Qgen(resultIndex,:) = met*SA_total*58;
         Buoyancy50(resultIndex,:) = dT146dC_mean*50;

        %%-----------------Envionmental parameters------------------------------
        Tair_output(resultIndex,:) = T_air-273.15;%[degC]
        MRT_output(resultIndex,:) = MRT-273.15;%[degC]
        RH_air_output(resultIndex,:) = RH_air;%[degC]
        To_output(resultIndex,:)=T_o-273.15;%[degC]
        To_mean_output(resultIndex,:)=To_mean-273.15;%[degC]
        v_air_output(resultIndex,:) = v_air;%[degC]
        met_output(resultIndex,:) = met;%[met]

        % Increment the result counter
        resultIndex = resultIndex + 1;
    end
```

```
    %%----------------Update the temperature for next loop----------------------------
-
    T = T_new;%[degC]
end

% To calculate the net heat storage
NetQ_storage_output = zeros(size(Q_storage_output));
for i = 1:length(Q_storage_output)
    NetQ_storage_output(i) = sum(Q_storage_output(1:i));
end
%%--------------------------------------------------------------------------------
%%----------------Writing output to the excel file-------------------------------
%%--------------------------------------------------------------------------------

TimeStamp = (TimeStamp/60)';
sheet1Table = table(TimeStamp, T_ms_output, T_output(:, 5), w_sk_output, CO_output,
...
    'VariableNames', {'TimeStamp','Mean skin temperature (°C)','Core temperature
(°C)', 'Skin wettedness (mean) (-)', 'CO_output (l/h)'});
sheet2Table = table(TimeStamp, Tair_output, MRT_output, To_mean_output, RH_air_output,
v_air_output, met_output, ...
    'VariableNames', {'TimeStamp', 'Air temperature (°C)', 'Mean radiant temperature
(°C)', 'Operating temperature (mean) (°C)', 'Relative humidity (-)',  'Air speed
(m/s)', 'Metabolic rate (met)'});
%%KR edit to include qcon, qrad, qgen (W), 10*buoyancy (dT+164*dC)
sheet3Table = table(TimeStamp, Qsk_sensible_total_output, Qsk_latent_total_output,
RES_output, Q_storage_output, Qconv_output, Qrad_output, Qgen,
Buoyancy50,'VariableNames', {'TimeStamp', 'Sensible heat loss from skin (W)', 'Latent
heat loss from skin (W)', 'Respiratory heat loss (W)', 'Total Heat storage
(W)','Convective heat rate (W)','Radiative heat rate (W)','Metabolic Rate
(W)','50*(dT+164dC)'});

sheet4Table = table(TimeStamp, T_output, ...
    'VariableNames', {'TimeStamp', 'Temperature (°C)'});
sheet5Table = table(TimeStamp, Skin_wettedness_output, ...
    'VariableNames', {'TimeStamp', 'Skin_wettedness (-)'});
sheet6Table = table(TimeStamp, SHL_sk_output, ...
    'VariableNames', {'TimeStamp', 'Sensible heat loss skin (W)'});
sheet7Table = table(TimeStamp, LHL_sk_output, ...
    'VariableNames', {'TimeStamp', 'Latent heat loss skin (W)'});
sheet8Table = table(TimeStamp, BF_output, ...
    'VariableNames', {'TimeStamp', 'Blood flow (l/hr)'});

sheets = {sheet1Table, sheet2Table, sheet3Table, sheet4Table, sheet5Table,
sheet6Table, sheet7Table, sheet8Table};
sheetNames = {'Summary', 'Ambient TCition', 'Heat exchange', 'Local temperature',
'Local skin wettedness','Local sensible heat loss','Local latent heat loss','Local
blood flow'};

%Write the tables to an Excel file
filename = 'dtdc.xlsx';
for i = 1:numel(sheets)
    writetable(sheets{i}, filename, 'Sheet', sheetNames{i});
end

%%--------------------------------------------------------------------------------
%%----------------Plot physiological parameters-------------------------------
%%--------------------------------------------------------------------------------
fig = figure('Units', 'inches', 'Position', [0, 0, 6, 4]);
```

```matlab
% Plot Core temperature
plot(TimeStamp, T_output(:, 5), 'LineWidth', 2, 'DisplayName', 'Core Temperature');
hold on;  % Hold the current axes to add more plots
plot(TimeStamp, T_ms_output, 'LineWidth', 2, 'DisplayName', 'Mean Skin Temperature');
hold off;  % Release the hold to prevent further additions to the same plot
xlabel('Time (min)');
ylabel('Temperature (°C)');
set(gca, 'FontSize', 12, 'FontName', 'Calibri');
legend('show');  % Show legend with dataset labels
ylim([34 39])

% Adjust layout
sgtitle('Physiological Measurements', 'FontSize', 14, 'FontName', 'Calibri');


%%-----------------------------------------------------------------------------
%%----------------Plot exterior fluxes -----------------------------
%%-----------------------------------------------------------------------------
fig = figure('Units', 'inches', 'Position', [0, 0, 6, 4]);

% Plot evaporation, convection, radiation, and heat generation

plot(TimeStamp, -Qsk_latent_total_output, 'LineWidth', 2, 'DisplayName', 'Evaporation 
(W)');
hold on;  % Hold the current axes to add more plots
plot(TimeStamp, -Qrad_output, 'LineWidth', 2, 'DisplayName', 'Radiation (W)');
plot(TimeStamp,-Qconv_output, 'LineWidth', 2, 'DisplayName', 'Convection (W)');
plot(TimeStamp, Qgen, 'LineWidth', 2, 'DisplayName', 'Heat Generation (W)');
plot(TimeStamp, Q_storage_output, 'LineWidth', 2, 'DisplayName', 'Heat Storage (W)');
plot(TimeStamp, Buoyancy50, 'LineWidth', 2, 'DisplayName', '(dT+146dC)*50');
hold off;  % Release the hold to prevent further additions to the same plot
xlabel('Time (min)');
ylabel('Heat rate (W)');
set(gca, 'FontSize', 12, 'FontName', 'Calibri');
legend('show');  % Show legend with dataset labels
ylim([-400 400])

% Adjust layout
sgtitle('Heat rates', 'FontSize', 14, 'FontName', 'Calibri');
```